\documentclass{ws-ijgmmp}
\usepackage{graphicx,amssymb,amsmath,amsfonts}
\usepackage{subfigure}
\usepackage{color}
\begin{document}

\markboth{Parthapratim Pradhan}{(Circular Orbits in the Taub-NUT and mass-less Taub-NUT Space-time)}

\title{Circular Orbits in the Taub-NUT and mass-less Taub-NUT Space-time }
\author{Parthapratim Pradhan\footnote{pppradhan77@gmail.com.}}

\address{Department of Physics, Hiralal Mazumdar Memorial College For Women,
Dakshineswar, Kolkata-700035,  India.}

\maketitle

\begin{history}
\received{Day Month Year}
\revised{Day Month Year}
\comby{Managing Editor}
\end{history}

\begin{abstract} 
In this work we study the equatorial causal geodesics of the Taub-NUT (TN) space-time in comparison
with \emph{mass-less} TN space-time. We emphasized both on the null circular geodesics and time-like 
circular geodesics.  From the effective potential diagram of null and time-like geodesics, we differentiate
the geodesics structure between TN spacetime and mass-less TN space-time. It has been shown that there is a 
key role of the NUT parameter to changes the shape of pattern of the potential well in the NUT spacetime in 
comparison with \emph{mass-less} NUT space-time.  We compared  the ISCO (innermost stable circular orbit), 
MBCO (marginally bound circular orbit) and CPO (circular photon orbit) of the said space-time with 
graphically in comparison with mass-less cases. Moreover, we compute the radius of ISCO, MBCO and CPO for 
\emph{extreme} TN black hole (BH). Interestingly, we show that these \emph{three radii} coincides with the Killing horizon i.e. 
the null geodesic generators of the horizon. Finally in Appendix-A, we compute the center-of-mass (CM) energy 
for TN BH and mass-less TN BH.  We show that in both cases the CM energy is finite. For \emph{extreme} NUT BH, we 
have found that the diverging nature of CM energy.  \emph{First}, we have observed that a non-asymptotic flat, 
spherically symmetric and stationary extreme BH is showing such feature.
\end{abstract}
\maketitle

\section{Introduction}
The Taub-NUT (Newman, Unti and Tamburino) space-time \cite{taub,nut} is a stationary, spherically symmetric and 
non-asymptotically flat solution of the vacuum Einstein equation in general theory 
of relativity. The space-time has topology $\Re \times S^3$ with Lorentzian signature \cite{mt}. 
The NUT space-time is described by two parameters: one is the mass parameter $M$ and 
another one is the NUT parameter $n$. There is  
no modification required in the Einstein-Hilbert action to accommodate the NUT charge 
\cite{zee} or ``dual mass'' \cite{sen81,sen86} or ``gravito-magnetic mass'' or ``gravito-magnetic monopole''
\cite{dr}. This dual mass is an intrinsic feature of general theory of relativity. The space-time contains 
closed time-like curve and null lines. It is a geodetically incomplete 
space-time \cite{mt}. Bonor \cite{bn69} has given a new interpretation of the NUT spacetime and it
describes `the field of a spherically symmetric mass together with a semi-infinite massless 
source of angular momentum along the axis of symmetry'. On the other hand, Manko and Ruiz 
\cite {manko05} analyzed the mass and angular momentum distributions in case of generalized NUT spacetime using 
Komar integral approach.

't Hooft and Polykov \cite{gh,ap} have demonstrated that the magnetic monopole present in 
certain non-Abelian  gauge theories. Zee \cite{zee} observed that there is an existence of a gravitational analog of 
Dirac's magnetic monopole \cite{dirac}. The author is also discussed regarding the mass quantization following 
the idea of Dirac quantization rule. He also claimed that there is certainly no experimental evidence of mass 
quantization. Moreover, he proposed that if mass is quantized there may have 
profound consequences in physics. For example, if a magnetic monopole moving around a nucleus then 
the mass quantization rule suggests that the binding energy of every level in the nucleus is also 
quantized. Friedman and Sorkin \cite{fs} observed that the gravito-pole may exist in topological 
solution. Dowker \cite{dk74} proposed that the NUT spacetime as a `gravitational dyon'.

The Euclidean version of the space-time is closely related to the dynamics of BPS (Bogomol'nyi-Prasad-Sommerfield) 
monopoles \cite{gm}. The experimental evidence of this dual mass has not been verified till now. There may be a possibilities 
of experimental evidences in near future and it was first proposed by Lynden-Bell and Nouri-Zonoz \cite{bell} in 1998. Letelier 
and Vieira \cite{lv} have observed that the manifestation of chaos for test particles moving in a TN space-time 
perturbed by dipolar halo using Poincare sections. The geodesics structure in Euclidean TN space-time has been 
studied in the Ref. \cite{vv}. 

The gravito-magnetic lensing effect in NUT space-time was first studied by Nouri-Zonoz et al. \cite{bell97} in 1997. 
They proved that all the geodesics in NUT spacetime confined to a cone with the opening angle $\delta$ defined by 
\begin{eqnarray}
 sin \delta &=& \frac{2n}{D \sqrt{1+\frac{4n^2}{D^2}}}
\end{eqnarray}
where $D=\frac{L}{E}$ is the impact factor \footnote{This parameter is defined in Eq. (\ref{n5}) for null circular 
geodesics.}. For small $\alpha$ and in the limit $\frac{2n}{D}<<1$, it should be 
\begin{eqnarray}
 \alpha \cong \frac{2n}{D}
\end{eqnarray}
It should also be noted that the opening angle is proportioal to the NUT parameter $n$.

Furthermore, they also examined the lensing of light rays passing through the NUT deflector. This properties 
modified the observed shape, size and orientation of a source. It has been also studied there that there is an extra shear due 
to the presence of the gravito-magnetic monopole, which changes the 
shape of the source. The same author also studied the electromagnetic waves in NUT space through the 
solutions of the Maxwell equations via Newman-Penrose null tetrad formalism to further deeper 
insight of the physical aspects of the dual mass. Since the TN space-time has  gravito-magnetic 
monopole that changes the structure of the accretion disk and might offer novel observational 
\footnote{It should be noted that due to many pathological properties of the TN spacetime such as closed time-like curves, one 
may think that it is not a physical solution to the Einstein field equations therefore in this sense the TN spacetime 
is unlikely to be physical.} prospects \cite{liu,chur}.

The maximal analytic extension or Kruskal like extension of the TN space-time shows 
that it has some unusual properties \cite{mkg}.  Maximal analytic extension is needed in order to understand 
the global properties of the space-time. Misner and Taub have shown that TN space is maximally analytic i.e. it has
no Hausdorff extension \cite{mt}. Whereas Hajicek \cite{ph} showed that the non-Hausdorff property occurs only on the 
Killing horizons and causes no geodesics to bifurcate.

Chakraborty and Majumdar \cite{cp} have derived the exact Lense-Thirrring precession (inertial frame dragging effect) in 
case of the TN space-time in comparison with the mass-less TN space-time. The \emph{mass-less 
dual mass} (i.e. TN space-time with $M=0$)  concept was first introduced by Ramaswamy and 
Sen \cite{sen81}. They also proved that `in the absence of gravitational radiation magnetic mass 
requires either that the metric be singular on a two dimensional world sheet or the space-time 
contain closed time-like lines, violating causality'. After that Ashtekar and Sen \cite{as} demonstrated 
that the consequences of magnetic mass in quantum gravity. They also proved that the dual mass implies 
the existence of `wire singularities' in certain potentials for Weyl curvature. Finally, Mueller and Perry 
\cite{mp} showed that the `mass quantization' rule regarding the NUT space-time.

In \cite{ghz}, the author has been studied $SU(2)$ time-dependent tensorial perturbations of Lorentzian TN 
space-time and proved that Lorentzian TN space-time is unstable.  Geodesics of accelerated circular orbits 
on the equatorial plane has been studied in detail of the NUT space using Frenet-Serret procedure \cite{bcmj}.

However, in the present work we wish to investigate  the complete geodesic structure of the TN space-time
in the equatorial plane. We compare the circular geodesics in the TN space-time with mass-less 
TN space-time and zero NUT parameter by \emph{analyzing the effective potential graphically  for
both null cases and time-like cases}. The presence of the dual mass can changes the geodesic structure in 
comparison with the mass-less dual mass and zero dual mass. This is clearly manifested in the effective 
potential diagram. We also differentiate \emph{graphically} the ISCO, MBCO and CPO  of the said space-time in comparison with 
the mass-less cases. Moreover, we examine the circular geodesics in the $L-r$ plane for different values of 
energy i.e. $E^2>1$, $E^2<1$ and $E^2=1$ in case of TN and mass-less TN spacetime, and plotted graphically. Furthermore, we have 
studied more exotic cases i.e. extreme cases, where we find surprising results that the radii of three important class of orbits 
namely, the radius of ISCO ($r_{isco}$), the radius of MBCO ($r_{mbco}$) and the the radius of CPO ($r_{cpo}$) are coincident with 
the horizon. From the best of my knowledge, this is the first we have observed in this work for any spherically symmetric, 
stationary and \emph{non-asymptotically} flat spacetime showing such feature.  Finally, we  compute the CM energy for these 
spacetimes and we have found that for non-extreme TN BH, the CM energy is finite whereas for \emph{extreme} TN 
BH the CM energy is diverging in nature.

The circular geodesics studied earlier by several authors \cite{chur,dr} but they have not been studied in more 
graphically. We here show the differences between the two spacetime with the 
mass parameter and  the mass-less parameter in visually.
It may be noted that circular orbits of arbitrary radii are not possible, there exists a minimum
radius below which no circular orbits are possible. The geodesic structure has been studied earlier for Schwarzschild 
BH \cite{sch}, Reissner Nordstr{\o}m (RN) BH \cite{sch}, Kerr-Taub-NUT (KTN) BH \cite{chur}  and Kerr-Newman-Taub-NUT (KNTN) 
more recently \cite{cqg}. In this work,  we have specialized on the cases when the 
parameter $a=Q=0$. By studying the geodesic structure, we can extract more information about the  back ground space-time. Different 
observables like Lense-Thirrring effect, gravitational time delay, gravitational bending of light etc. all are the 
phenomenon related to the geodesic structure of the space-time. This is one of the 
major motivation behind to study them. 

The structure of the manuscript is as follows. In section \ref{nut}, we discuss the basics  of TN BH. In 
section \ref{egtn}, we  study the equatorial geodesic properties of the said BH. Section \ref{ex} devoted to study the 
extreme TN BH.  The conclusions are given in the section \ref{dis}.

\section{\label{nut} The TN Space-time:}
The metric is given by \cite{taub,mkg,mis,miller}
\begin{eqnarray}
ds^2 &=& -{\cal H}(r) \, \left(dt+2n\cos\theta d\phi\right)^2+ \frac{dr^2}{{\cal H}(r)}+\left(r^2+n^2\right) \left(d\theta^2
+\sin^2\theta d\phi^2 \right) ~,\label{tn1}\\
{\cal H}(r) &=& 1-\frac{2(M r+n^2)}{r^2+n^2} ~.
\end{eqnarray}
where, $M$ denotes the gravito-electric mass or ADM mass  and $n$ denotes
the gravito-magnetic mass or dual mass or magnetic mass of the space-time. It is clearly 
evident that there are two types of singularities are  present in the metric (\ref{tn1}). 
One is at ${\cal H}(r)=0$ which give us the Killing horizons or BH horizons:
\begin{eqnarray}
r_{\pm} &=& M\pm\sqrt{M^2+n^2}\,\, \mbox{and}\,\,  r_{+}> r_{-} ~,\label{tn2}
\end{eqnarray}
$r_{+}$ is called event horizon  and $r_{-}$  is called Cauchy horizon. 

\begin{figure}
\begin{center}
{\includegraphics[width=0.45\textwidth]{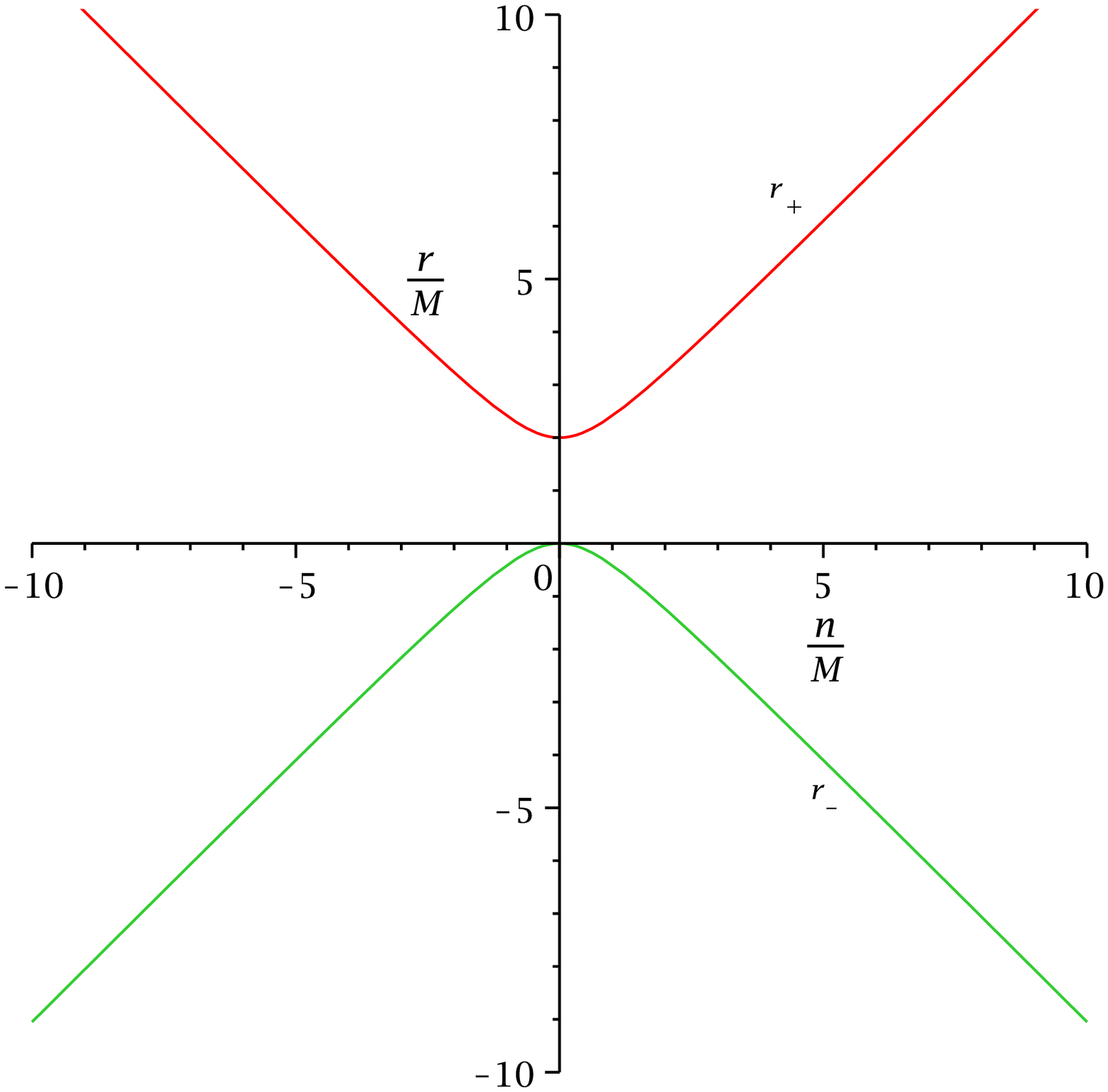}}
{\includegraphics[width=0.45\textwidth]{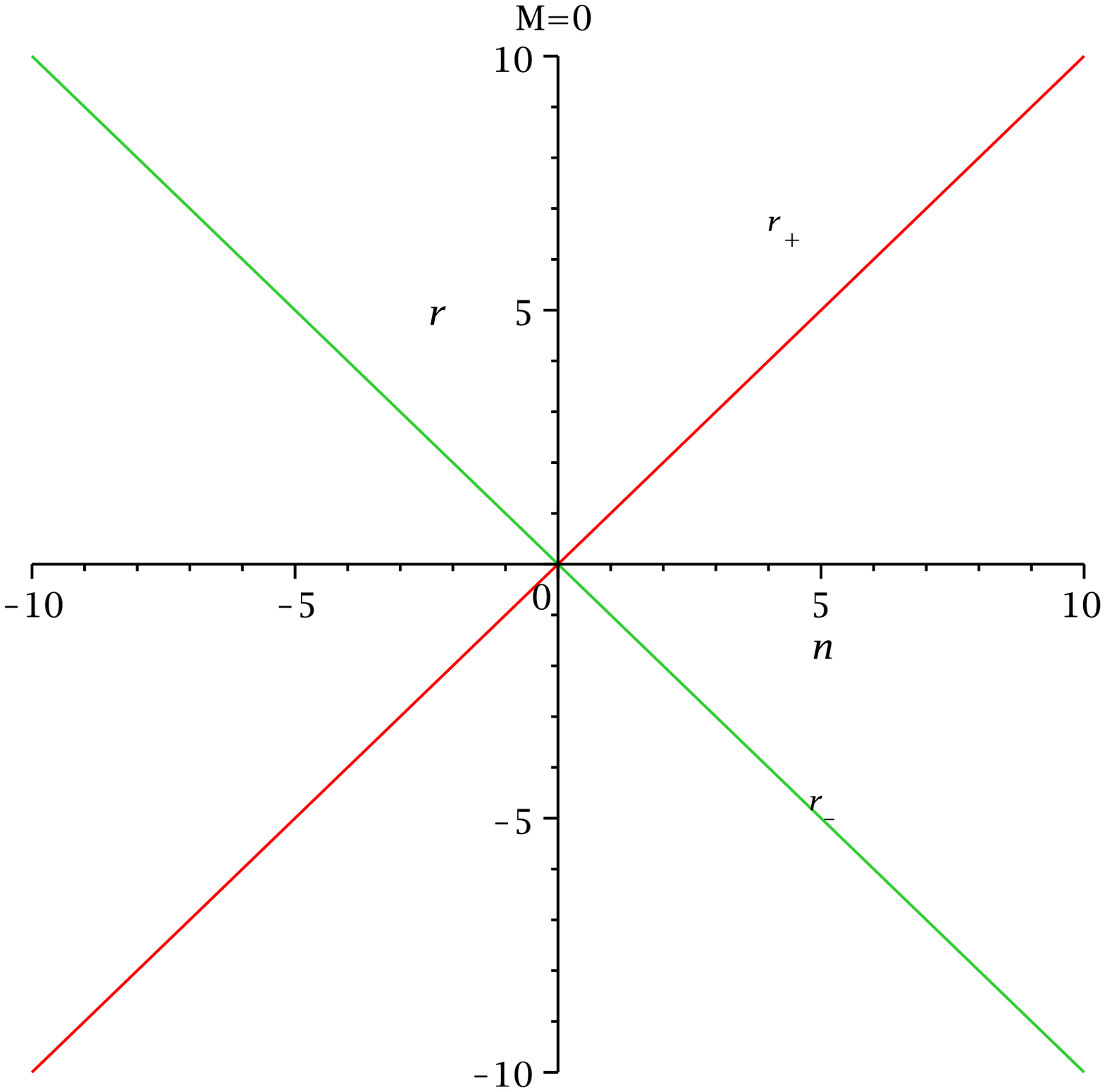}}
\end{center}
\caption{The figure depicts the horizon structure of TN and mass-less TN black hole.
\label{eq1}}
\end{figure}

From Fig. \ref{eq1}, we can see the horizon structure of TN and mass-less TN BH. There is a 
qualitative difference between two horizon structure with NUT parameter and without NUT 
parameter. 
In the limit $n=0$, one obtains the Schwarzschild BH. Interestingly, when $M=0$, we get 
mass-less TN space-time and the horizons at 
\begin{eqnarray}
r_{\pm} &=& \pm n\,\,\mbox{and}\,\,  r_{+}> r_{-} ~.\label{tn3}
\end{eqnarray}
The other type of singularity occurs at $\theta=0$ and $\theta=\pi$,  where the determinant of the metric component 
vanishes. Misner\cite{mis} first demonstrated  that in order to remove the apparent singularities at $\theta=0$ and 
$\theta=\pi$ , $t$ must be identified modulo $8\pi n$. Provided that $r^2+n^2 \neq 2(Mr+n^2)$. 
It should be noticed here that the NUT parameter actually  measures deviation from the asymptotic flatness 
at infinity which may be manifested in the off-diagonal components of the metric and this is happening due to
presence of the Dirac-Misner type of singularity. 

When, 
\begin{eqnarray}
M^2 + n^2 \geq 0 ~.\label{ineq}
\end{eqnarray}
the TN metric describes a BH, otherwise it has a naked  singularity. When $M^2 +n^2=0$, we find  extreme TN BH.

\begin{figure}
\begin{center}
{\includegraphics[width=0.45\textwidth]{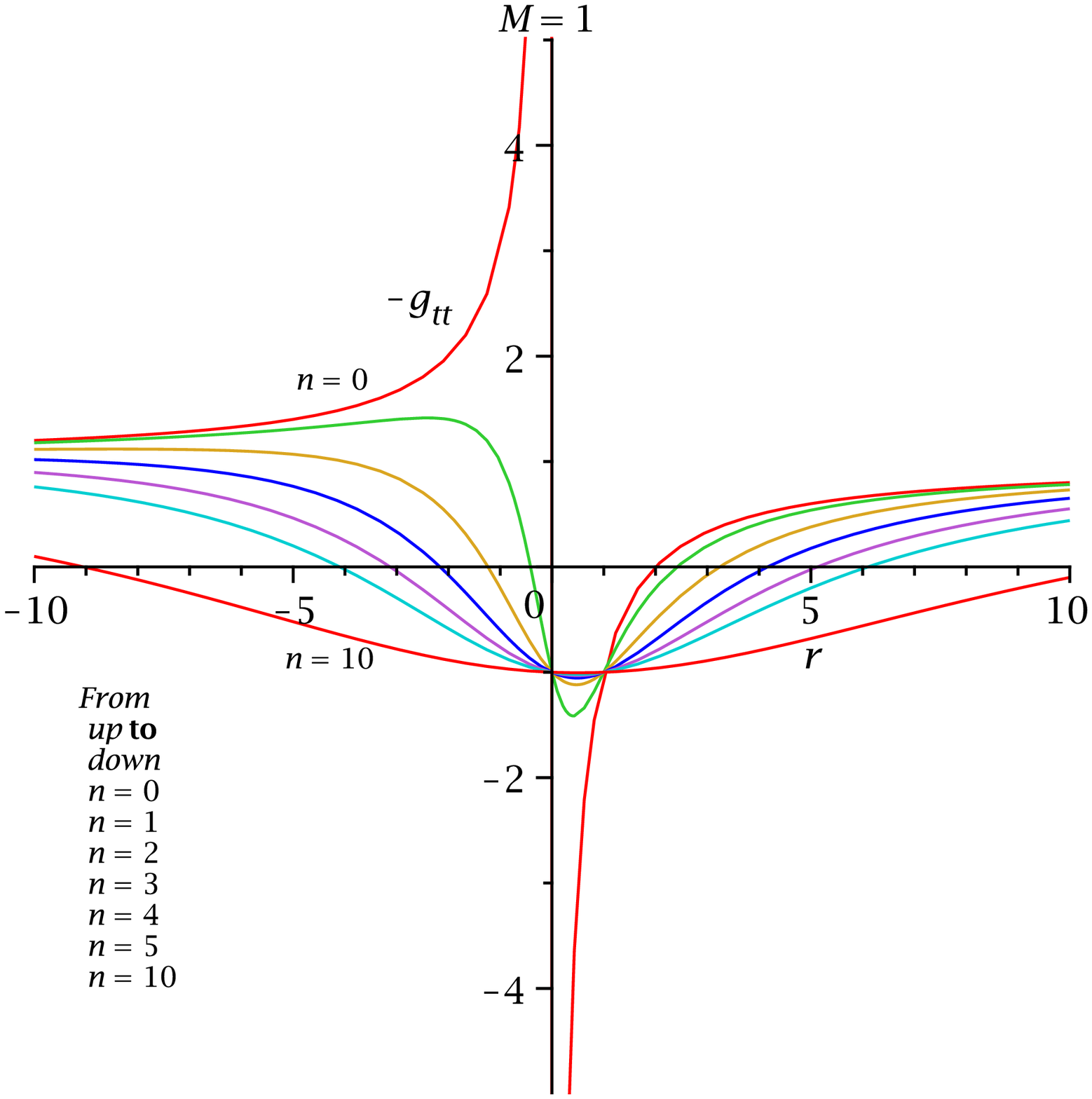}}
{\includegraphics[width=0.45\textwidth]{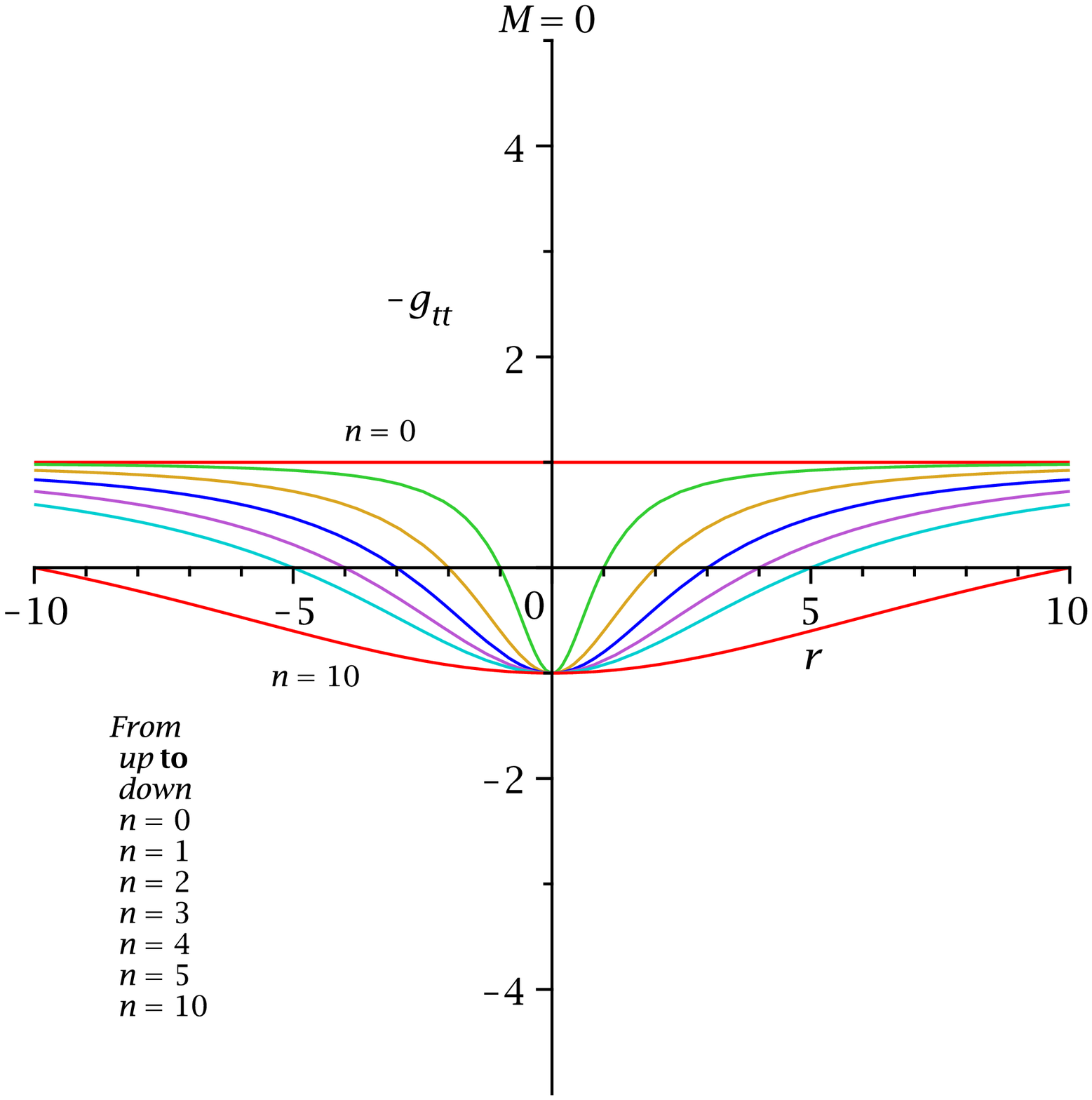}}
\end{center}
\caption{The figure shows the variation  of $-g_{tt}$  with $r$ for TN and  massless TN BH.
\label{eq2}}
\end{figure}

\begin{figure}
\begin{center}
{\includegraphics[width=0.45\textwidth]{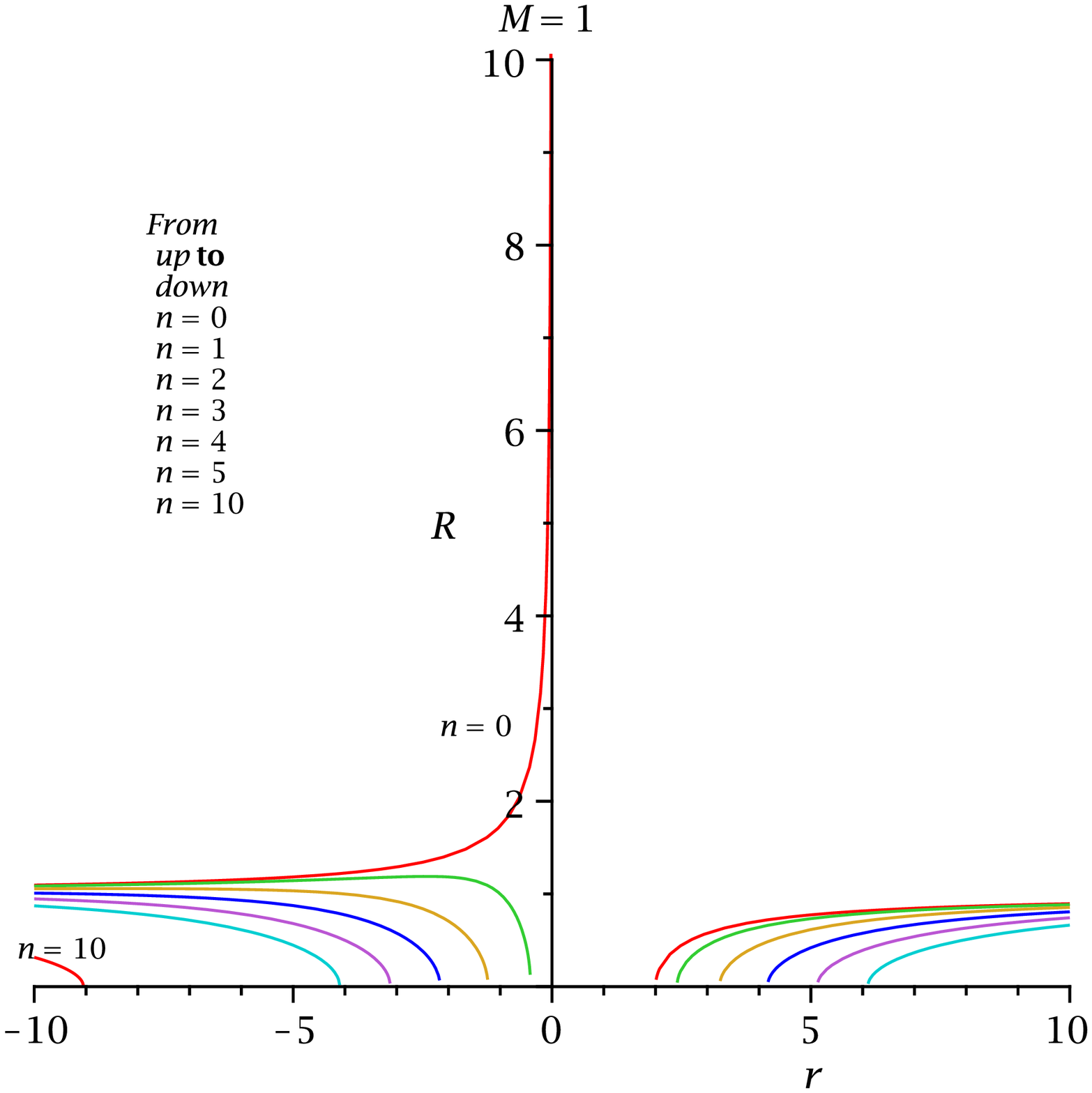}}
{\includegraphics[width=0.45\textwidth]{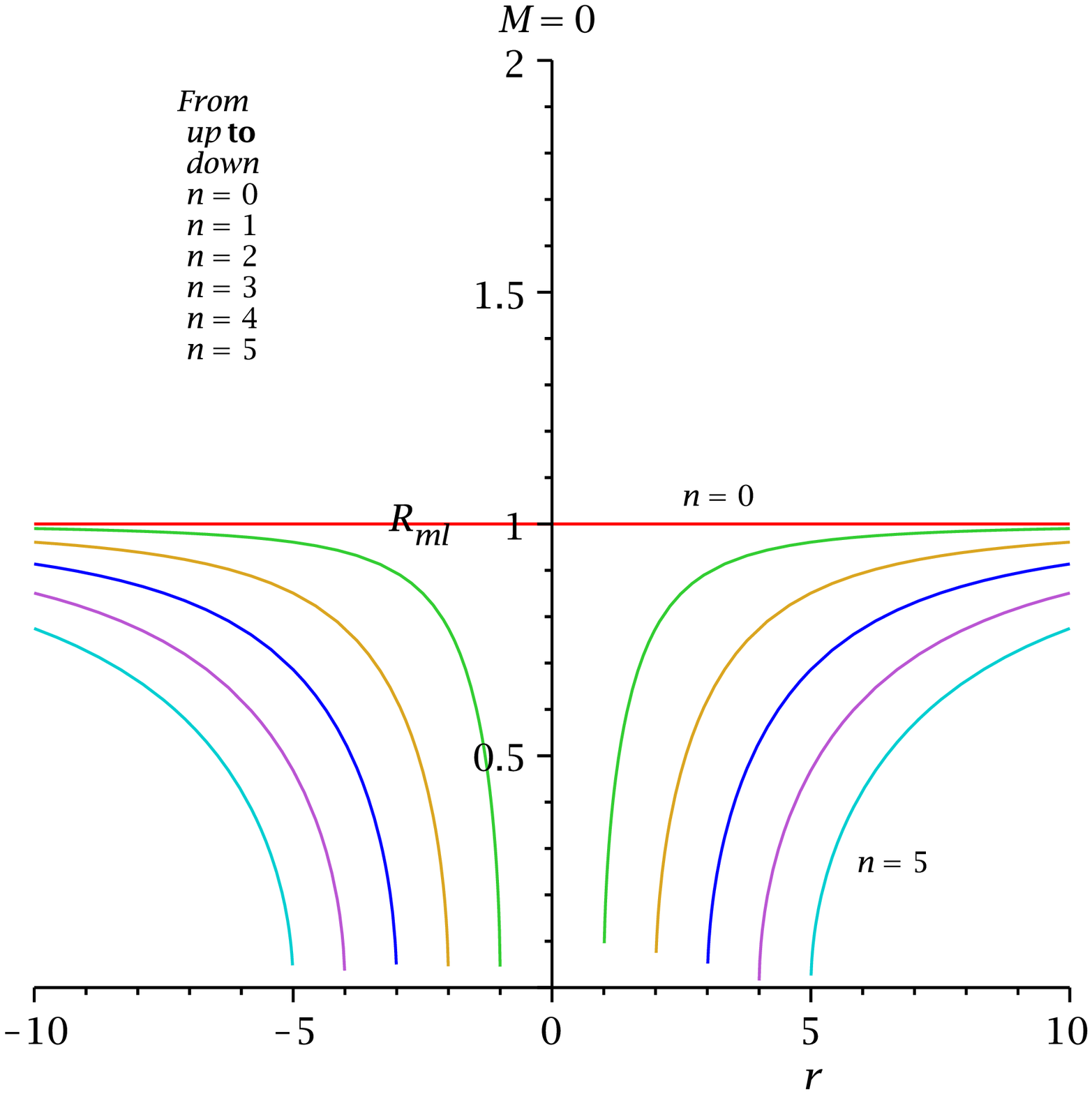}}
\end{center}
\caption{The figure shows the variation  of red-shift factor with $r$ for TN and  massless TN BH.
\label{eqrs}}
\end{figure}

The characteristics of the variation of $-g_{tt}$  with radial coordinate is shown in Fig. \ref {eq2}. 
The ``red-shift factor'' ($\mathcal{R}$) \cite{mtw} for TN BH is given by
\begin{eqnarray}
\mathcal{R} &=& \frac{d\tau}{dt}= \frac{1}{u^{t}}=\sqrt{\frac{r^2-2Mr-n^2}{r^2+n^2}}  ~. \label{rf}
\end{eqnarray}
and  for mass-less TN BH, it is 
\begin{eqnarray}
\mathcal{R}_{ml} &=& \sqrt{\frac{r^2-n^2}{r^2+n^2}}  ~. \label{rf1}
\end{eqnarray}
The  variation of redshift factor  with radial coordinate is shown in Fig. \ref {eqrs}. 

\begin{figure}
\begin{center}
{\includegraphics[width=0.45\textwidth]{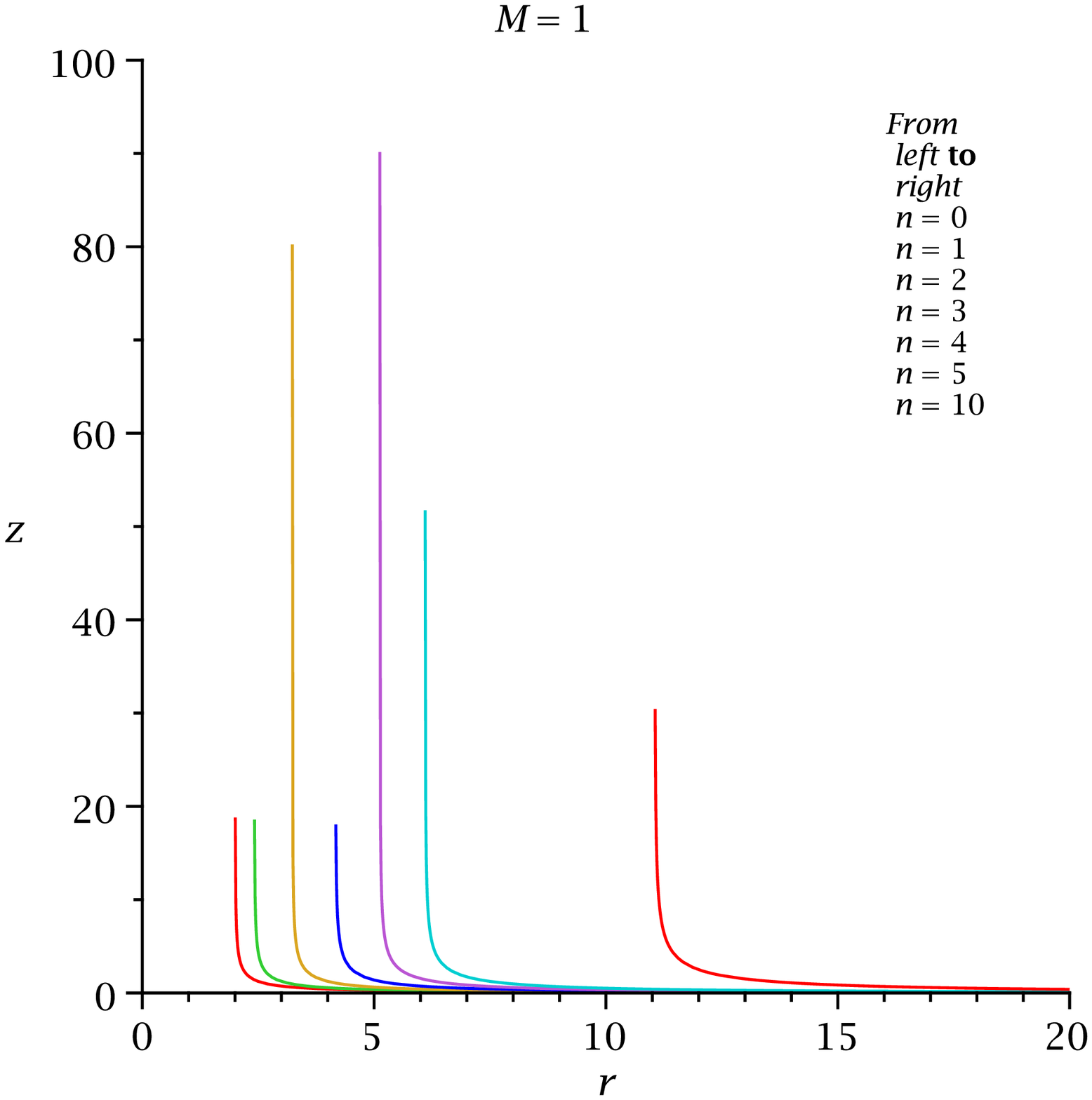}}
{\includegraphics[width=0.45\textwidth]{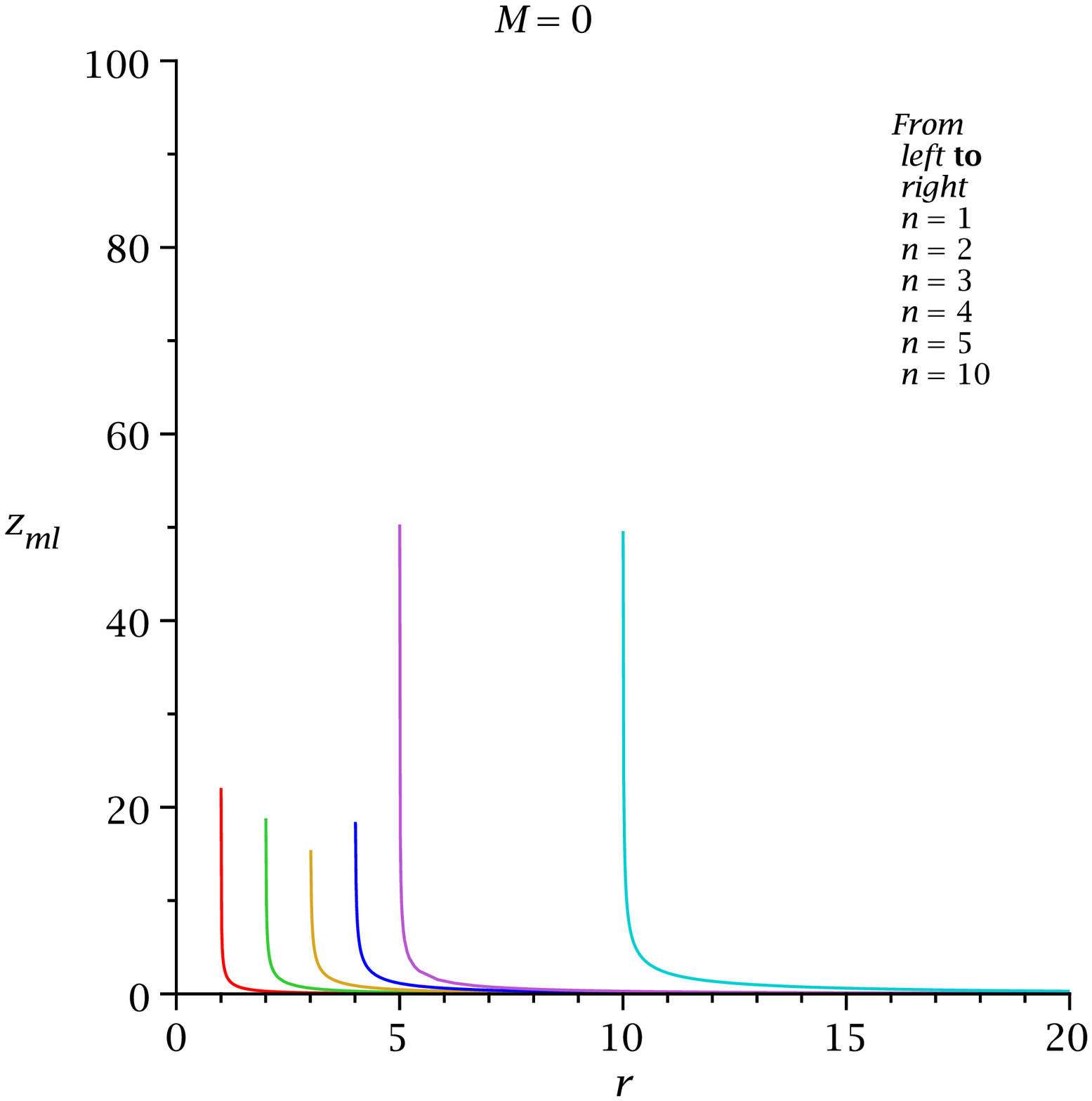}}
\end{center}
\caption{The figure shows the variation  of red-shift with $r$ for TN and  massless TN BH.
\label{eqz}}
\end{figure}

The ``red-shift''($\textbf{z}$) \cite{mtw}  is given by
\begin{eqnarray}
\textbf{z} &\equiv& \frac{\Delta \lambda}{\lambda}=\frac{\lambda_{received}-\lambda_{emitted}}{\lambda_{emitted}}
=u^{t}-1=\sqrt{\frac{r^2+n^2}{r^2-2Mr-n^2}}-1  ~. \label{red}
\end{eqnarray}
For mass-less TN BH,  it is 
\begin{eqnarray}
\textbf{z}_{ml} &\equiv& \sqrt{\frac{r^2+n^2}{r^2-n^2}}-1  ~. \label{red1}
\end{eqnarray}
The  variation of redshift  with radial coordinate is depicted in Fig. \ref {eqz}. 

The thermodynamic properties of TN BH could be found in detail in \cite{jetp} and the BH temperature of ${\cal H}^{+}$
\footnote{ ${\cal H}^{+}$ denotes event horizon of the BH.}
was calculated there 
\begin{eqnarray}
T_{+} &=& \frac{r_{+}-r_{-}}{ 8 \pi \left(Mr_{+}+n^2 \right)} 
.~\label{temp}
\end{eqnarray}
In the limit $r_{+}=r_{-}$, $T_{+}=0$, this indicates that there must exists \emph{extreme} TN BH. 
The geodesic properties have been studied in Sec. \ref{ex}.

\section{\label{egtn} Equatorial circular geodesics of the TN BH:}
Since the equatorial plane is the good location where we can see the causal characteristics of the 
geodesics so in this section we would like to study them. It should be noted that $r=r_{0}$, a constant is 
called circular geodesics. Also, the study of geodesics in the TN space-time is governed by the laws of 
conservation of energy and angular momentum because the space-time is independent of the coordinates 
$t$ and $\phi$. The TN space-time possesses time-like isometry generated by the time-like Killing vector 
$\xi \equiv \partial_t$ whose projection along the four velocity ${\bf u}$  of geodesics: $\xi \cdot {\bf u} = -E$, 
is conserved along such geodesics and the another conserved quantity is angular momentum followed 
by the relation $L \equiv \zeta \cdot {\bf u}$  (where $\zeta \equiv \partial_{\phi}$). Where $\zeta$ is the 
space-like Killing vector field due to the rotational isometry.

Using these conditions together with the normalization of the four velocity, one can easily derived 
the radial equation for TN BH  on the $\theta=\frac{\pi}{2}$ plane:
\begin{eqnarray}
\dot{r}^{2}=E^{2}-V_{eff}=E^{2}-{\cal H}(r)\left(\frac{L^{2}}{r^2+n^2}-\epsilon \right)~.\label{tnr}
\end{eqnarray}
where the  effective potential \cite{mt} is
\begin{eqnarray}
V_{eff} &=& {\cal H}(r) \left(\frac{L^{2}}{r^2+n^2}-\epsilon \right) ~.\label{vrn}
\end{eqnarray}
Here, $\epsilon=-1$ for time-like geodesics, $\epsilon=0$ for light-like geodesics and $\epsilon=+1$ for 
space-like geodesics.

\subsection{\label{cpotn} Lightrays orbit:}
\begin{figure}
\begin{center}
{\includegraphics[width=0.45\textwidth]{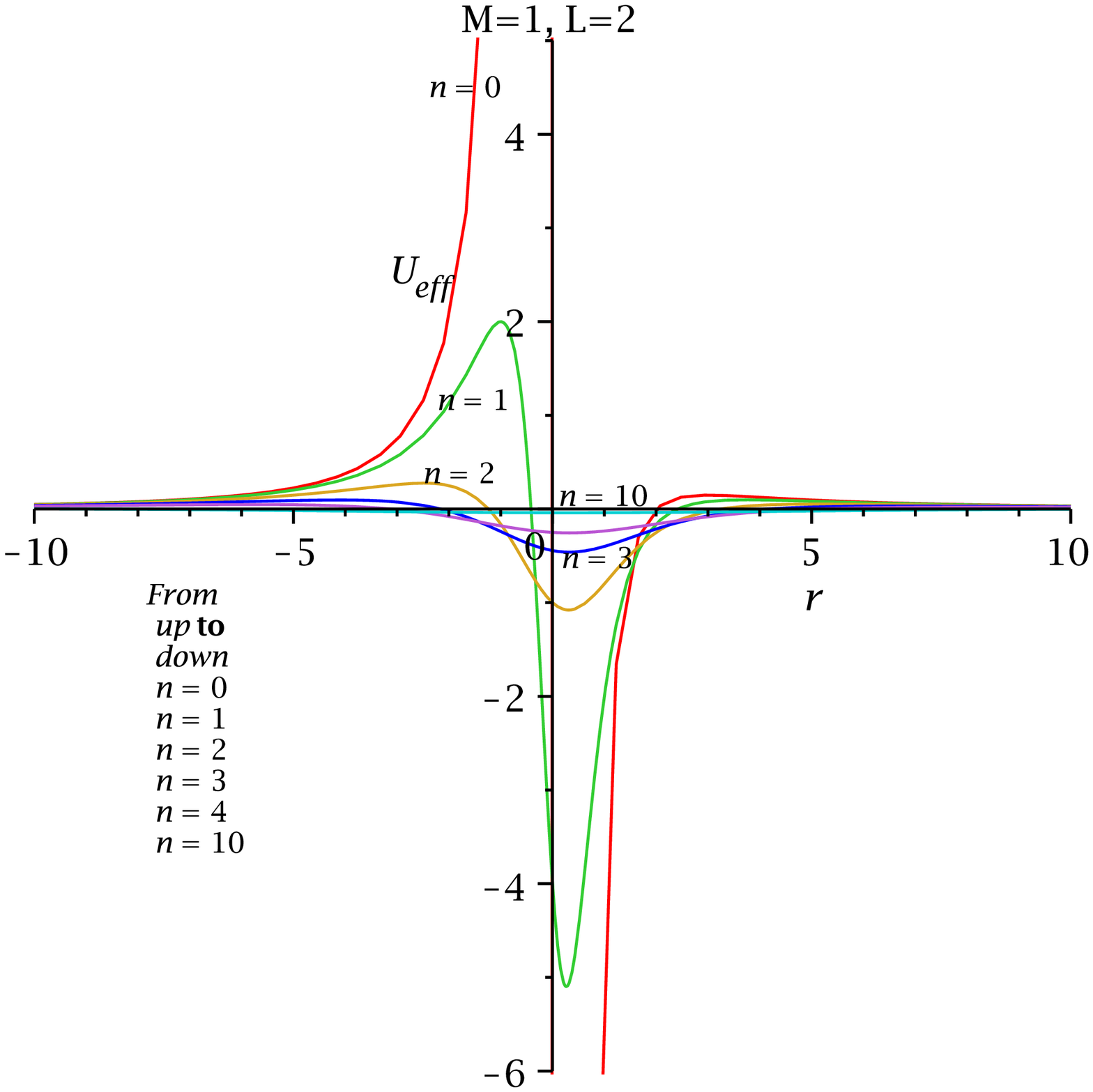}}
{\includegraphics[width=0.45\textwidth]{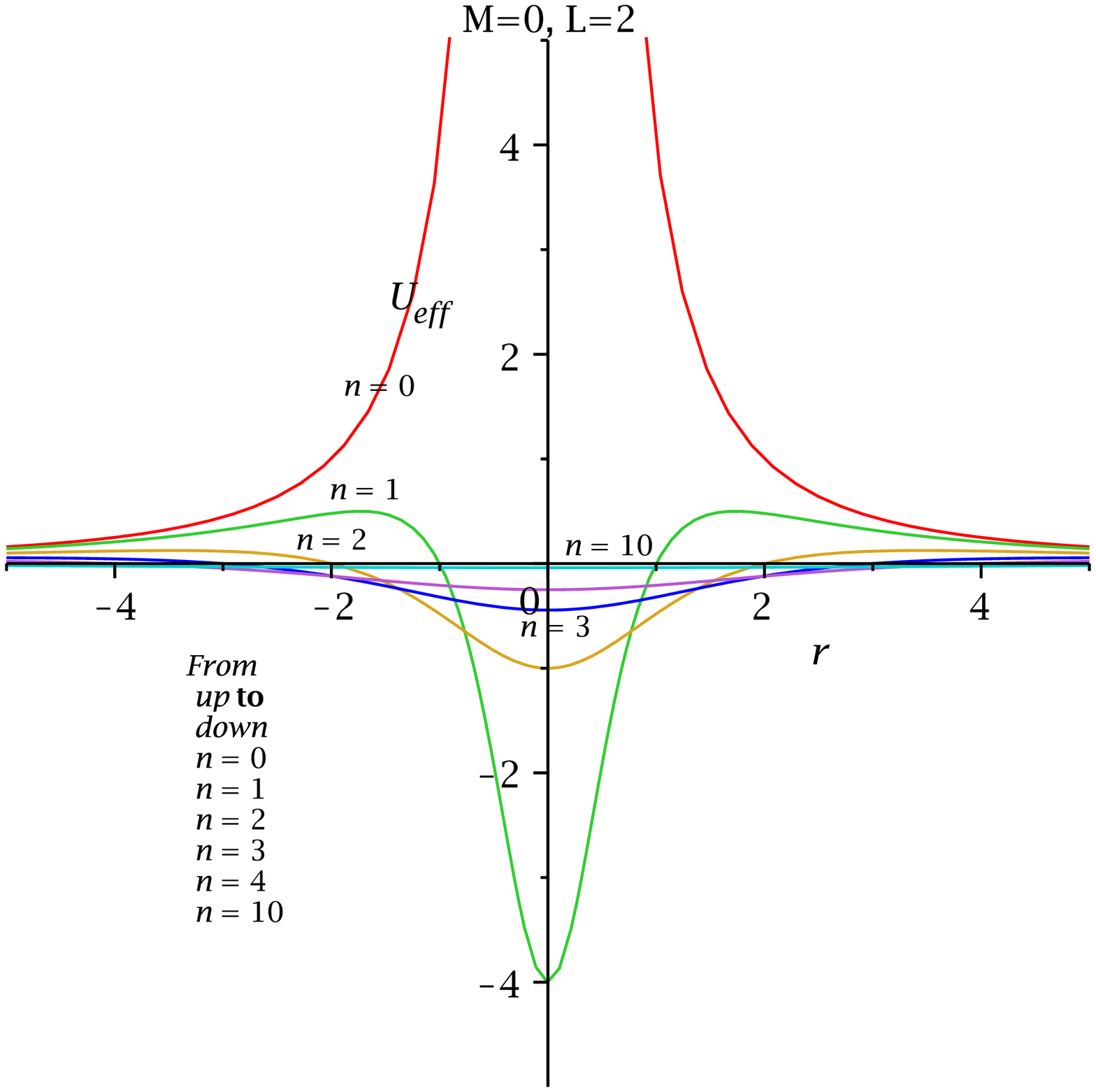}}
\end{center}
\caption{The figure shows the variation  of $U_{eff}$  with $r$ for TN BH and mass-less TN BH.
\label{nu1}}
\end{figure}

\begin{figure}
\begin{center}
{\includegraphics[width=0.45\textwidth]{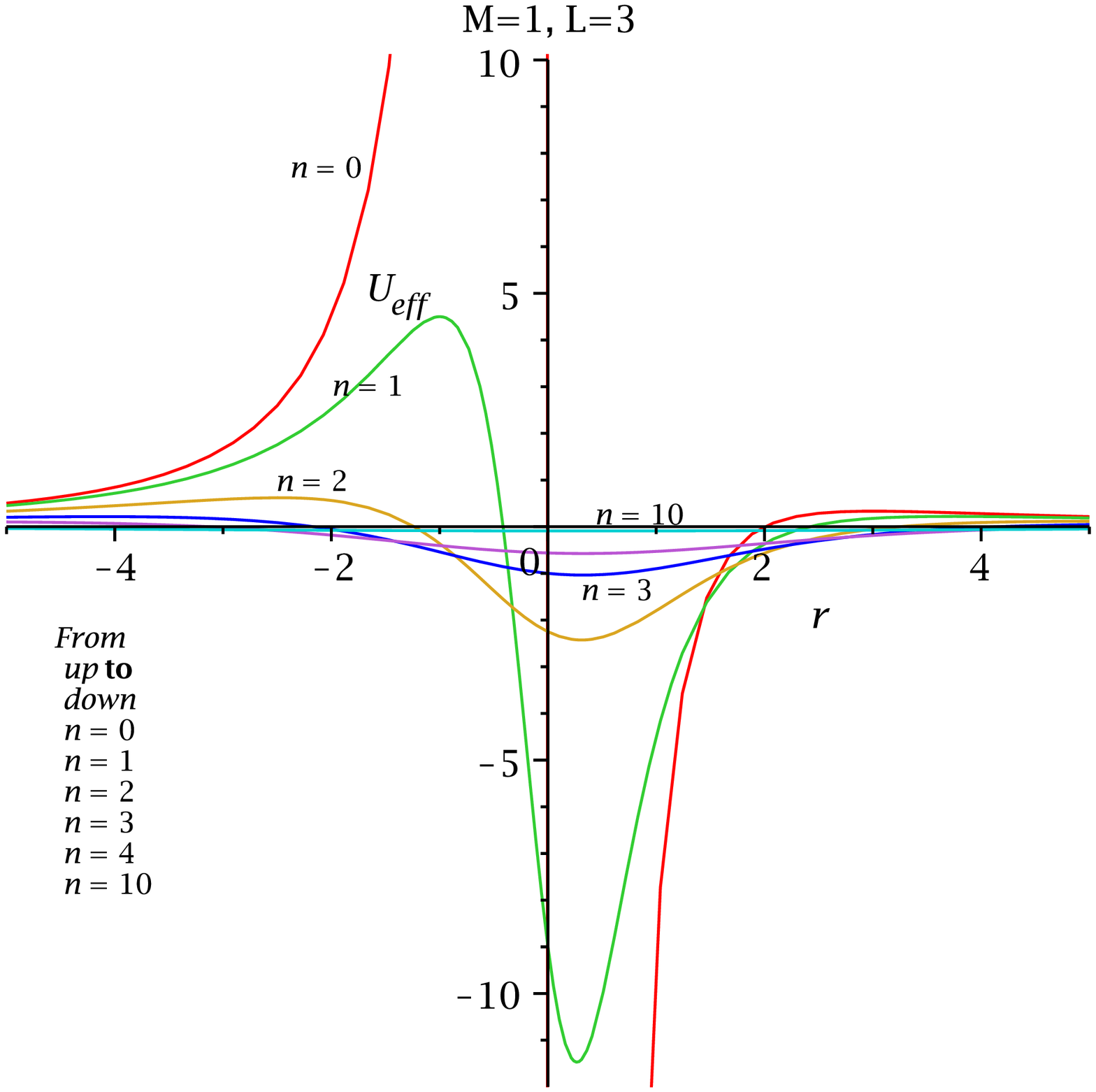}}
{\includegraphics[width=0.45\textwidth]{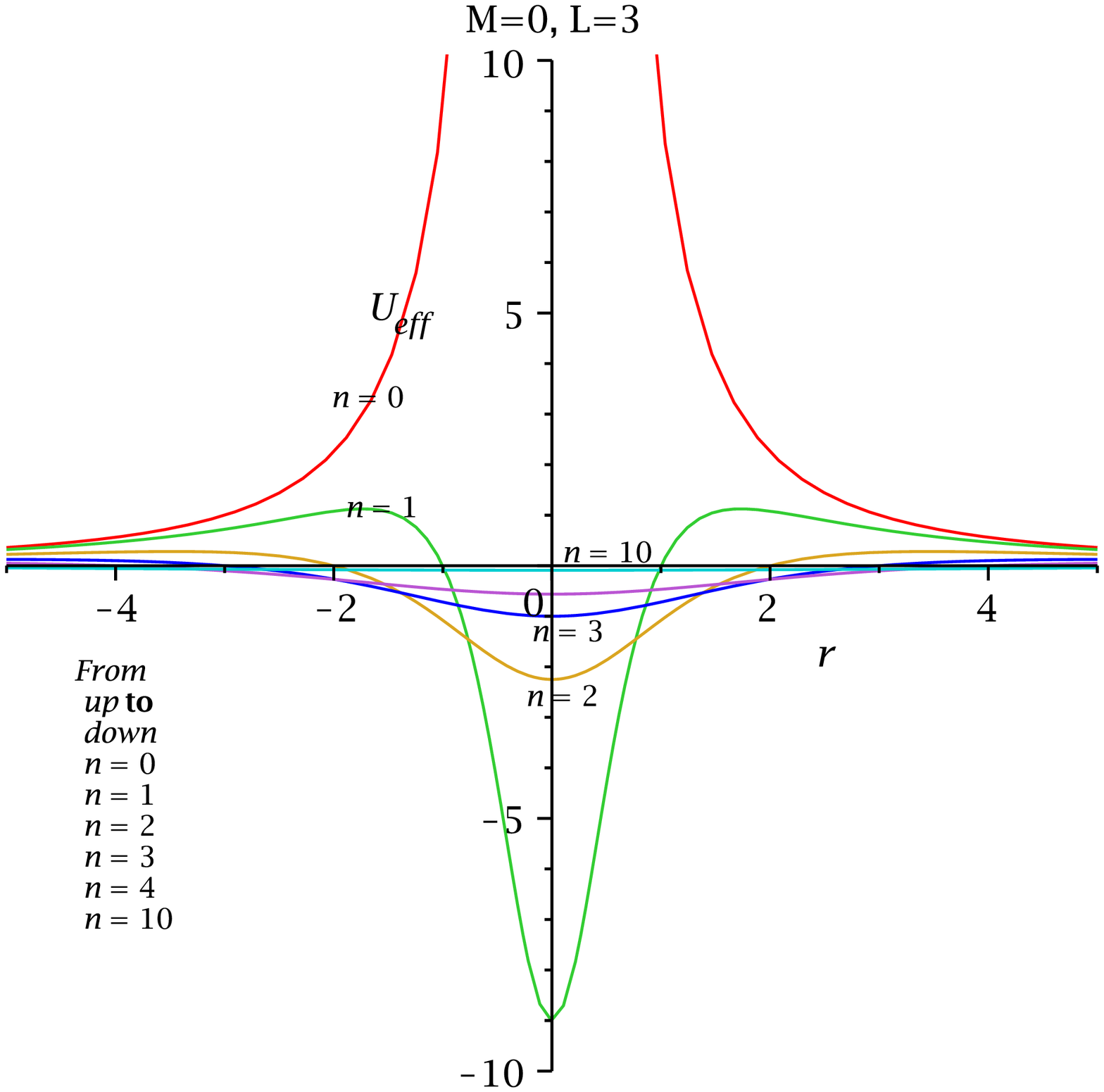}}
\end{center}
\caption{The figure shows the variation  of $U_{eff}$  with $r$ for TN BH and mass-less TN BH
for different values of $n$.
\label{nu2}}
\end{figure}

\begin{figure}
\begin{center}
{\includegraphics[width=0.45\textwidth]{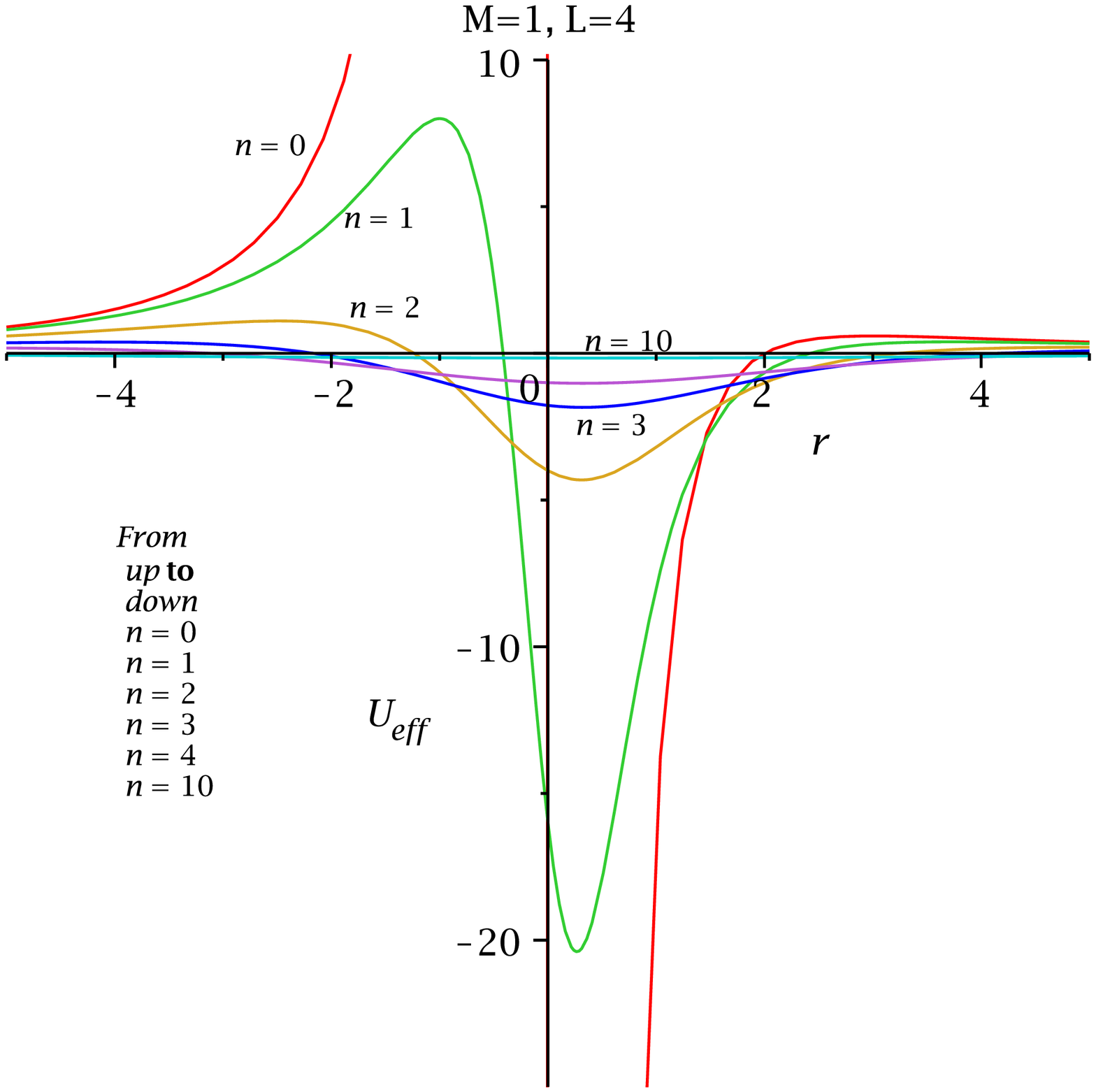}}
{\includegraphics[width=0.45\textwidth]{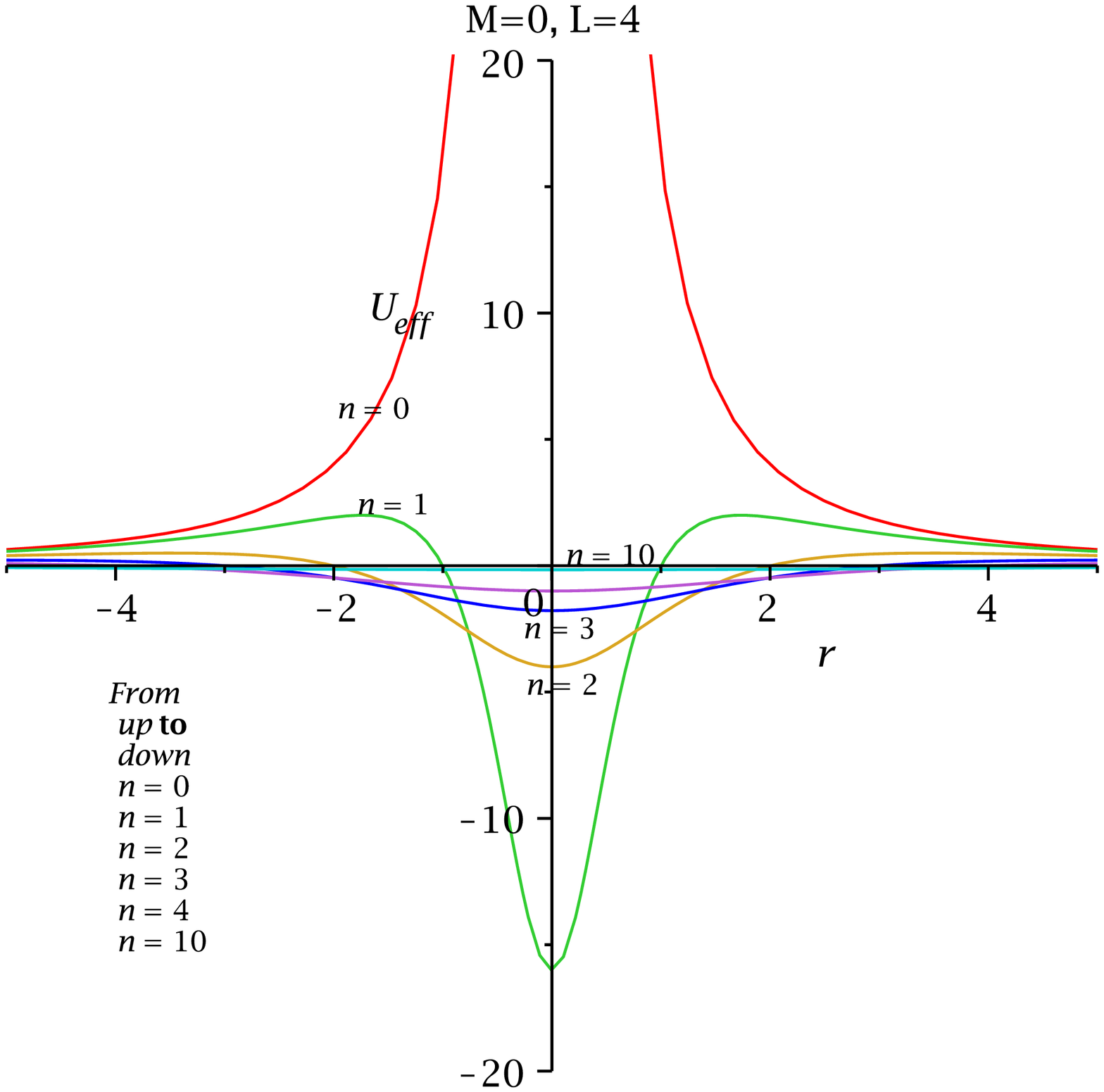}}
\end{center}
\caption{The figure shows the variation  of $U_{eff}$  with $r$ for TN BH and mass-less TN BH.
\label{nu3}}
\end{figure}

\begin{figure}
\begin{center}
{\includegraphics[width=0.45\textwidth]{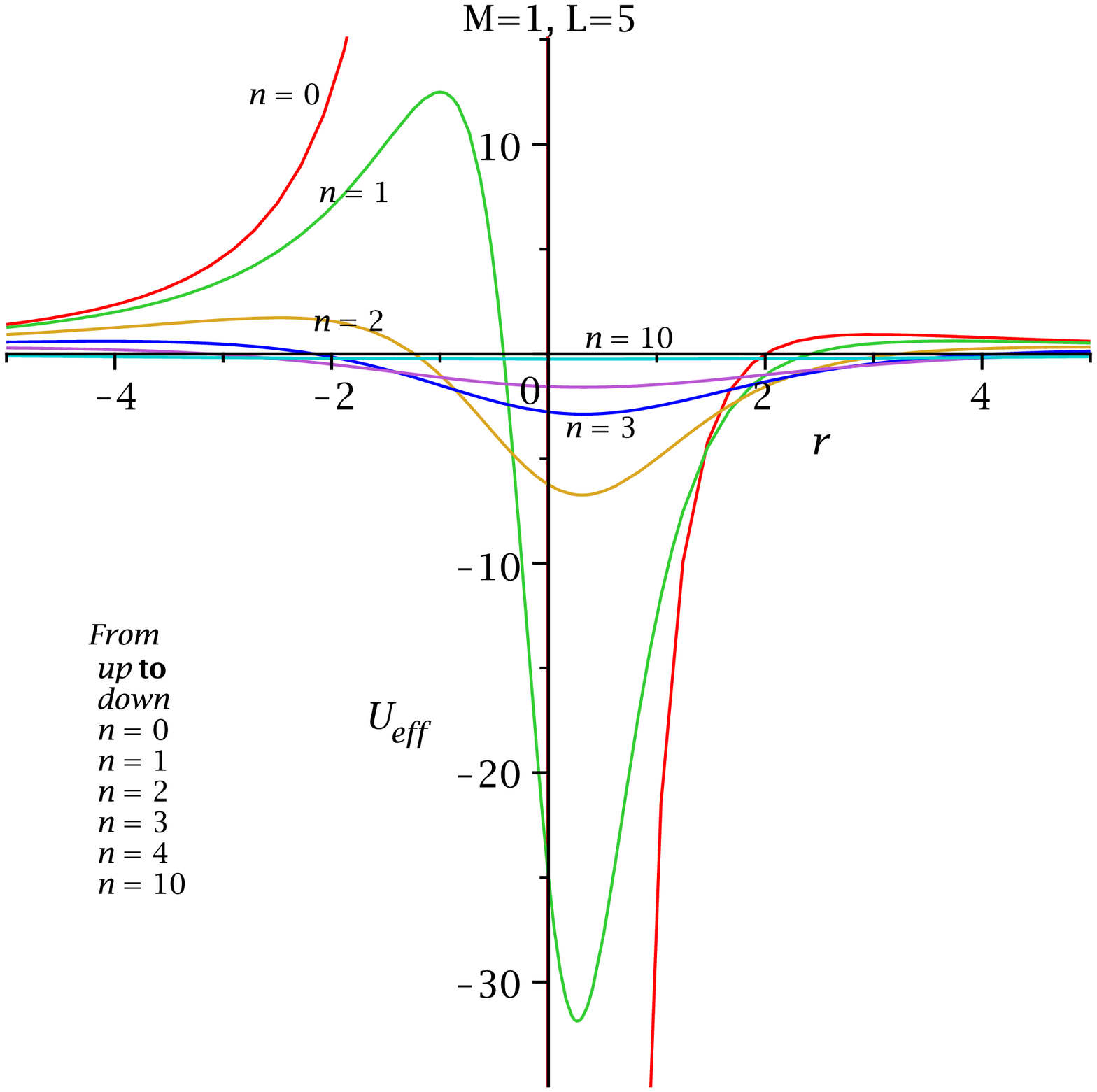}}
{\includegraphics[width=0.45\textwidth]{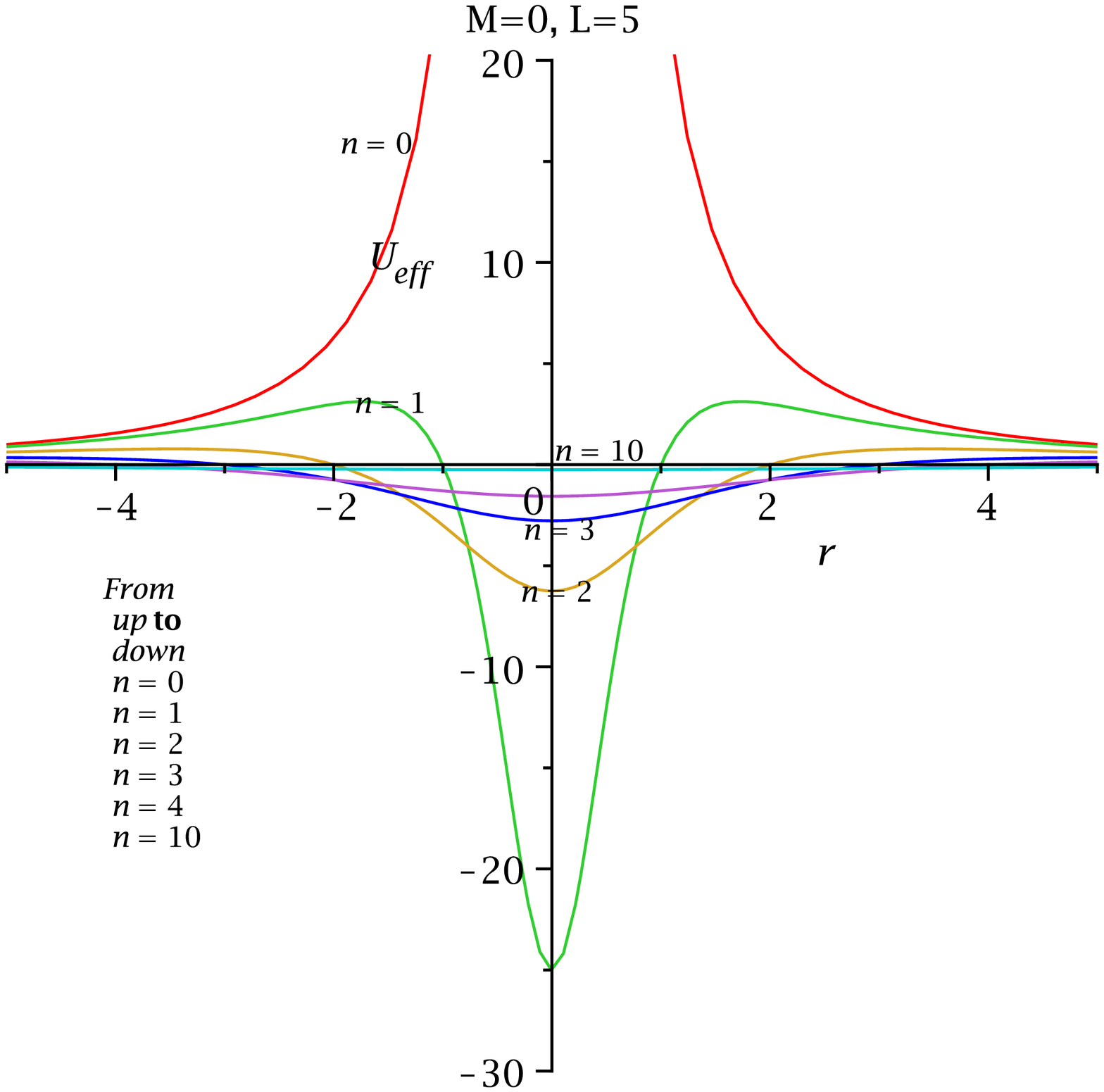}}
\end{center}
\caption{The figure shows the variation  of $U_{eff}$ with $r$ for TN BH and mass-less TN BH.
\label{nu4}}
\end{figure}

\begin{figure}
\begin{center}
{\includegraphics[width=0.45\textwidth]{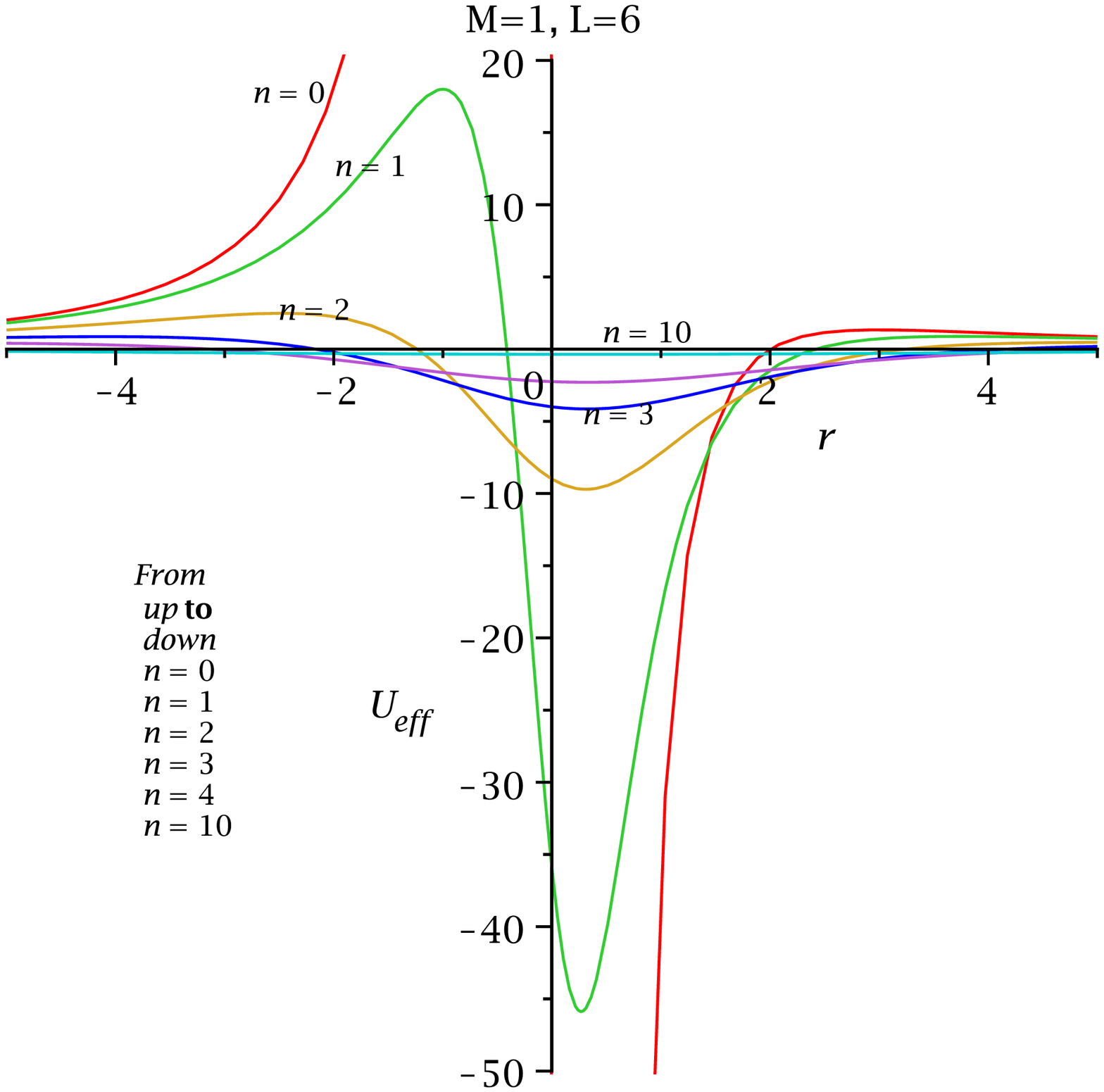}}
{\includegraphics[width=0.45\textwidth]{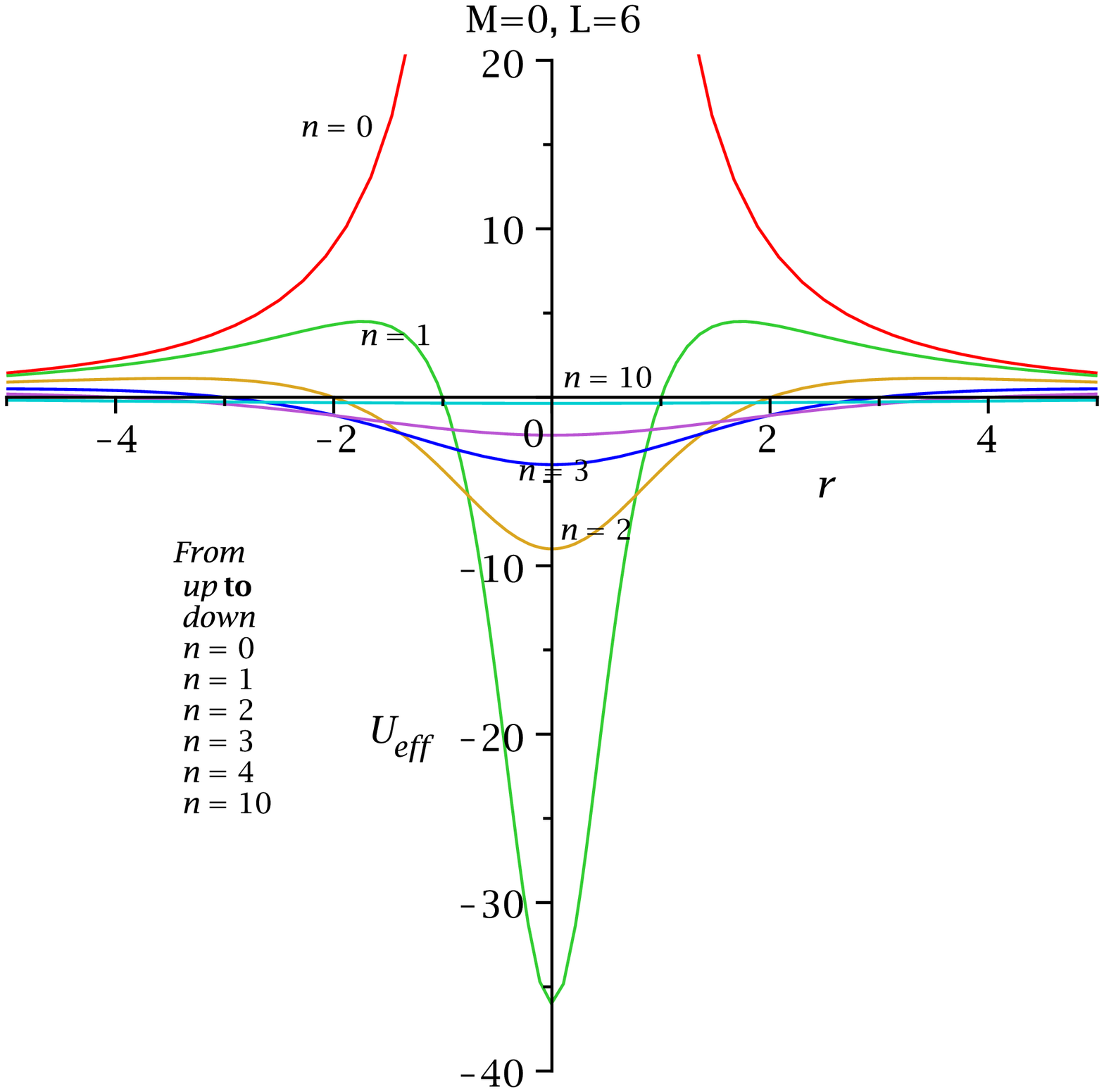}}
\end{center}
\caption{The figure shows the variation  of $U_{eff}$ with $r$ for TN BH and mass-less TN BH.
\label{nu5}}
\end{figure}

\begin{figure}
\begin{center}
{\includegraphics[width=0.45\textwidth]{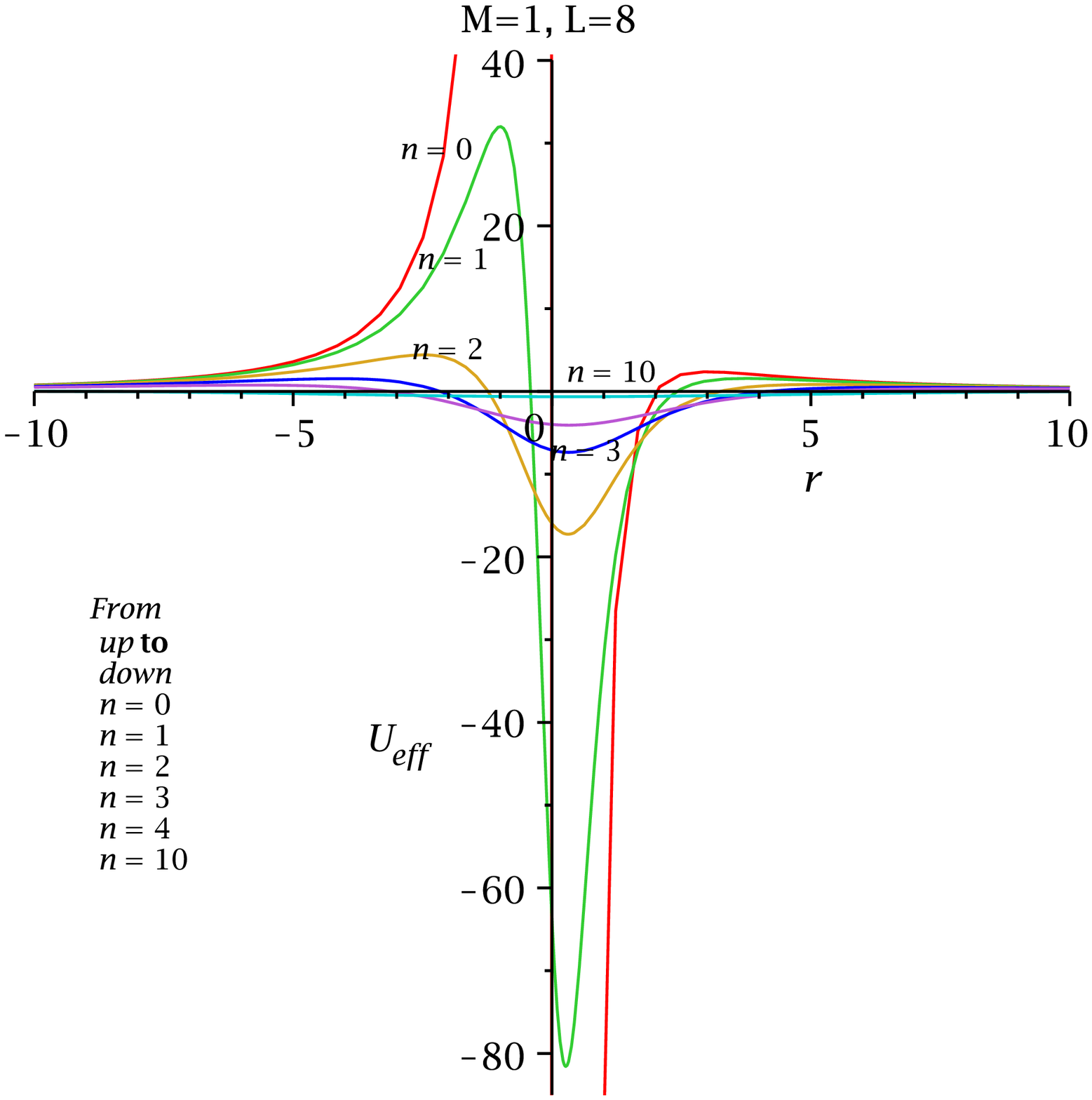}}
{\includegraphics[width=0.45\textwidth]{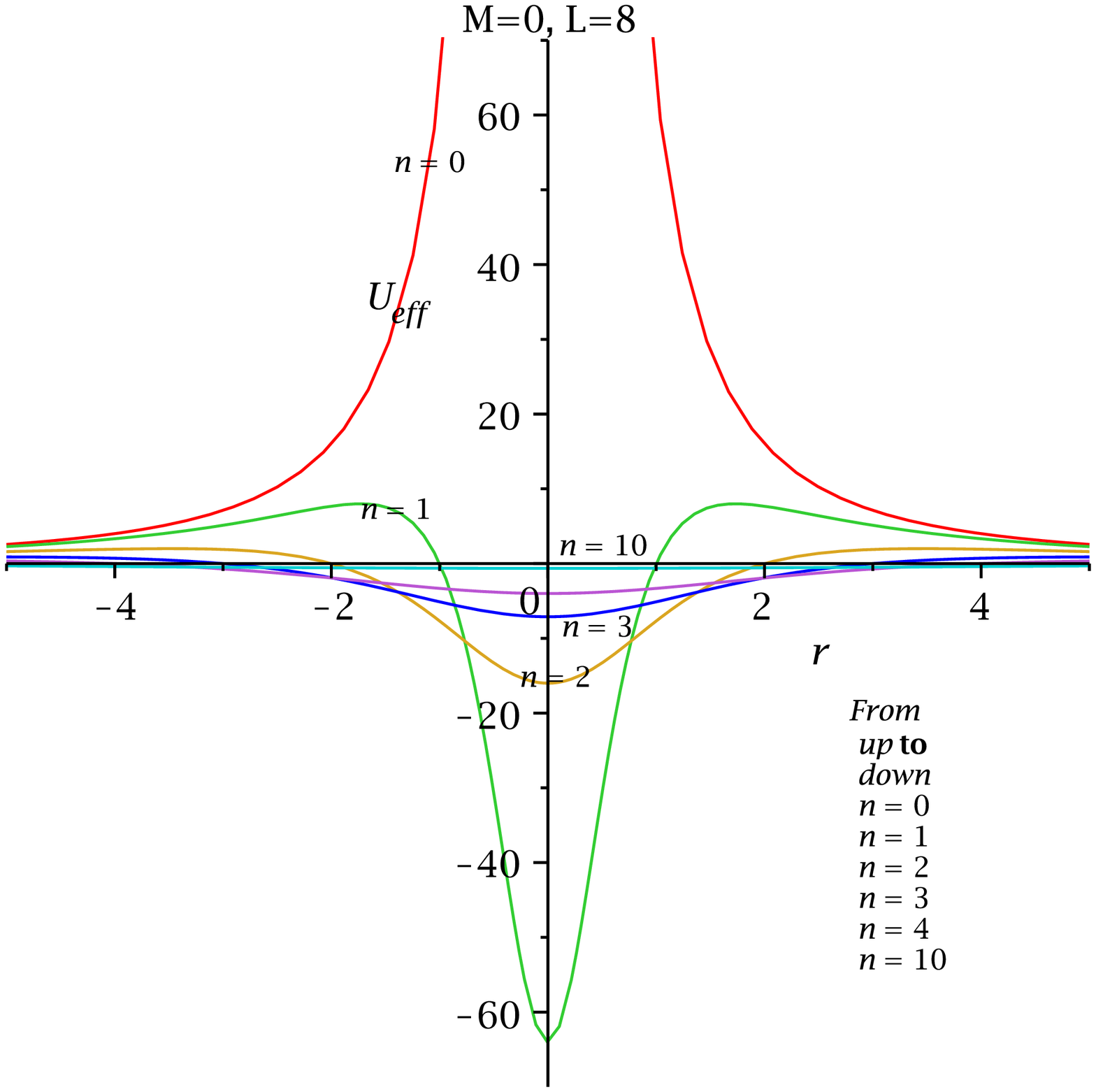}}
\end{center}
\caption{The figure shows the variation  of $U_{eff}$ with $r$ for TN BH and mass-less TN BH.
\label{nu6}}
\end{figure}

\begin{figure}
\begin{center}
{\includegraphics[width=0.45\textwidth]{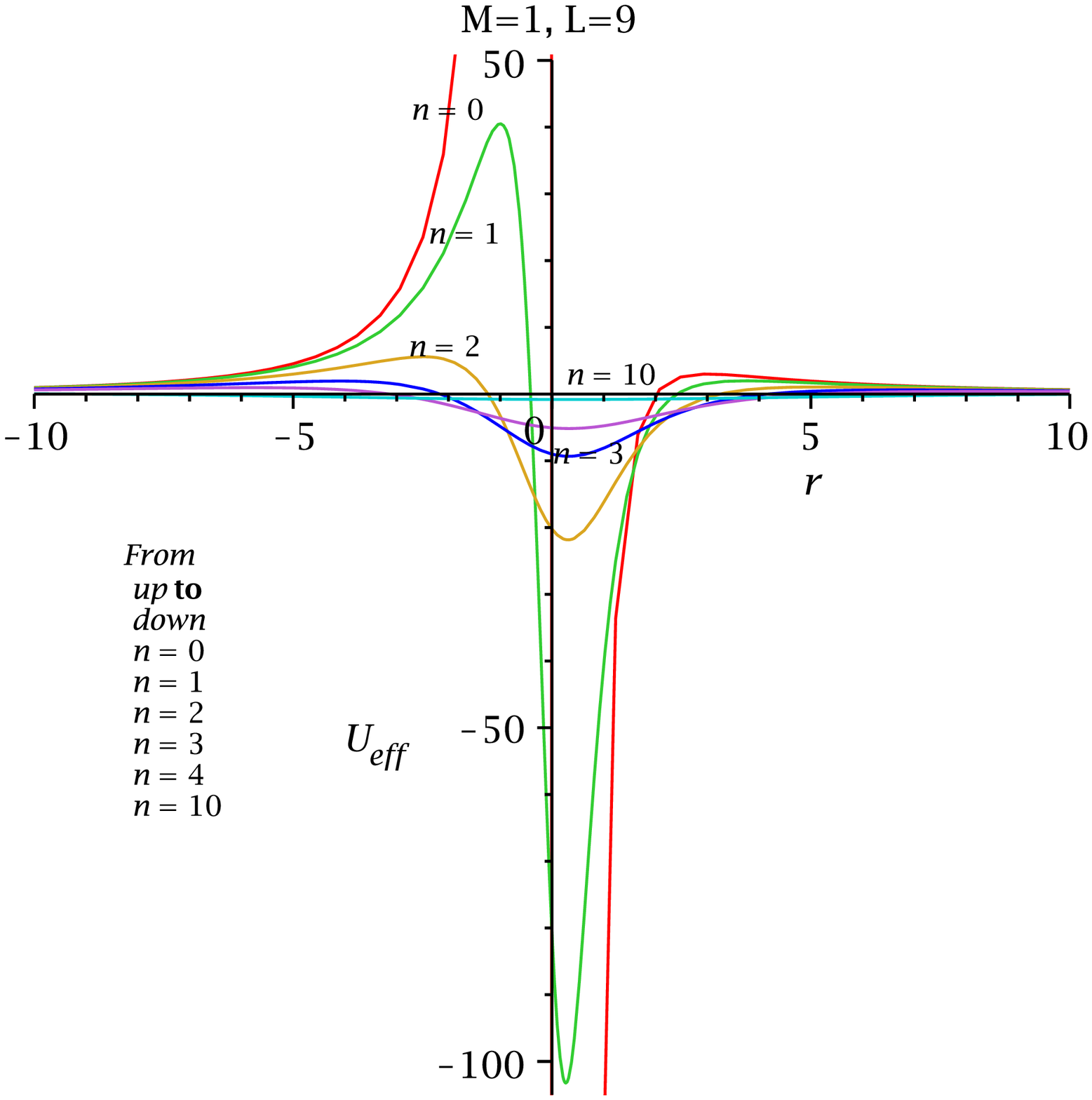}}
{\includegraphics[width=0.45\textwidth]{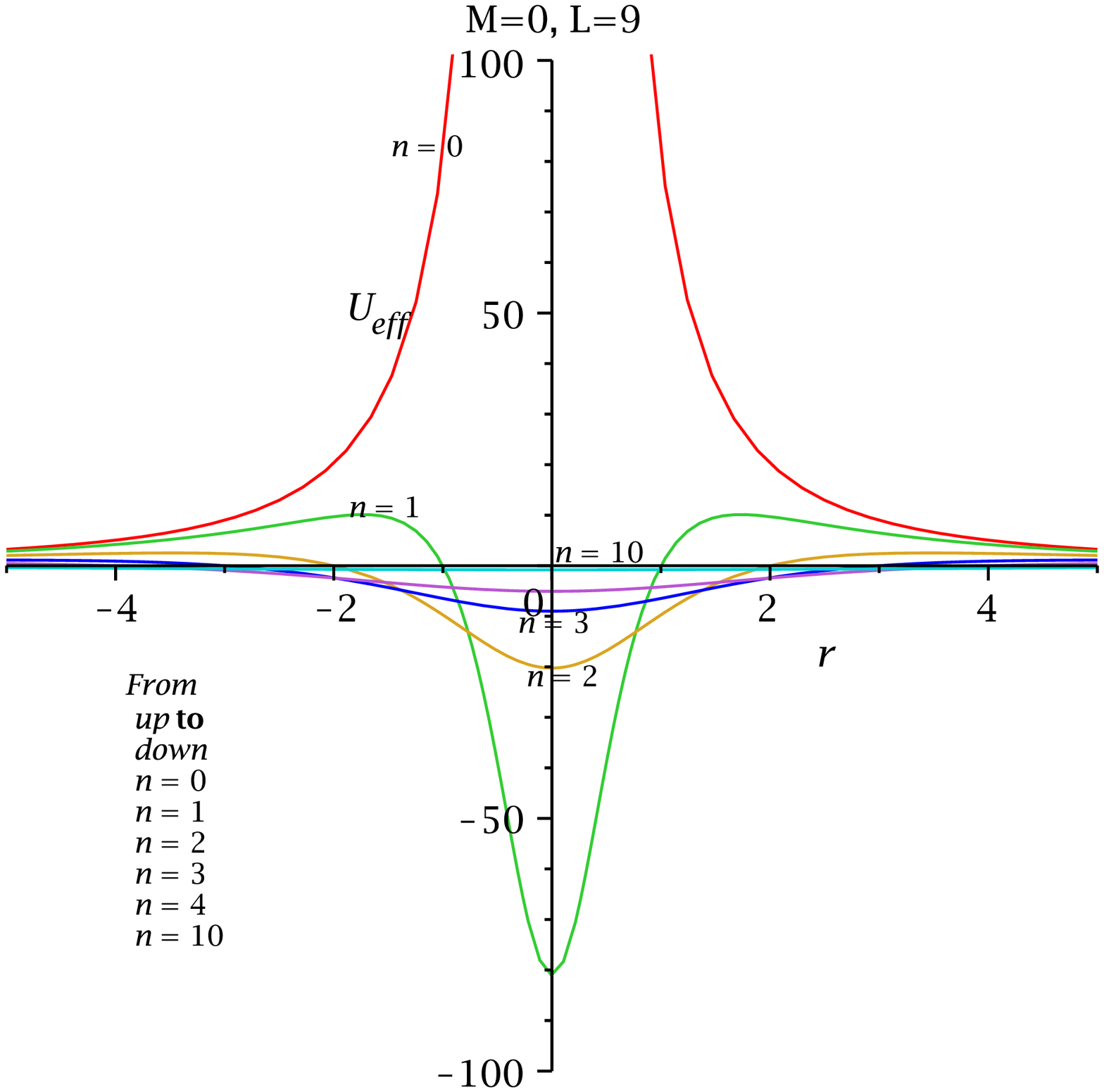}}
\end{center}
\caption{The figure shows the variation  of $U_{eff}$  with $r$ for TN BH and mass-less TN BH
\label{nu7}}
\end{figure}

\begin{figure}
\begin{center}
{\includegraphics[width=0.45\textwidth]{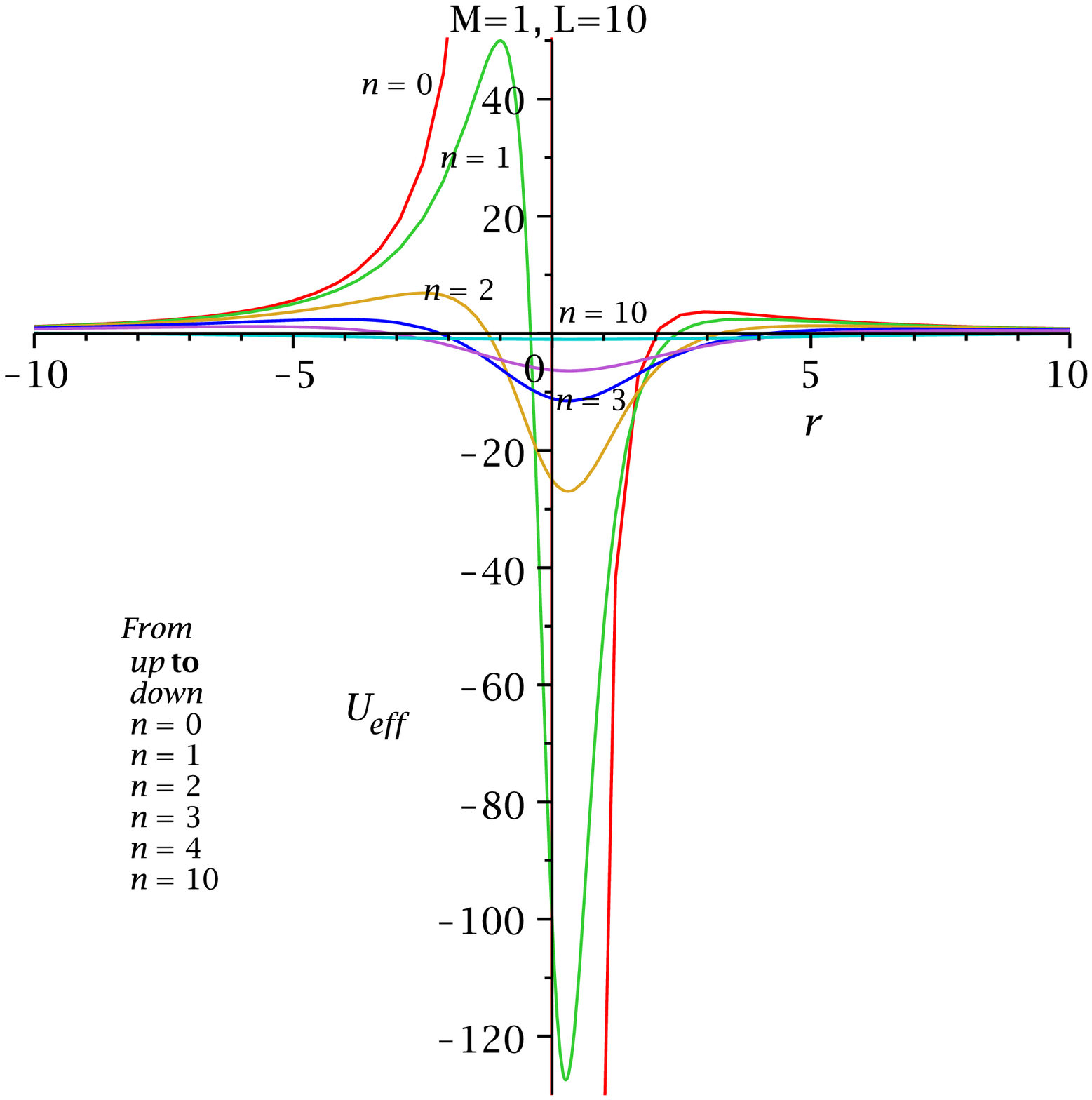}}
{\includegraphics[width=0.45\textwidth]{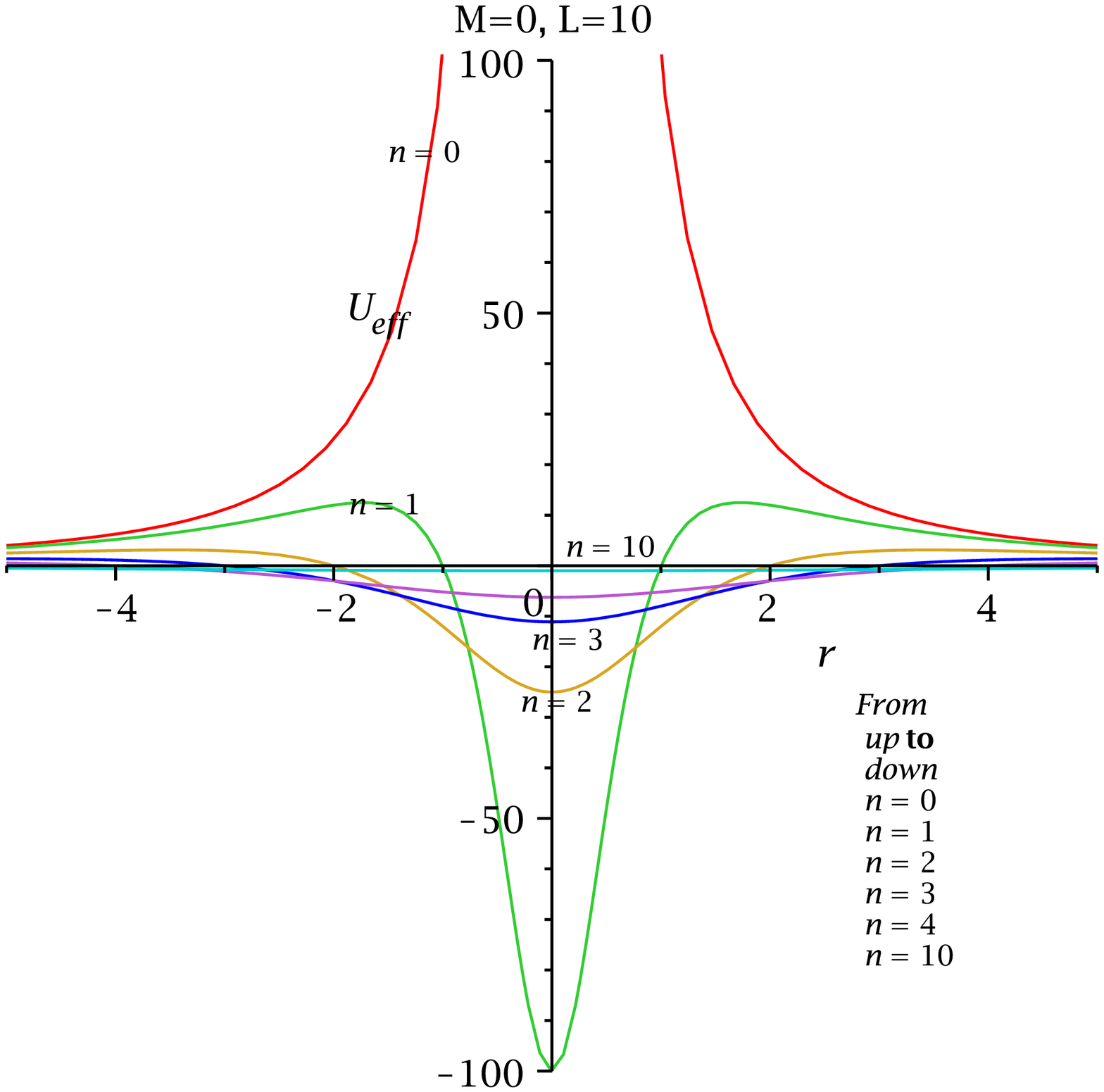}}
\end{center}
\caption{The figure shows the variation  of $U_{eff}$  with $r$ for TN BH and mass-less TN BH.
\label{nu8}}
\end{figure}

For light rays orbit, the effective potential becomes
\begin{eqnarray}
U_{eff} &=& \frac{L^{2}}{r^2+n^2}\left(\frac{r^2-2Mr-n^2}{r^2+n^2}  \right)  ~.\label{n1}
\end{eqnarray}
First we see the behaviour of the test particle in the potential well diagram.
In Fig. \ref{nu1}, Fig. \ref{nu2}, Fig. \ref{nu3}, Fig. \ref{nu4}, Fig. \ref{nu5}, Fig. \ref{nu6}, Fig. \ref{nu7} and Fig. \ref{nu8}, 
we show how the effective potential for photon changes for different values of angular momentum parameter and 
NUT parameter. From the effective potential diagram, it has been observed that the presence of \emph{dual mass} 
parameter effectively changes the shape of the potential well in comparison with absence of the 
\emph{dual mass} parameter. The structure of the potential well also changes in the 
presence of ADM mass parameter and in the absence of ADM mass parameter.
Now by introducing the impact parameter $D=\frac{L}{E}$, one can reparametrization of any null geodesics described 
by the parameter $L$ and $E$. $D$ be the angular momentum of null geodesics when it is reparametrized to have unit 
energy. Now the plot of $D$ with $r$ gives the information of radial motion for null geodesics. 

\begin{figure}
\begin{center}
{\includegraphics[width=0.45\textwidth]{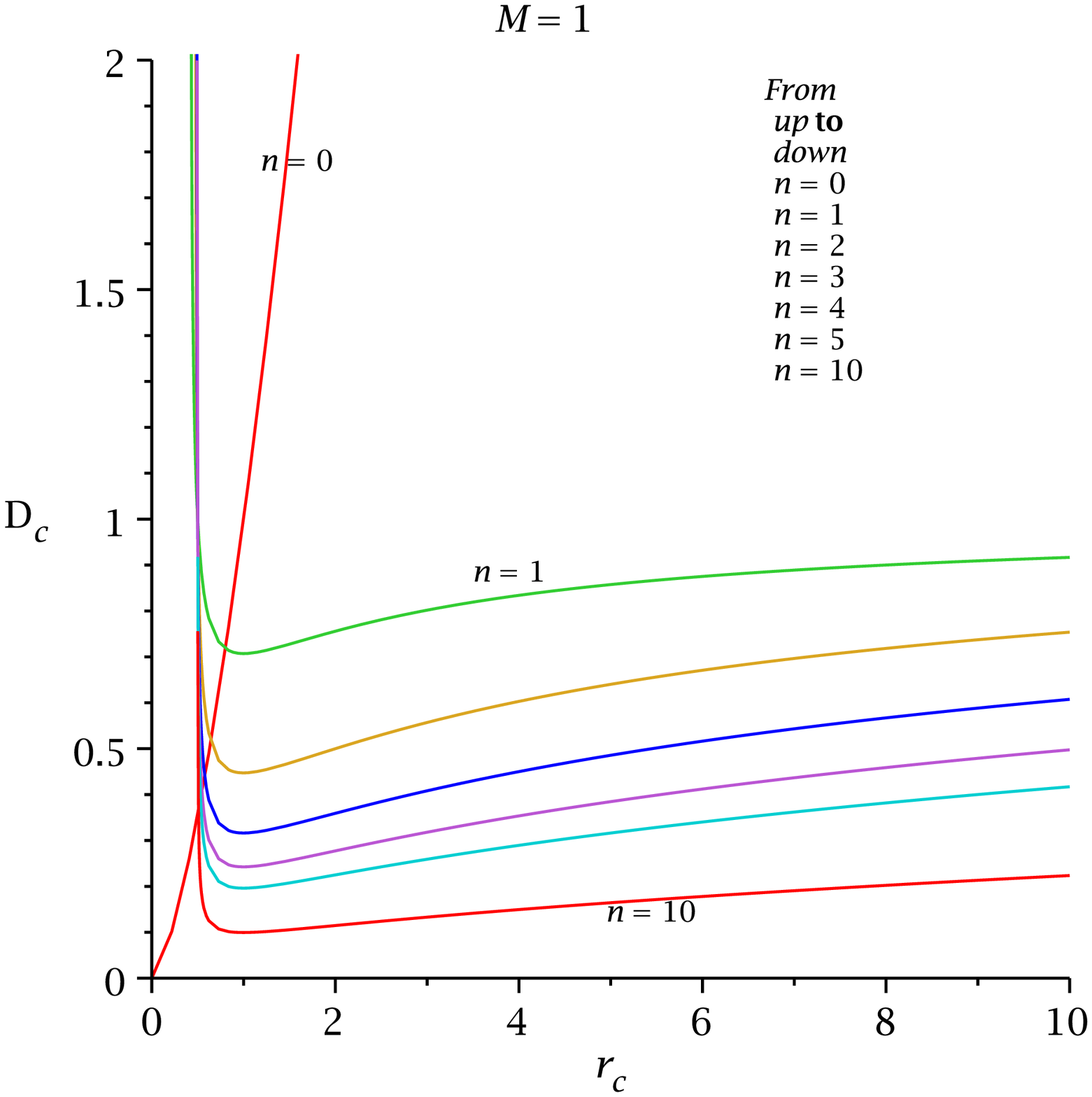}}
{\includegraphics[width=0.45\textwidth]{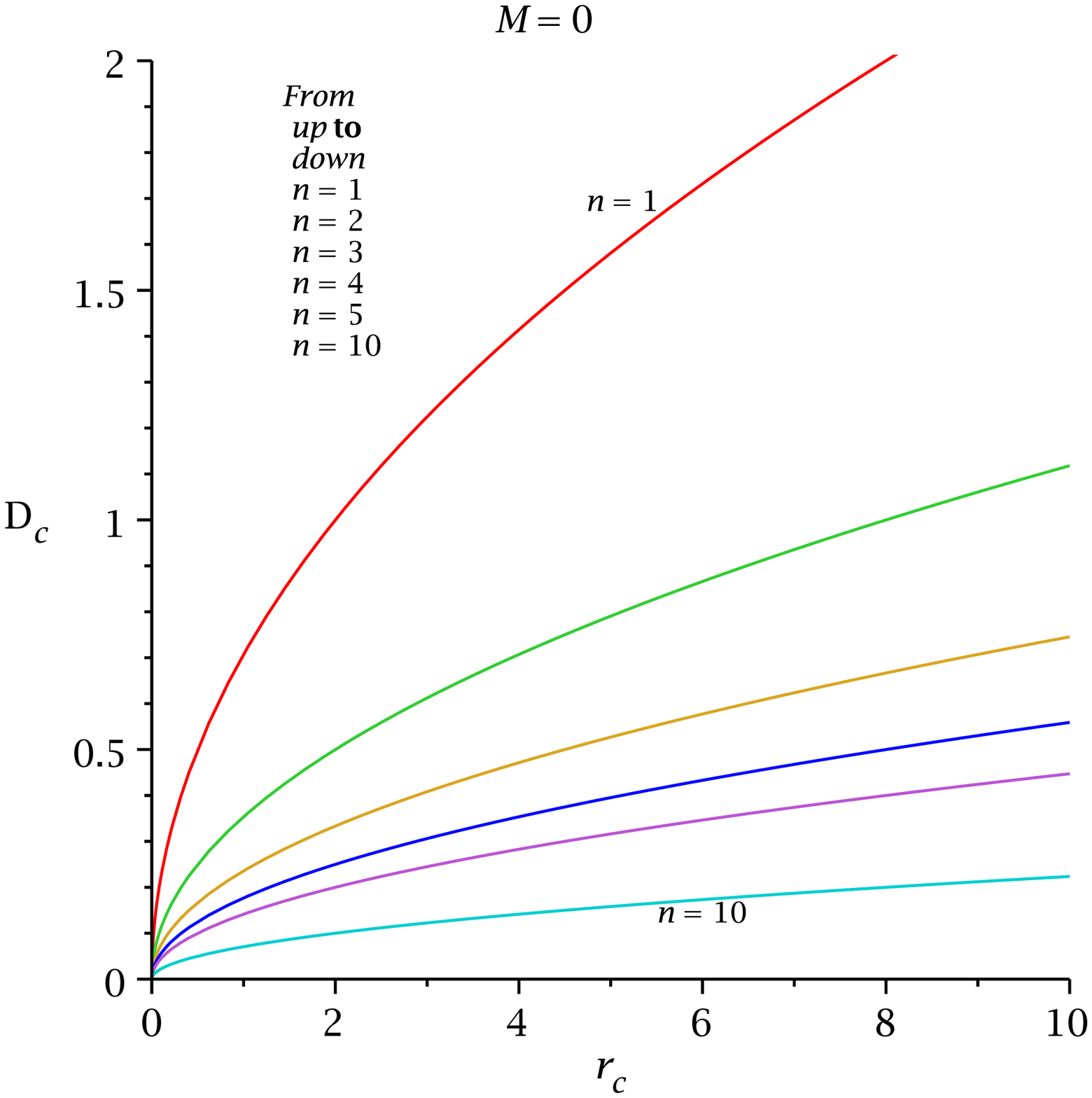}}
\end{center}
\caption{The figure shows the variation  of $D_{c}$  with $r_{c}$ for TN BH and mass-less TN BH.
\label{nu10}}
\end{figure}

The equations evaluating the radius  $r_{c}$ of the unstable circular photon orbit at $E=E_{c}$ and $L=L_{c}$
by introducing the impact parameter $D_{c}=\frac{L_{c}}{E_{c}}$ are
\begin{eqnarray}
r_{c}^{2}+n^2+\left(\frac{2Mr_{c}+n^2-r_{c}^2}{(r_{c}^2+n^2)^2}\right)D_{c}^2 &=& 0  ~.\label{n3}\\
r_{c}-\left[\frac{Mr_{c}^2-Mn^2+2n^2r_{c}}{(r_{c}^2+n^2)^2}\right]D_{c}^2 &=& 0 ~.\label{n4}
\end{eqnarray}
From Eq. (\ref{n4}), we find
\begin{eqnarray}
D_{c} &=& \pm \sqrt{\frac{r_{c}(r_{c}^2+n^2)^2}{Mr_{c}^2+2n^2r_{c}-Mn^2}} ~.\label{n5}
\end{eqnarray}
The behaviour of the impact parameter can be shown from the Fig. \ref{nu10}.
Putting the Eq. (\ref{n5}) in Eq. (\ref{n3}), we obtain the equation of CPO \cite{chur,cqg}:
\begin{eqnarray}
r_{c}^3-3Mr_{c}^2-3n^2r_{c} +Mn^2 &=& 0  ~.\label{n6}
\end{eqnarray}
For mass-less TN BH \cite{chur}, the root of the Eq. becomes 
\begin{eqnarray}
r_{c}  &=& \pm \sqrt{3}n  ~.\label{n7}
\end{eqnarray}
The variation of CPO with $r_{c}$ for TN BH and mass-less TN BH could be found in the Fig. \ref{nu10}.
\begin{figure}
\begin{center}
{\includegraphics[width=0.45\textwidth]{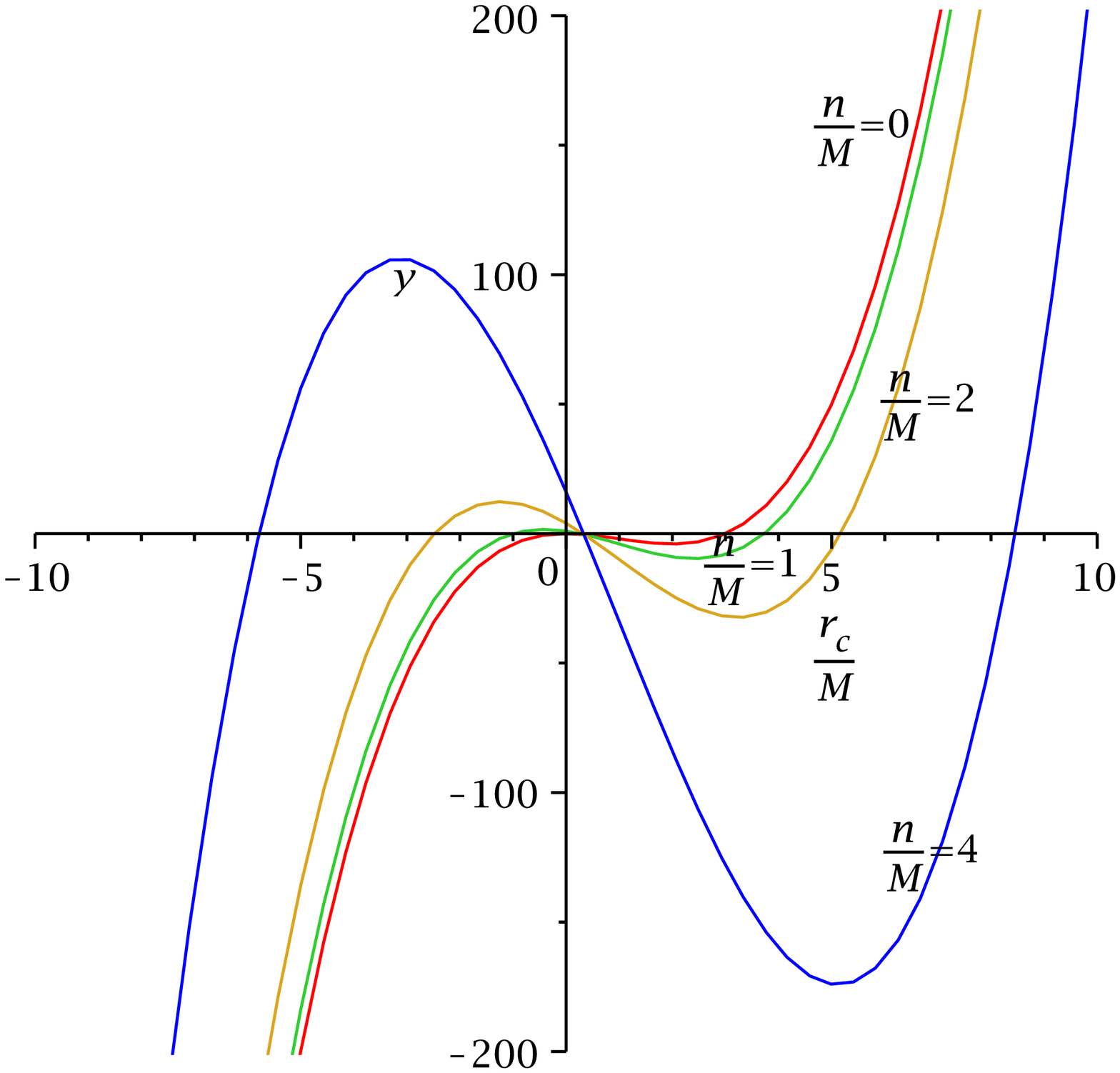}}
{\includegraphics[width=0.45\textwidth]{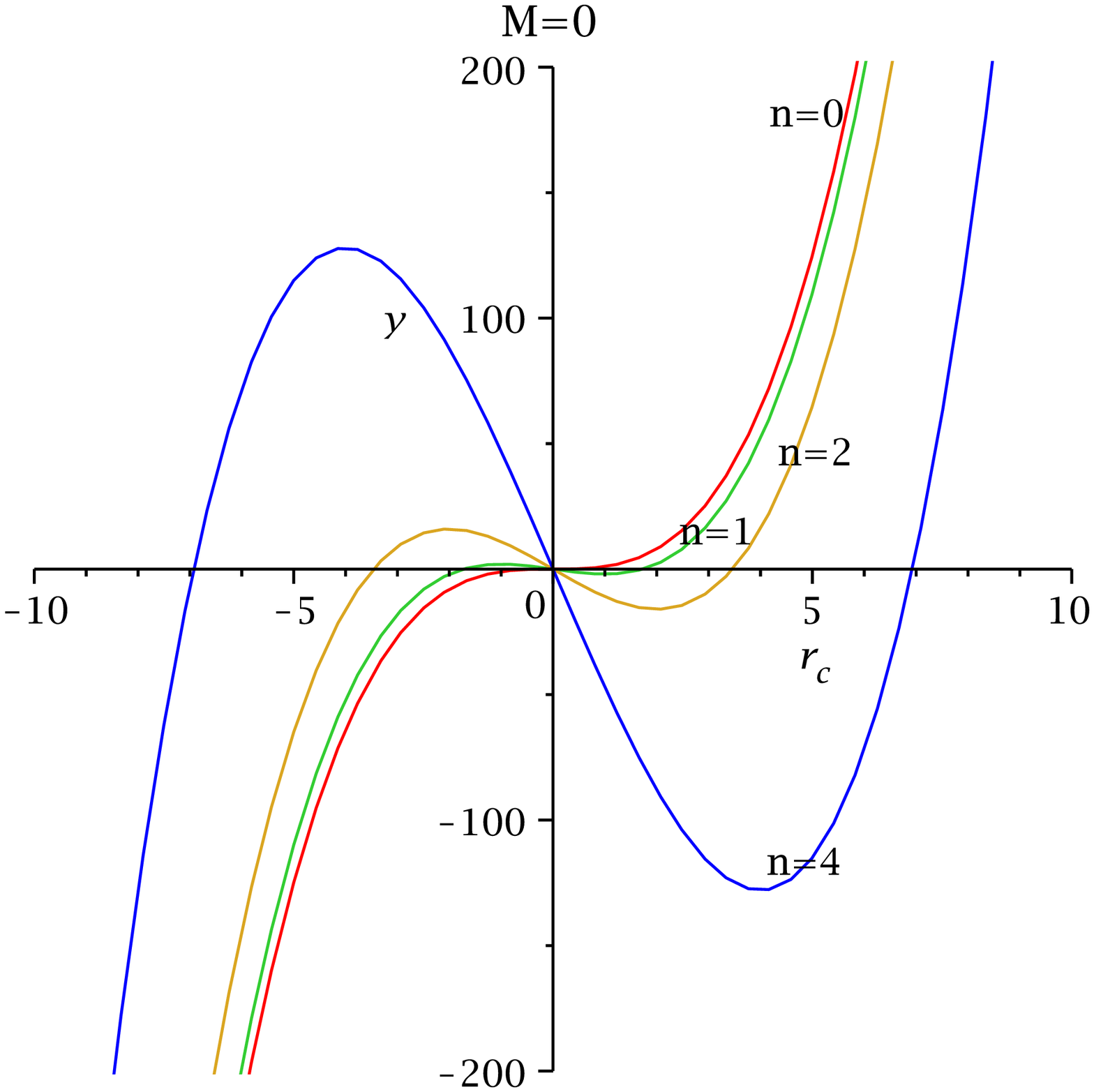}}
\end{center}
\caption{The stability threshold is defined by the positive real root of the cubic
$y=r_{c}^3-3Mr_{c}^2-3n^2r_{c} +Mn^2=0$. For mass-less cases, $y=r_{c}^3-3n^2r_{c}=0$. 
For $n=0$ this root has the value $r_{c}=3M$. 
\label{nu11}}
\end{figure}

The gravitational bending of light analyzed  in \cite{bell97} and the bending angle on the cone is computed 
in \cite{bell97} $\alpha=\frac{4M}{D}$. In terms of opening angle it should be $\alpha=\frac{2M\delta}{n}$. 
For \emph{massless TN spacetime, this angle reduces to zero value}.

\subsection{\label{cto} Circular Time-like Geodesics:}
\begin{figure}
\begin{center}
{\includegraphics[width=0.45\textwidth]{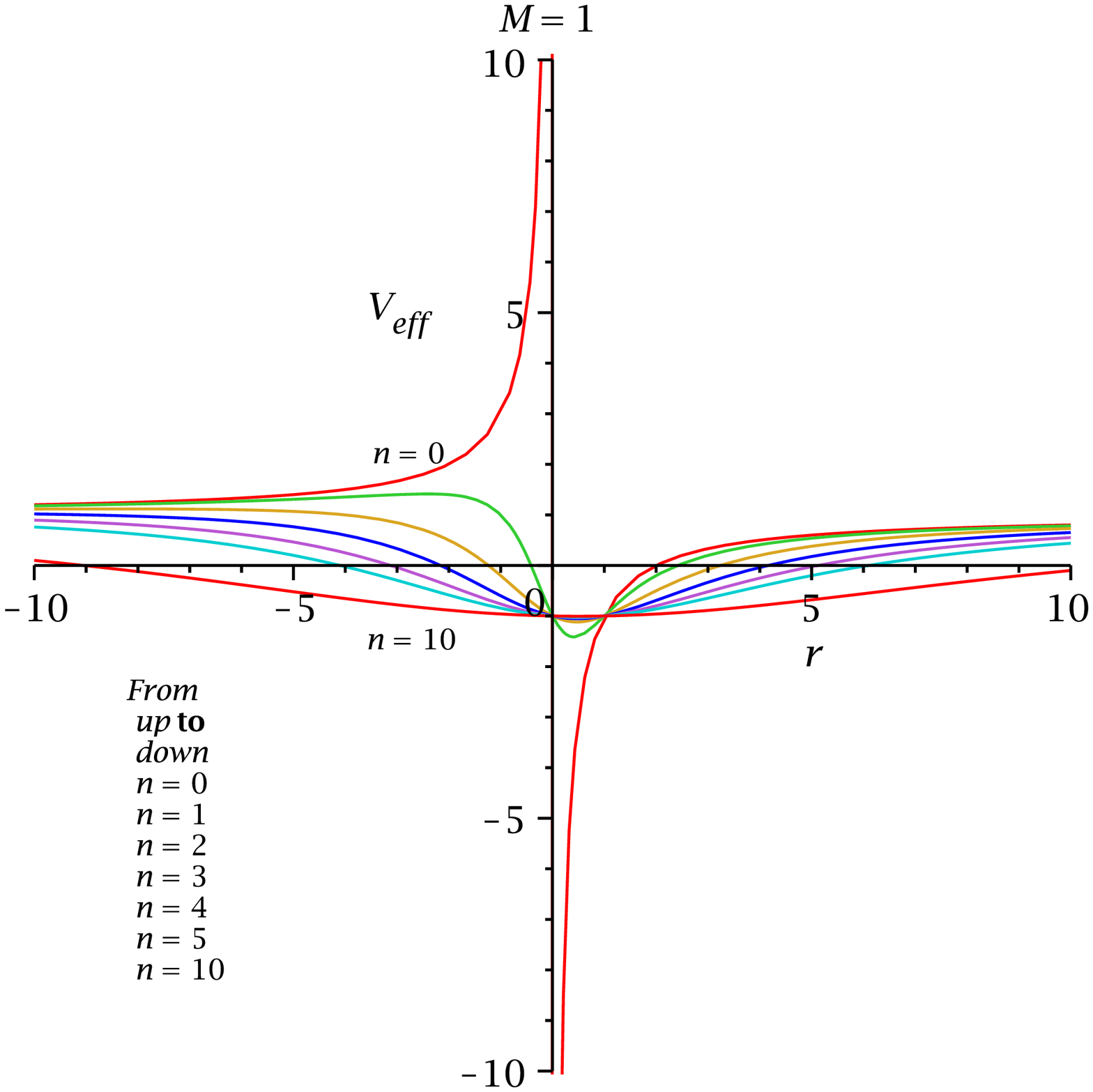}}
{\includegraphics[width=0.45\textwidth]{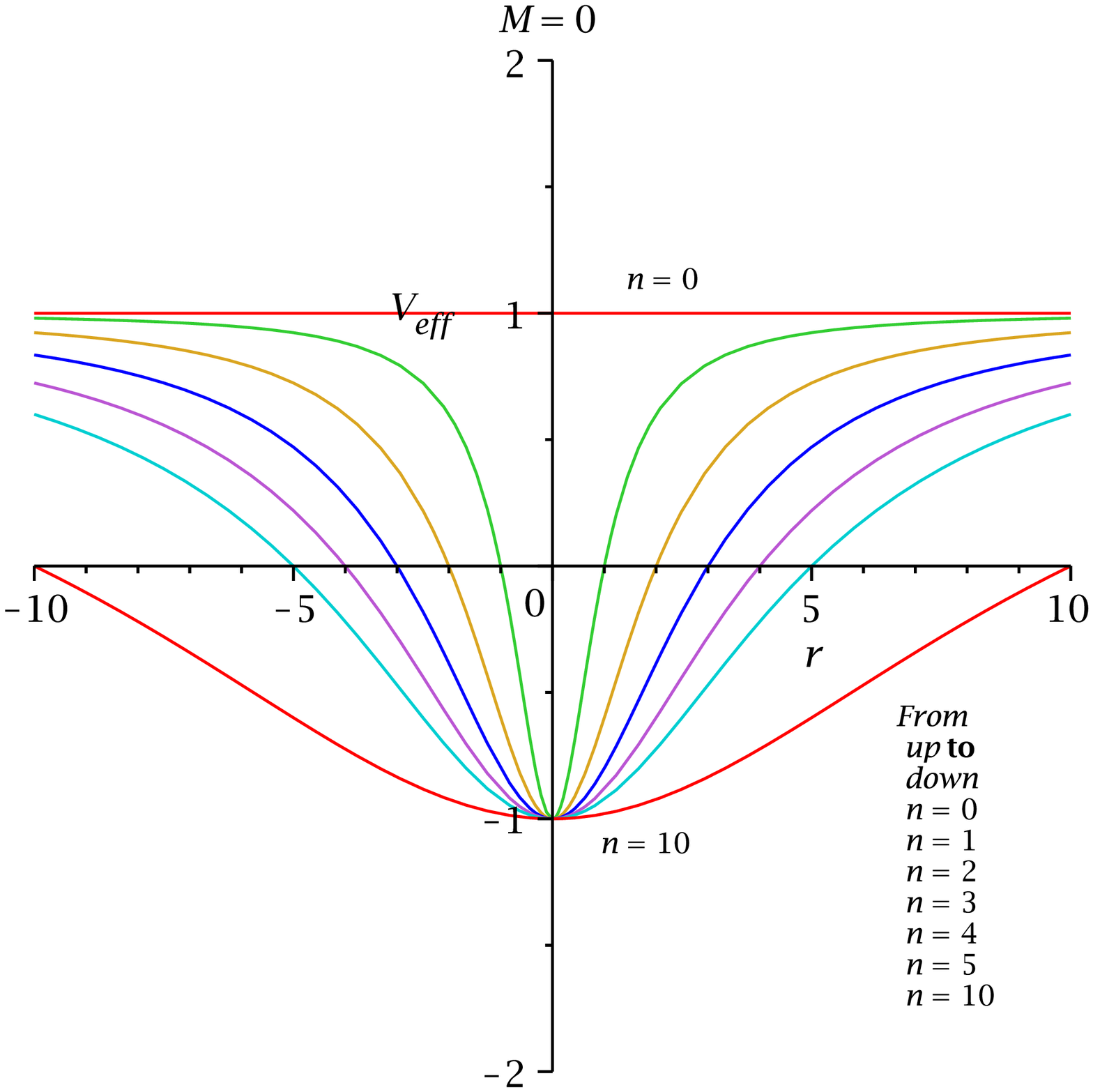}}
\end{center}
\caption{The figure shows the variation  of $V_{eff}$  with $r$ for TN BH and mass-less TN BH.
\label{t1}}
\end{figure}

For time-like geodesic we have to set $\epsilon=-1$ then the effective potential  becomes 
\begin{eqnarray}
V_{eff} &=& \left(\frac{r^2-2Mr-n^2}{r^2+n^2}\right) \left(1+\frac{L^{2}}{r^2+n^2}\right) ~.\label{n8}
\end{eqnarray}
The qualitative behaviour of the test particle can be obtain by studying this potential. First we consider 
the zero angular momentum geodesics for this the effective potential reduces to 
\begin{eqnarray}
V_{eff} &=& \left(\frac{r^2-2Mr-n^2}{r^2+n^2}\right)  ~.\label{n9}
\end{eqnarray}
One can observe the qualitative behaviour of the geodesics in the presence of dual mass and with out dual mass 
and also the mass-lees parameter. This can be seen from the Fig. \ref{t1}.
In this plot, we can see that the presence of the dual mass parameter deforms the shape of the 
radial effective potential in comparison with zero dual mass parameter. This behaviour  
can be seen from the mass-less case also.

\begin{figure}
\begin{center}
{\includegraphics[width=0.45\textwidth]{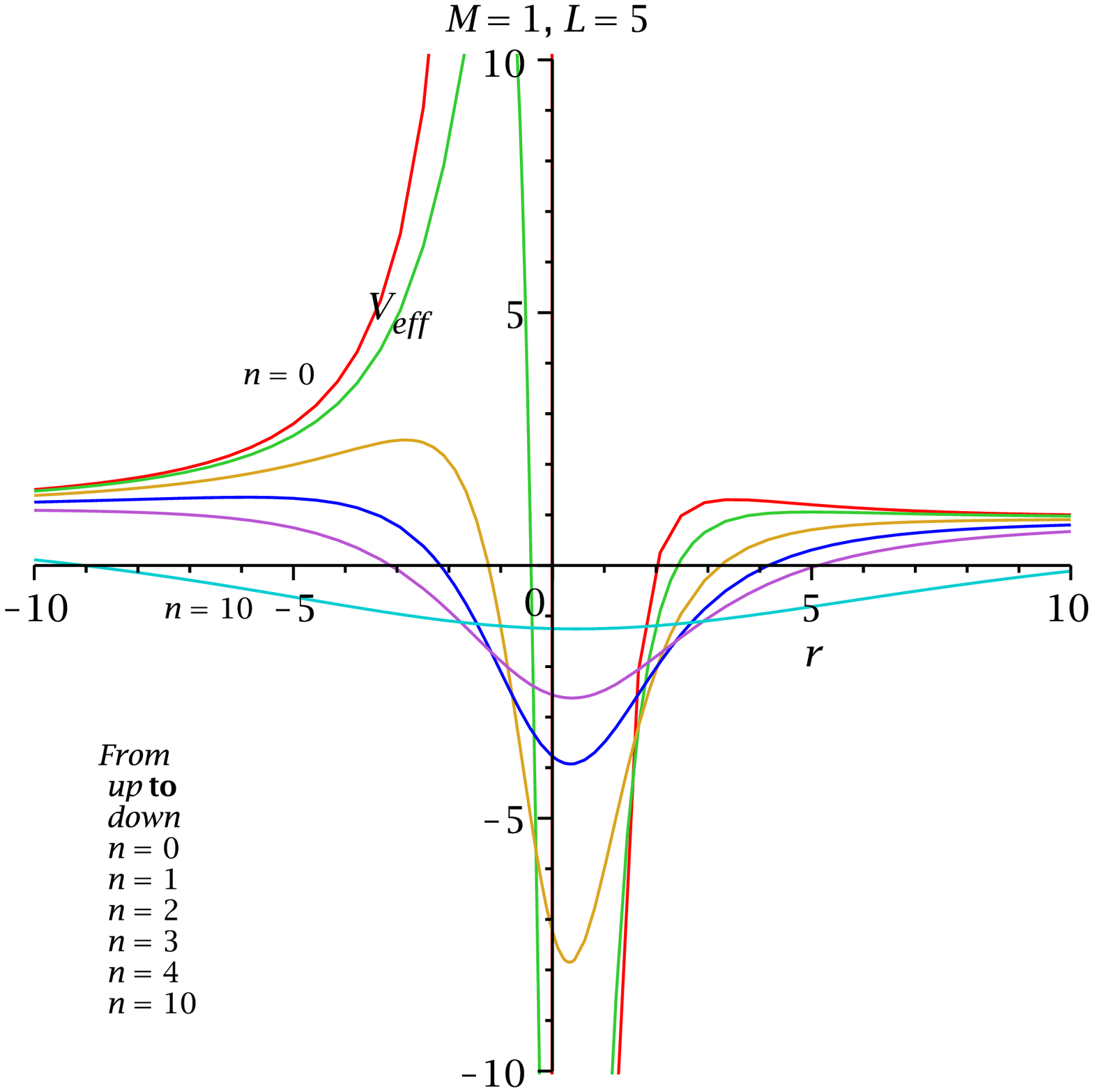}}
{\includegraphics[width=0.45\textwidth]{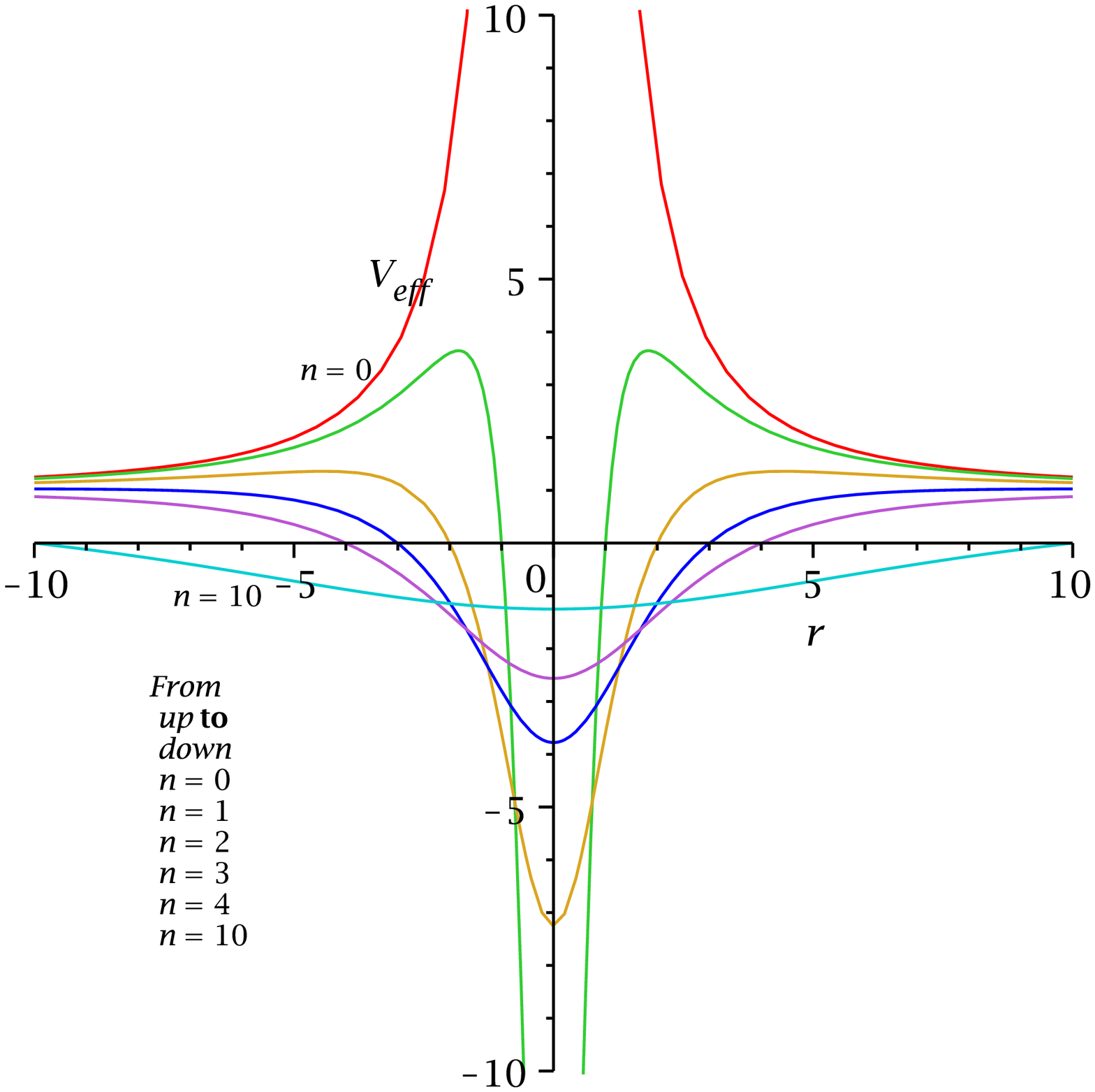}}
\end{center}
\caption{The figure shows the variation  of $V_{eff}$  with $r$ for TN BH and mass-less TN BH.
\label{t5}}
\end{figure}

\begin{figure}
\begin{center}
{\includegraphics[width=0.45\textwidth]{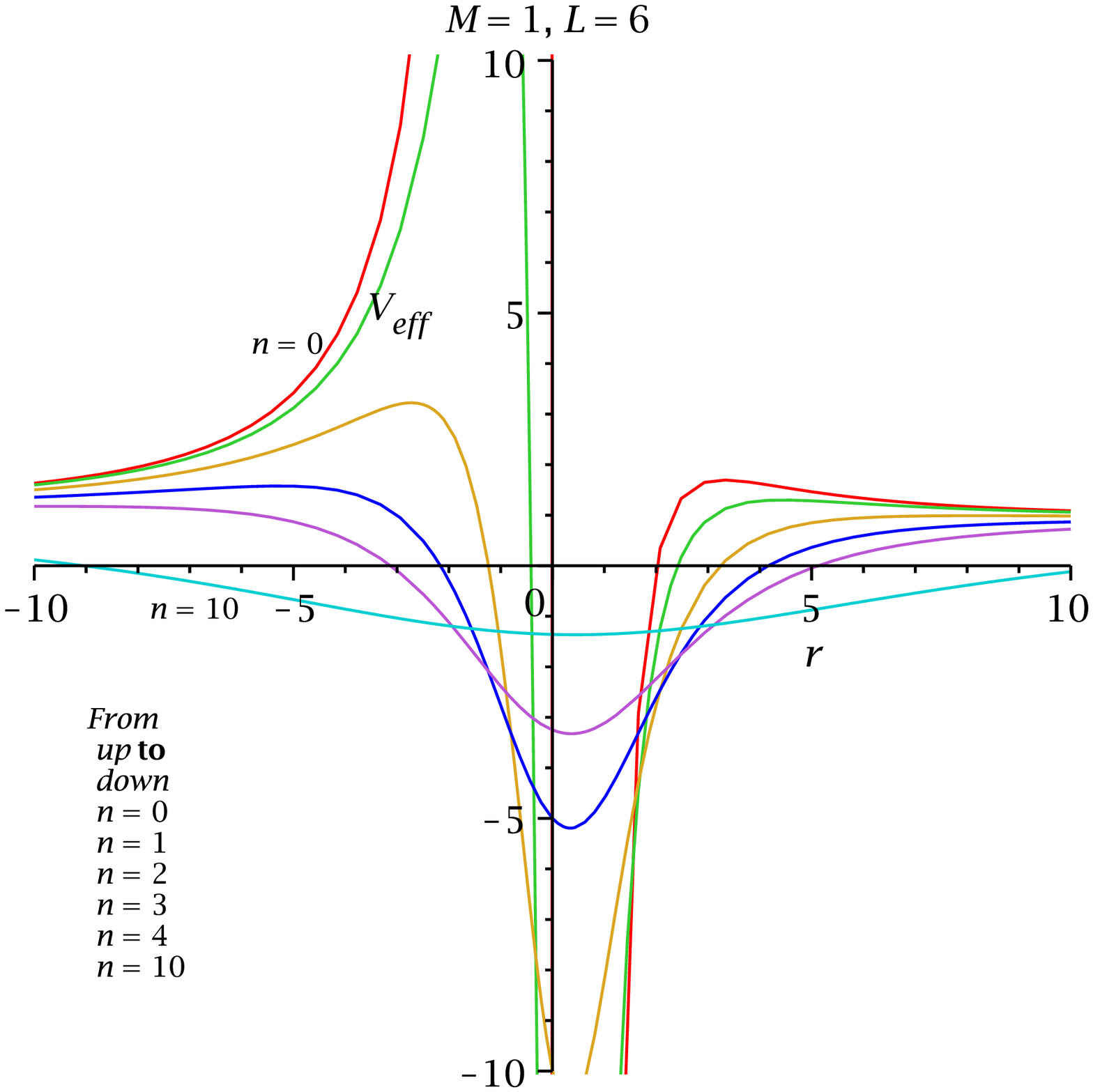}}
{\includegraphics[width=0.45\textwidth]{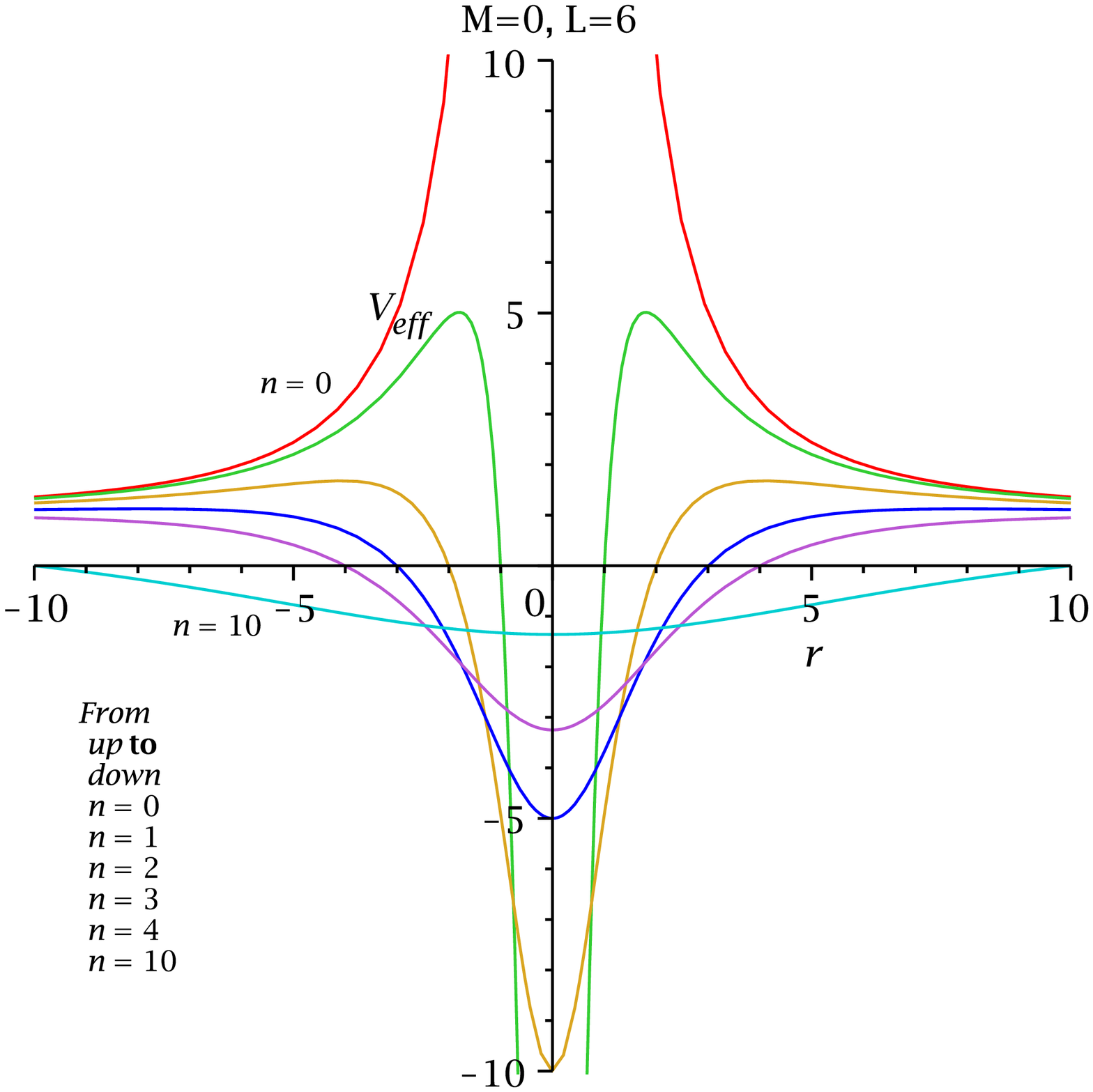}}
\end{center}
\caption{The figure shows the variation  of $V_{eff}$  with $r$ for  TN BH and mass-less TN BH.
\label{t6}}
\end{figure}

\begin{figure}
\begin{center}
{\includegraphics[width=0.45\textwidth]{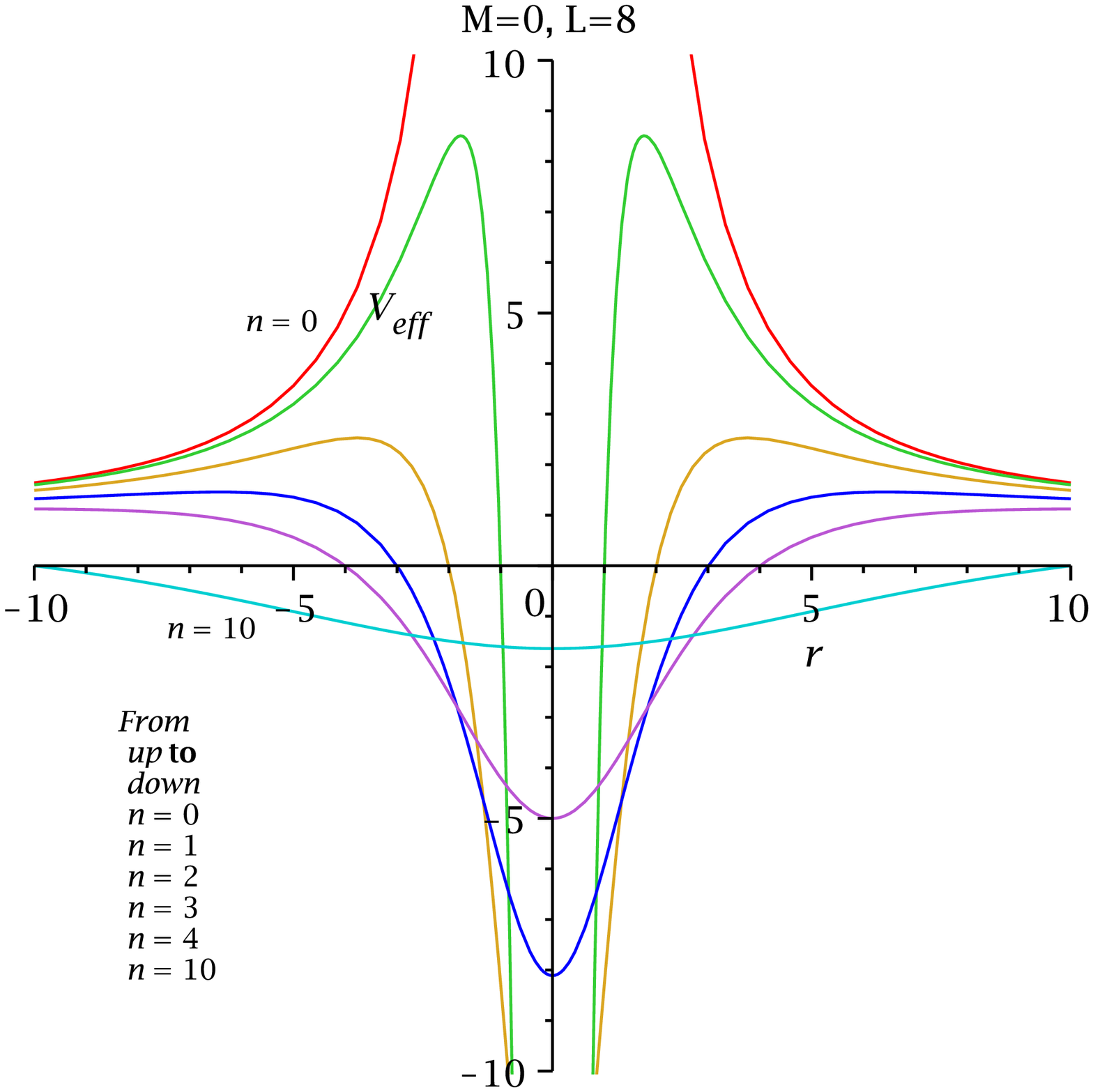}}
{\includegraphics[width=0.45\textwidth]{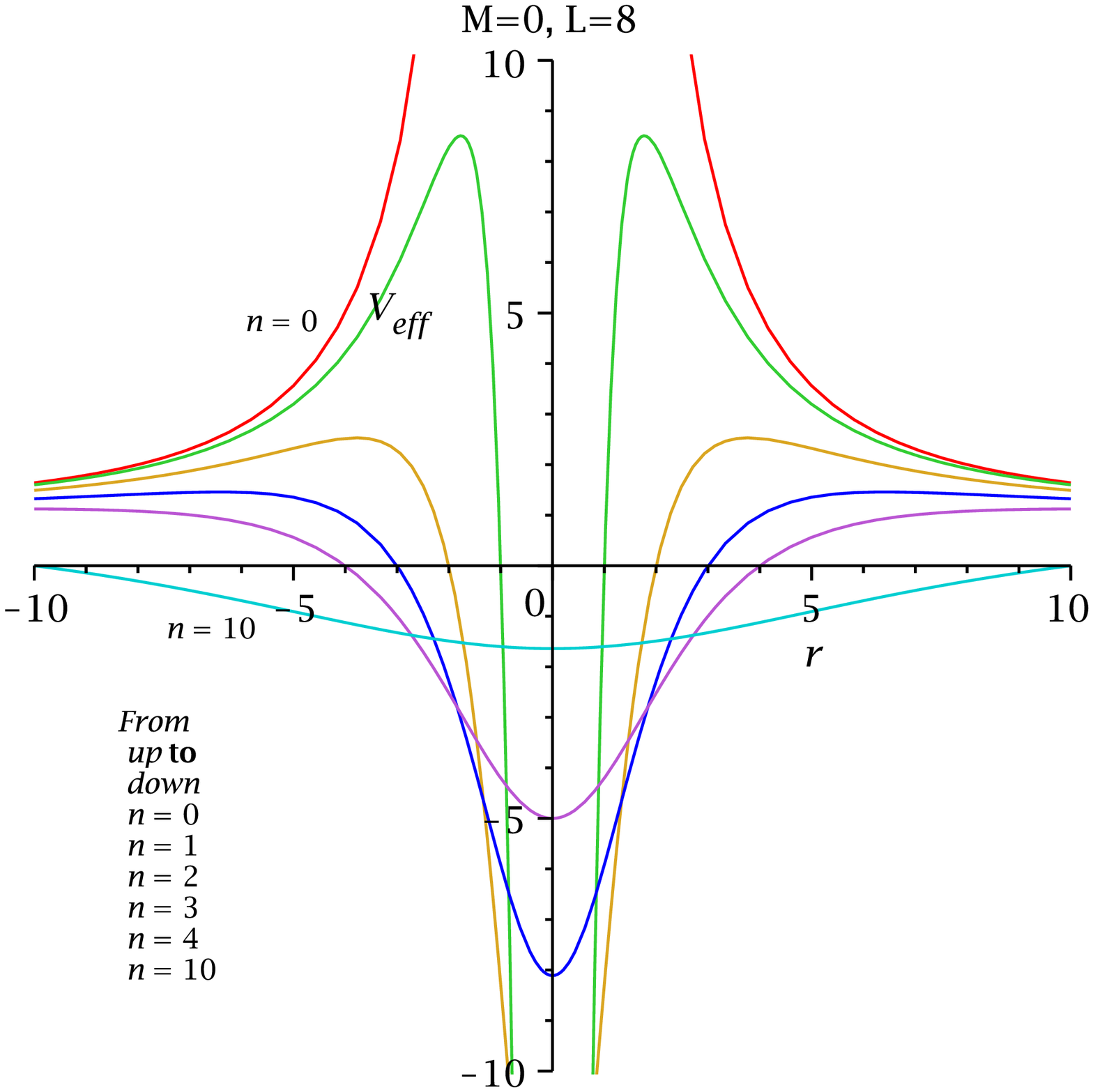}}
\end{center}
\caption{The figure shows the variation  of $V_{eff}$  with $r$ for TN BH and mass-less TN BH.
\label{t7}}
\end{figure}

\begin{figure}
\begin{center}
{\includegraphics[width=0.45\textwidth]{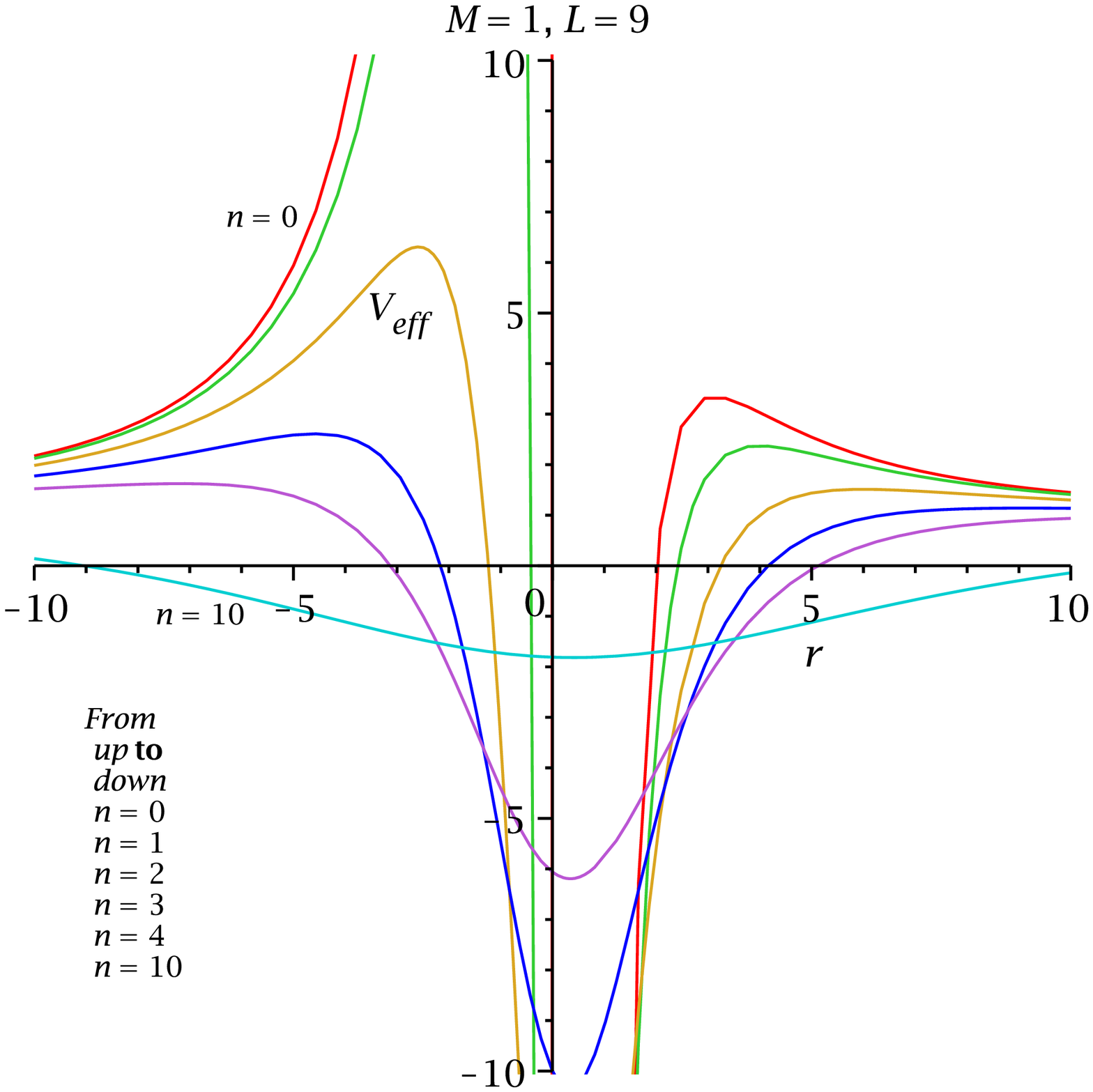}}
{\includegraphics[width=0.45\textwidth]{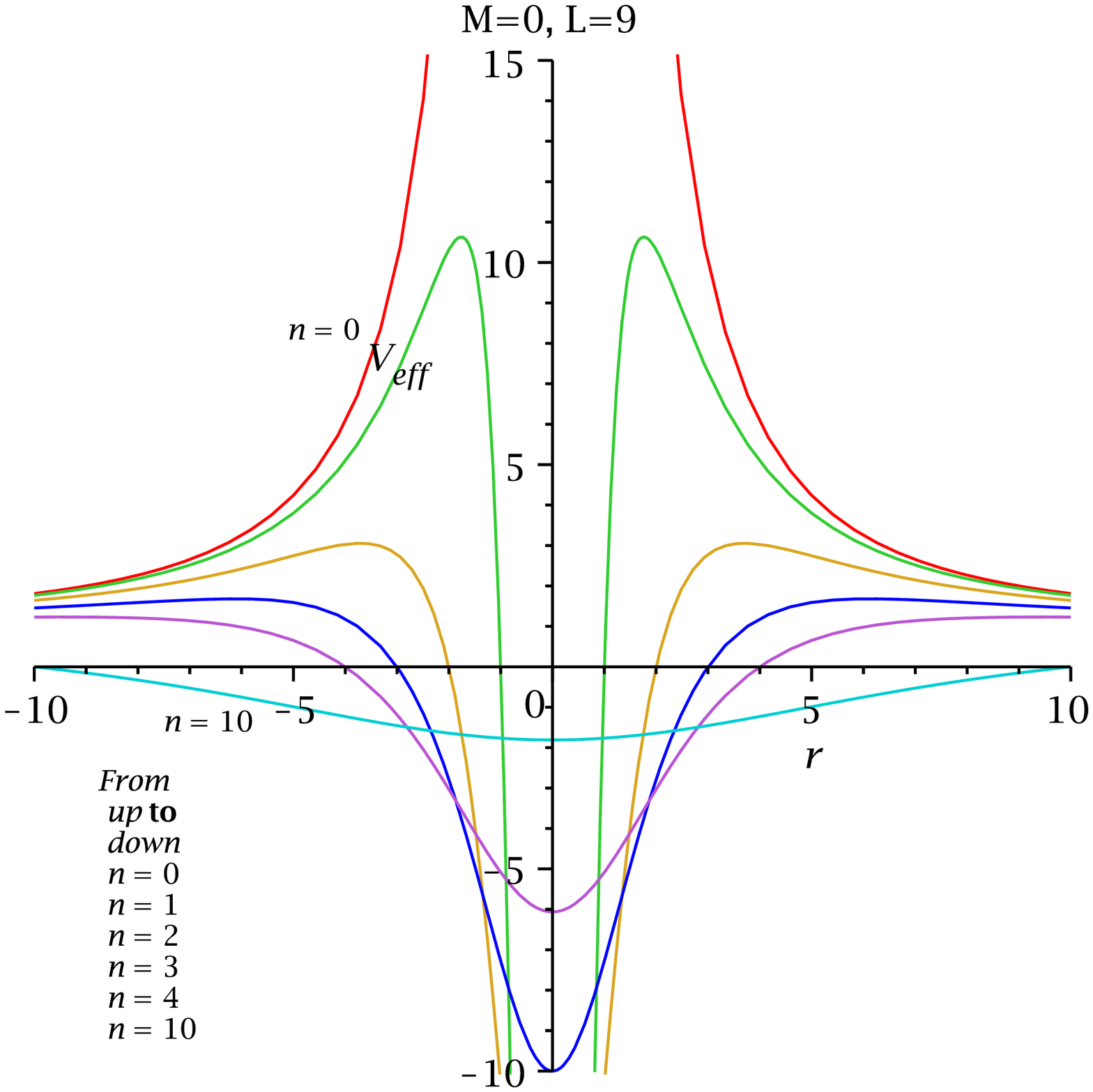}}
\end{center}
\caption{The figure shows the variation  of $V_{eff}$  with $r$ for TN BH and mass-less TN BH.
\label{t8}}
\end{figure}

\begin{figure}
\begin{center}
{\includegraphics[width=0.45\textwidth]{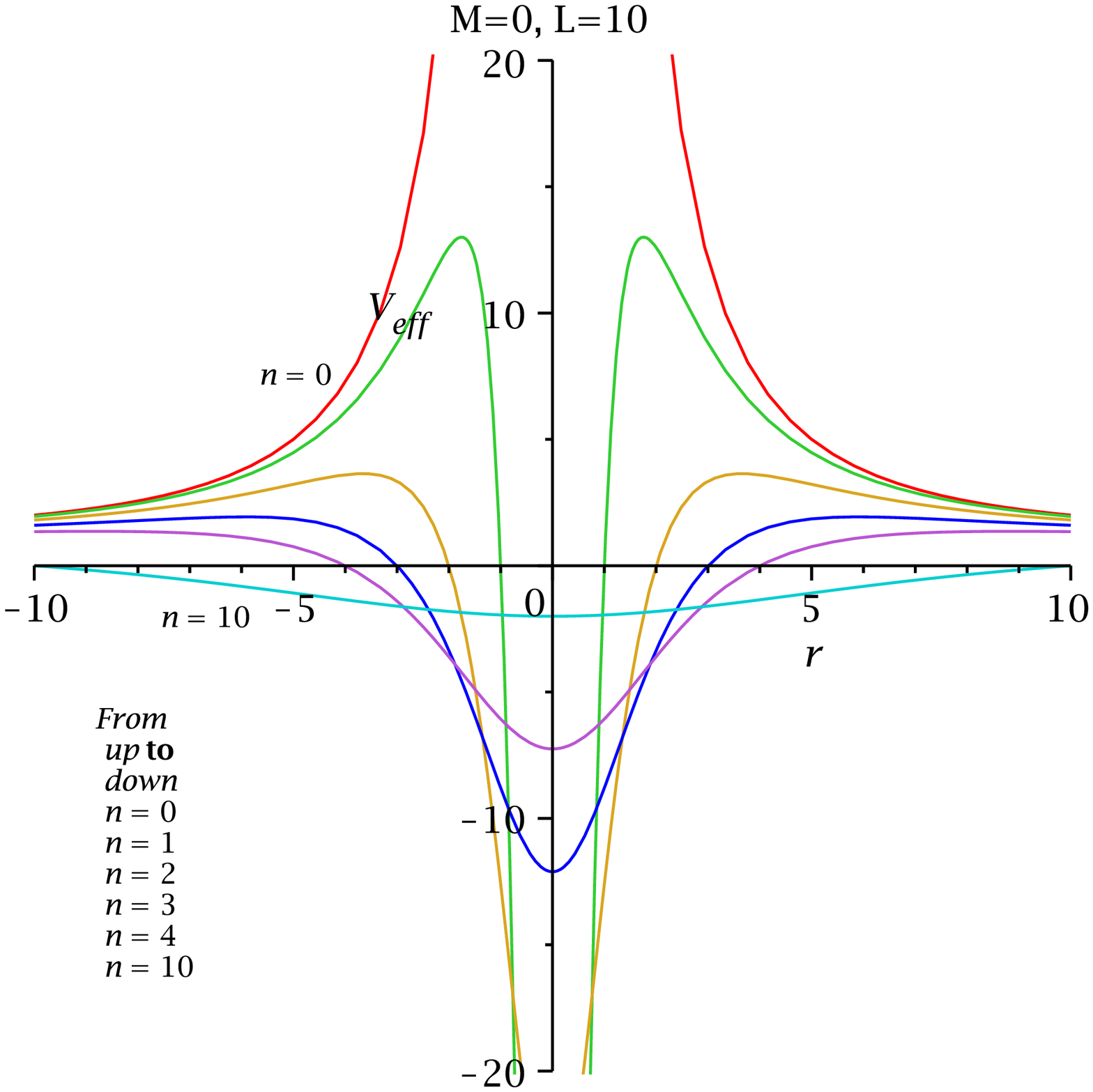}}
{\includegraphics[width=0.45\textwidth]{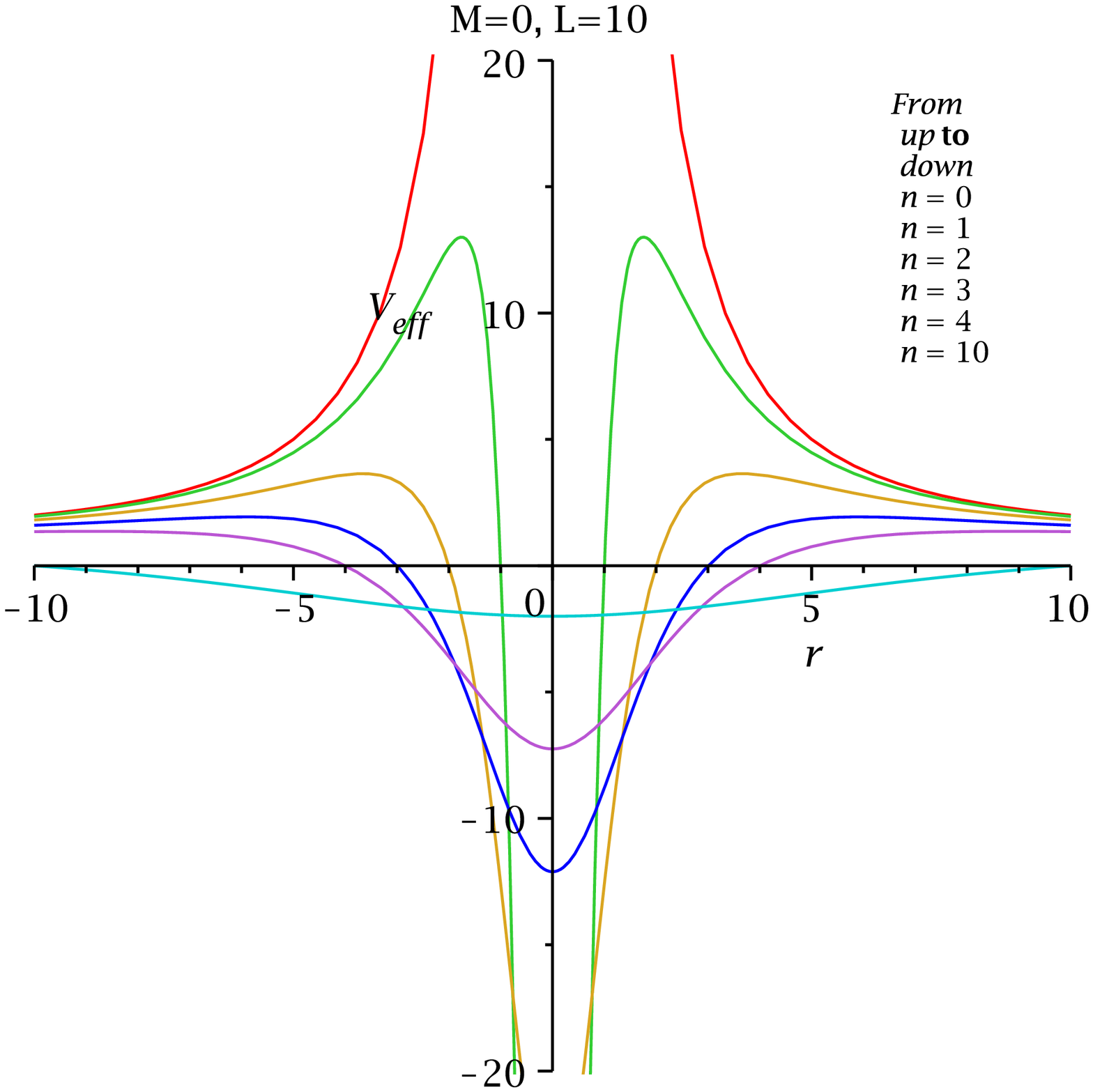}}
\end{center}
\caption{The figure shows the variation  of $V_{eff}$  with $r$ for TN BH and mass-less TN BH.
\label{t9}}
\end{figure}

For $L\neq 0$, the behaviour of the test particle in the potential well could be seen from the following 
Fig. \ref{t2}, Fig. \ref{t3}, Fig. \ref{t4}, Fig. \ref{t5}, Fig. \ref{t6}, Fig. \ref{t7}, Fig. \ref{t8} and  
Fig. \ref{t9}.
\begin{figure}
\begin{center}
{\includegraphics[width=0.45\textwidth]{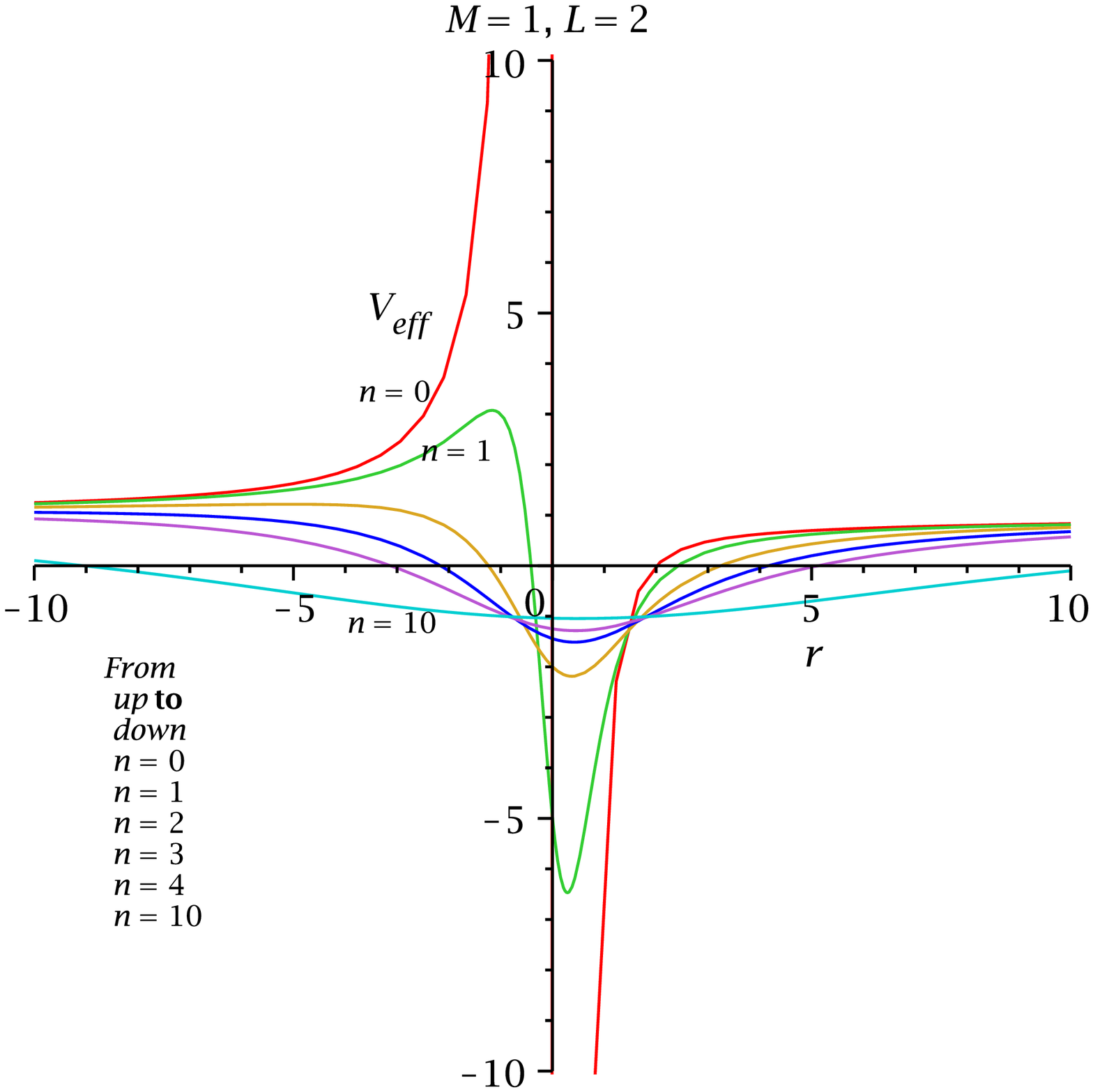}}
{\includegraphics[width=0.45\textwidth]{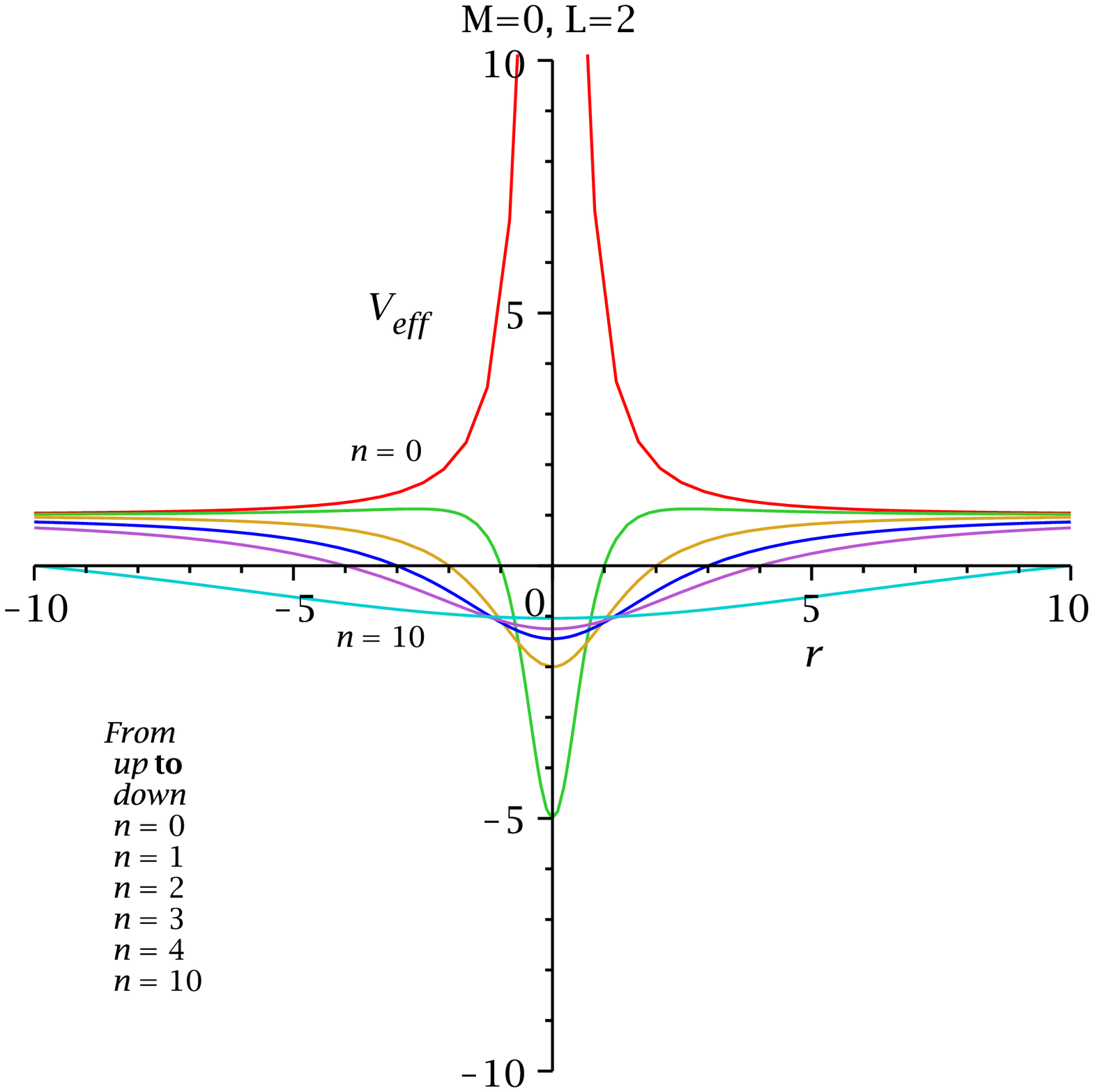}}
\end{center}
\caption{The figure shows the variation  of $V_{eff}$  with $r$ for TN BH and mass-less TN BH.
\label{t2}}
\end{figure}
\begin{figure}
\begin{center}
{\includegraphics[width=0.45\textwidth]{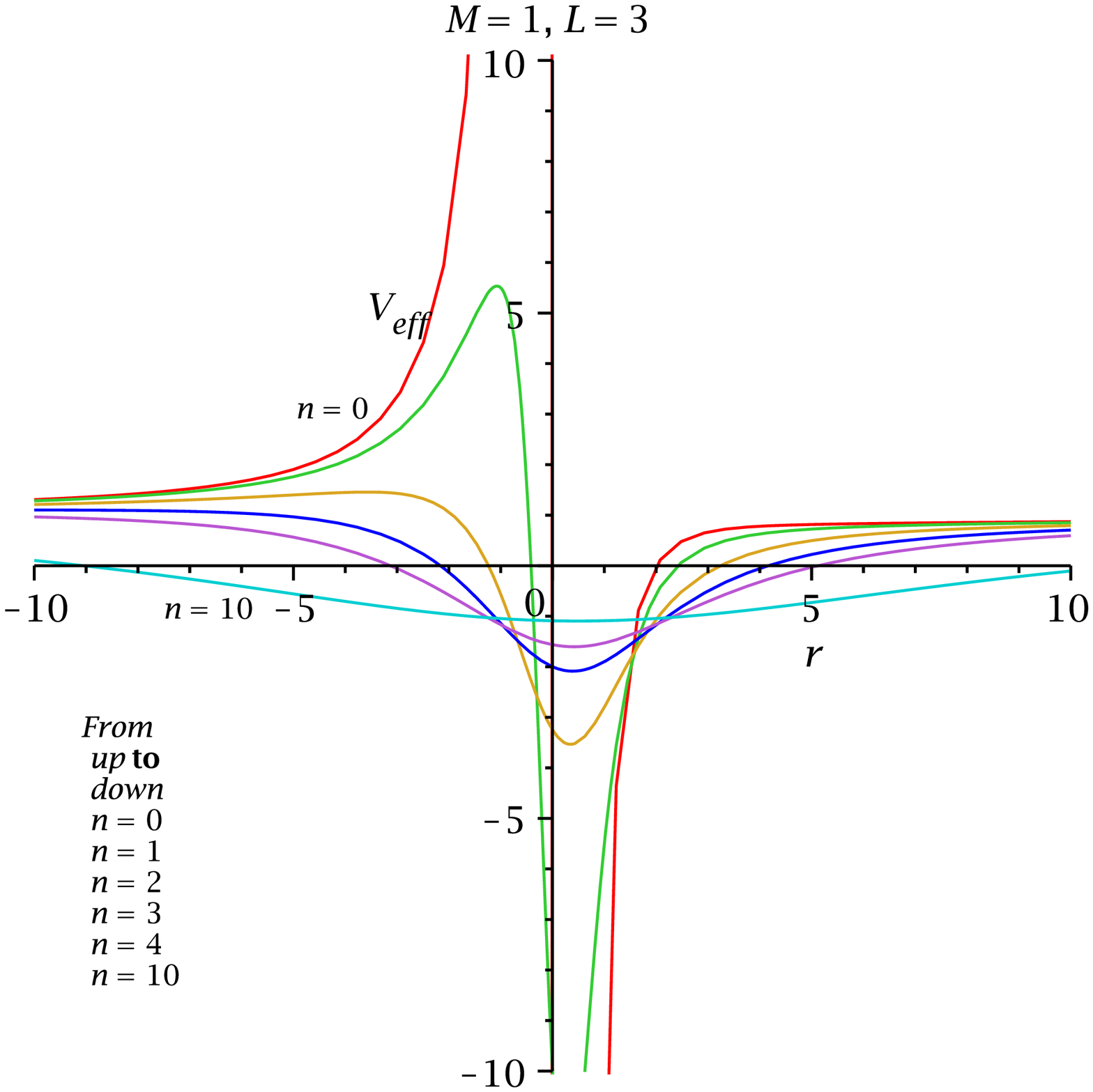}}
{\includegraphics[width=0.45\textwidth]{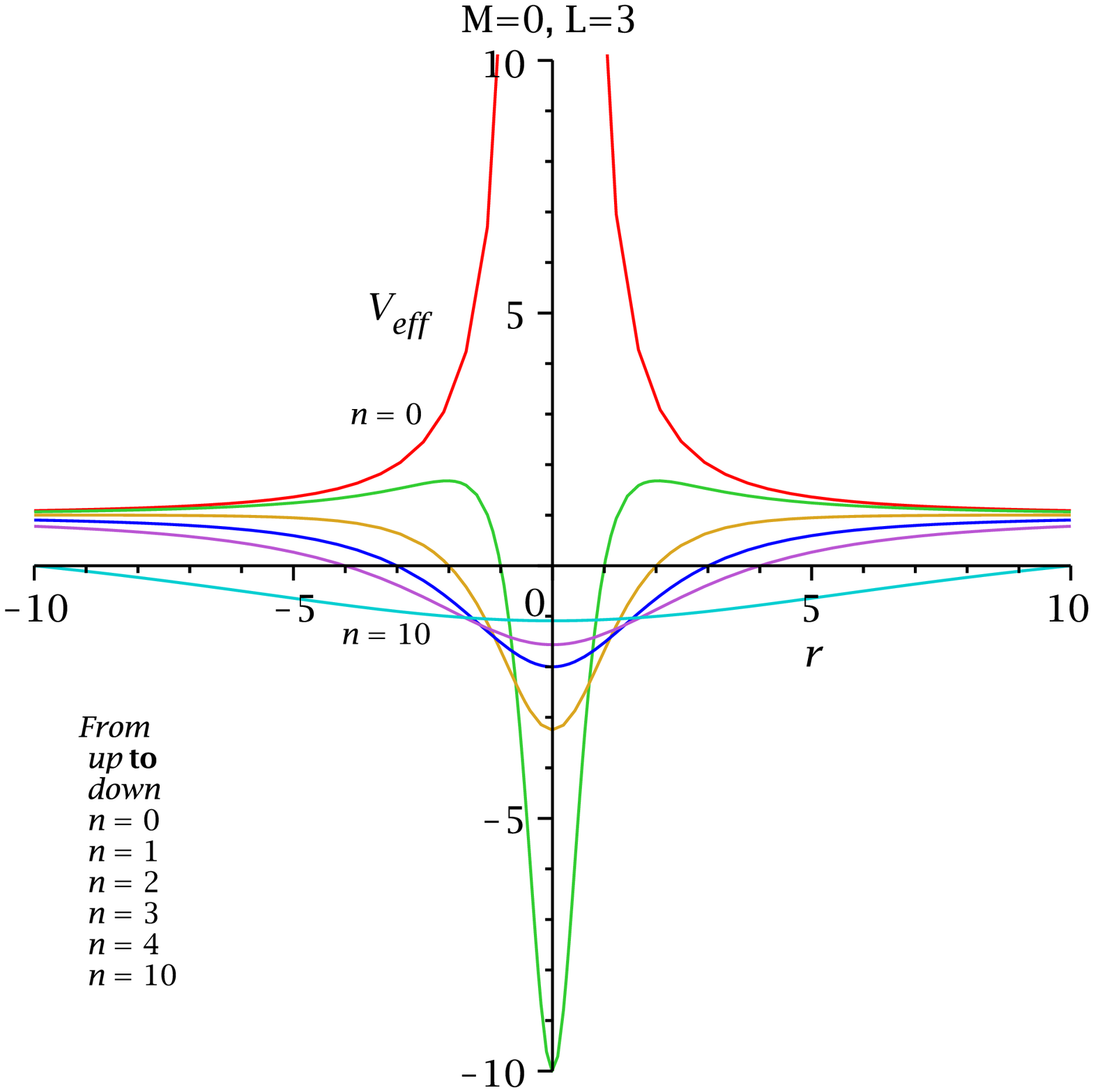}}
\end{center}
\caption{The figure shows the variation  of $V_{eff}$  with $r$ for  TN BH and mass-less TN BH.
NUT parameter.
\label{t3}}
\end{figure}
\begin{figure}
\begin{center}
{\includegraphics[width=0.45\textwidth]{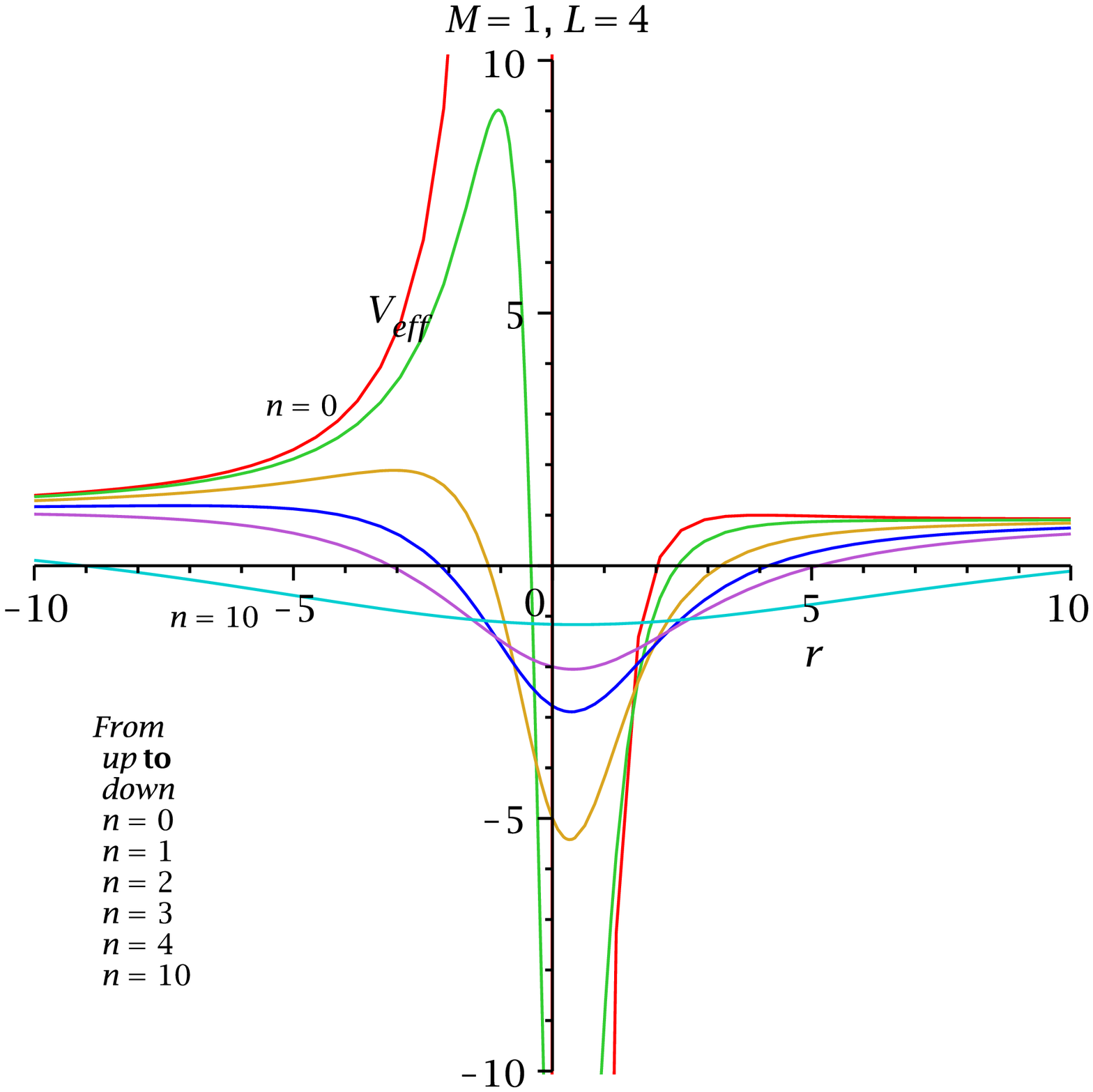}}
{\includegraphics[width=0.45\textwidth]{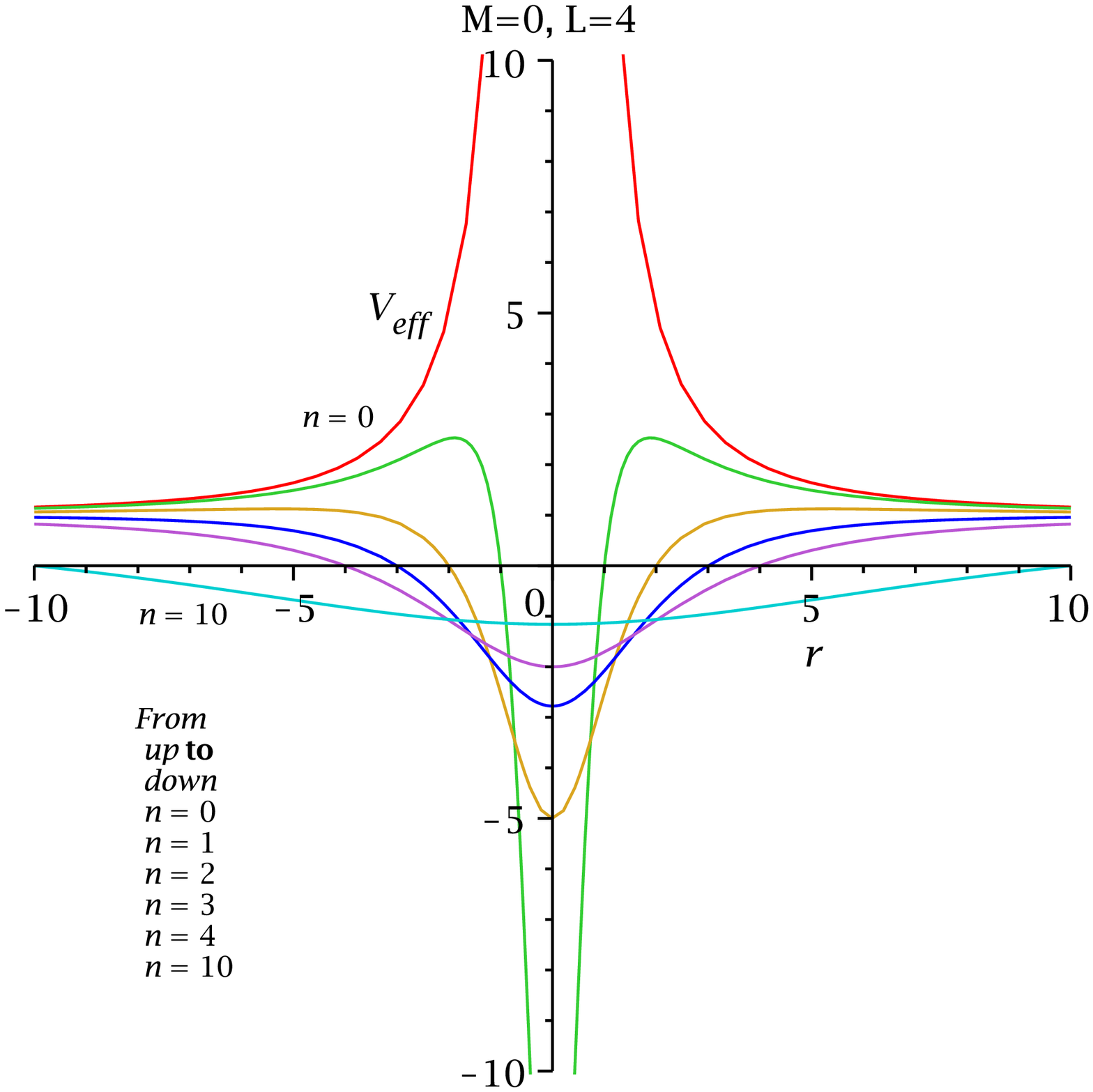}}
\end{center}
\caption{The figure shows the variation  of $V_{eff}$  with $r$ for  TN BH and mass-less TN BH.
\label{t4}}
\end{figure}
In the above figures, we have seen that the radial dependency of the effective potential with 
dual mass parameter, without dual mass parameter and $M=0$. 
When $n=0$, the effective potential at large radial distance does not change 
much more with the increasing of $L$. When we introduced the NUT parameter,
the shape of the effective potential \emph{deforms} in comparison with NUT less case and it also changes 
for different values of angular momentum parameter.
Finally, when we increase the value of $n$ the height of the potential well decreases. This work has been 
done earlier for the parameter $a$, $Q$ and $n$ \cite{cqg}. This work is specially for 
the parameter values $a=Q=0$ and for mass-less cases which has been not 
studied previously.

\begin{figure}
\begin{center}
{\includegraphics[width=0.45\textwidth]{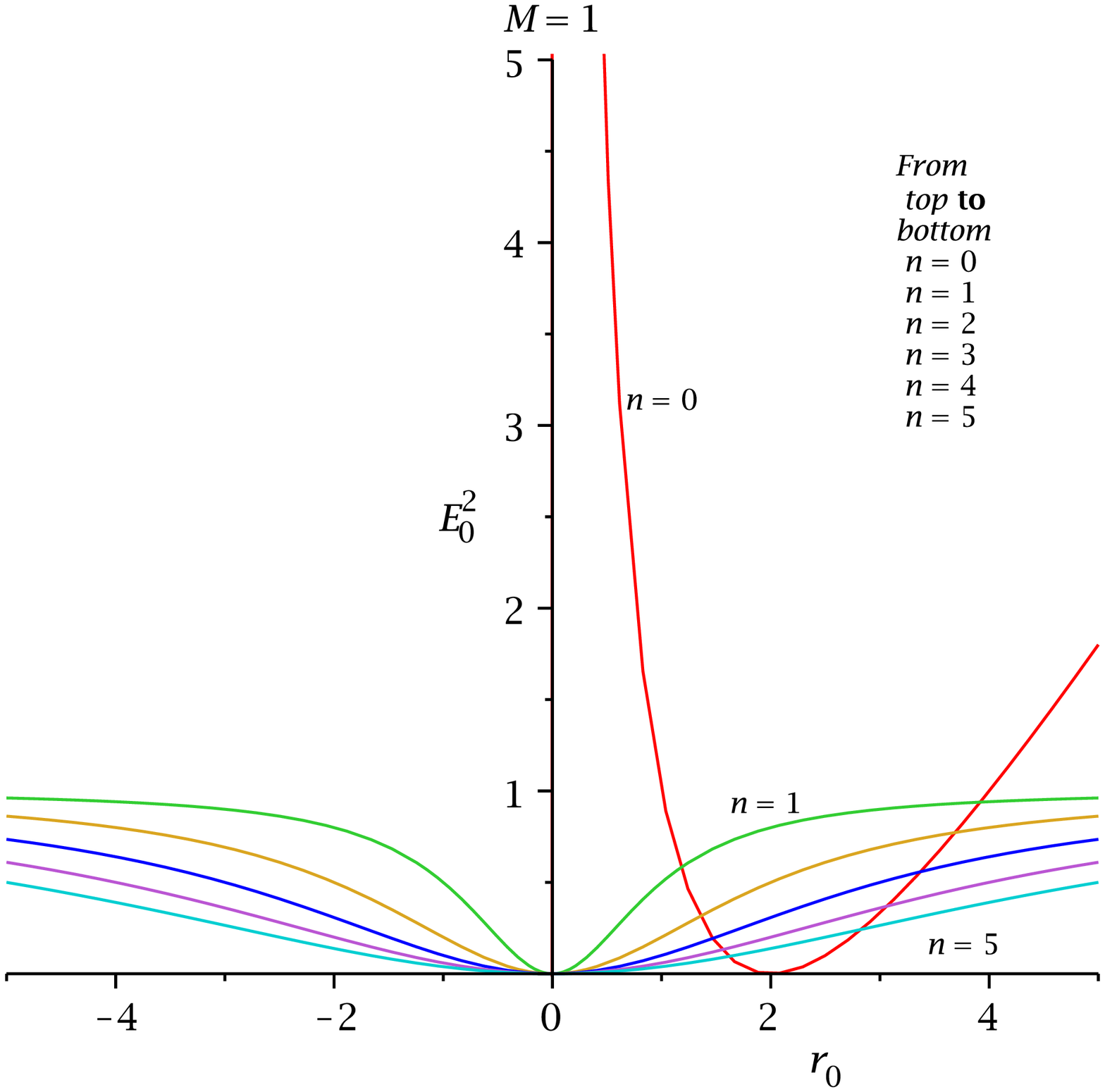}}
{\includegraphics[width=0.45\textwidth]{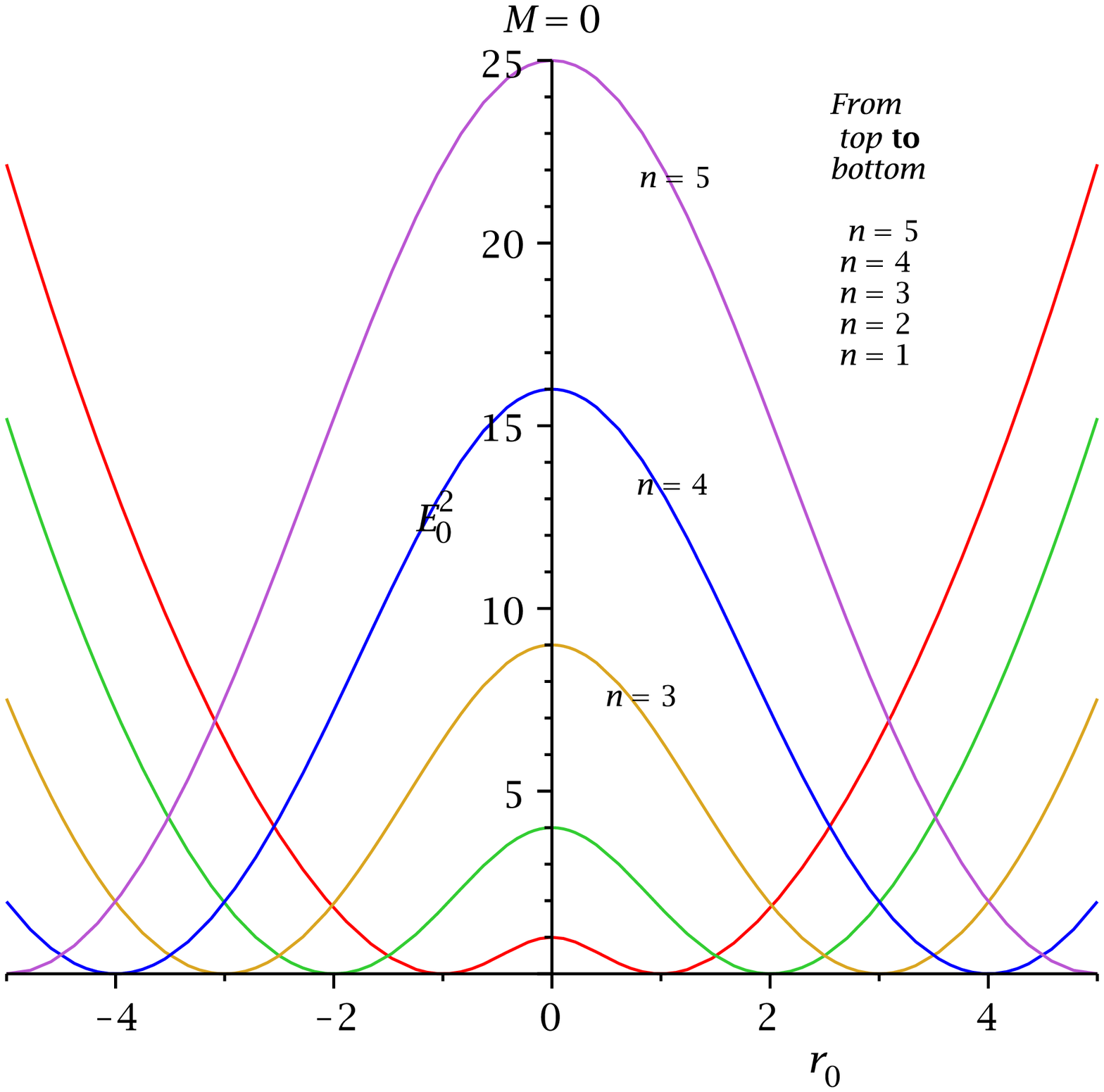}}
\end{center}
\caption{The figure depicts the variation  of $E_{0}^2$  with $r_{0}$ for TN BH and mass-less TN BH.
\label{ff}}
\end{figure}

\begin{figure}
\begin{center}
{\includegraphics[width=0.45\textwidth]{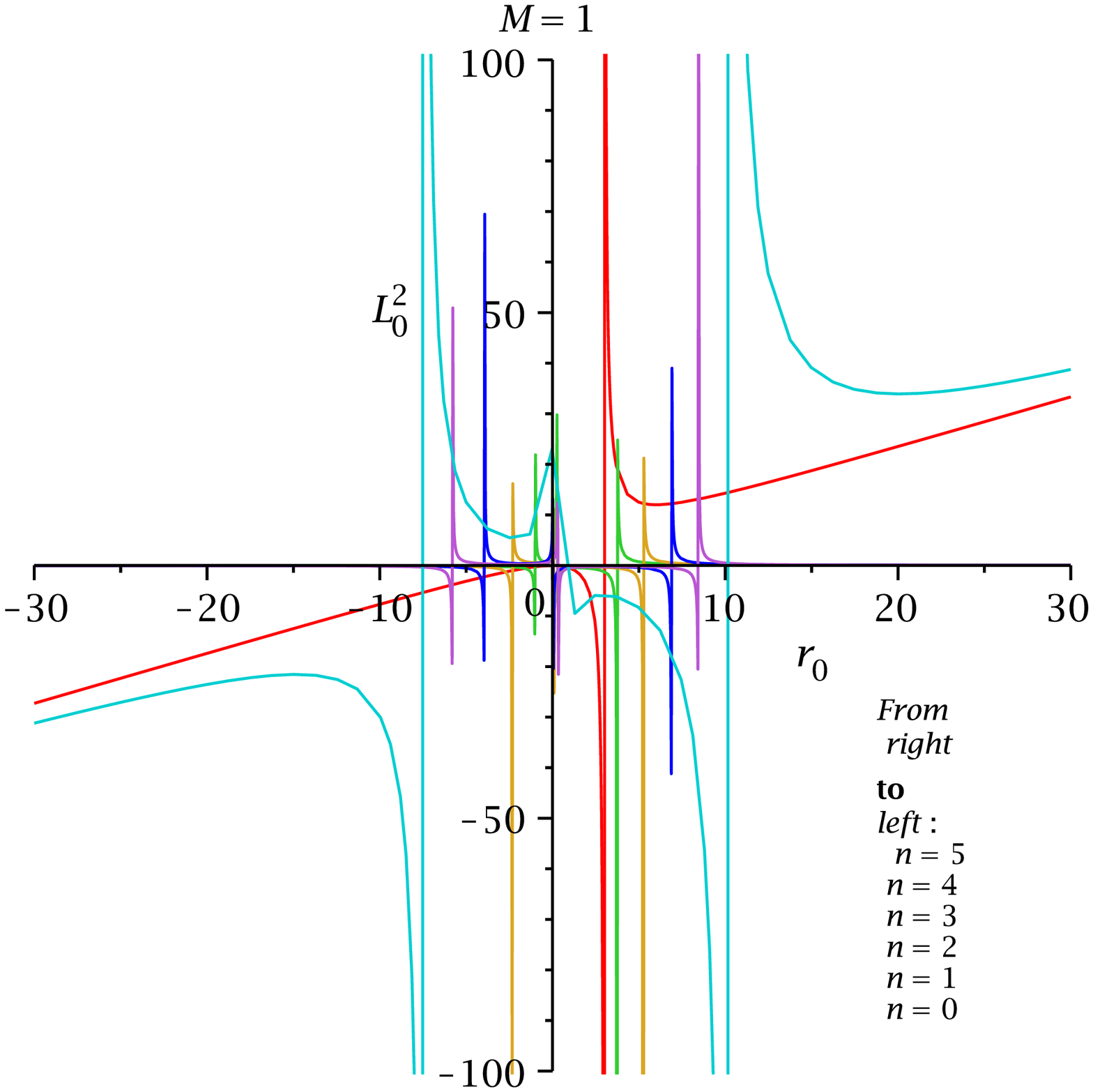}}
{\includegraphics[width=0.45\textwidth]{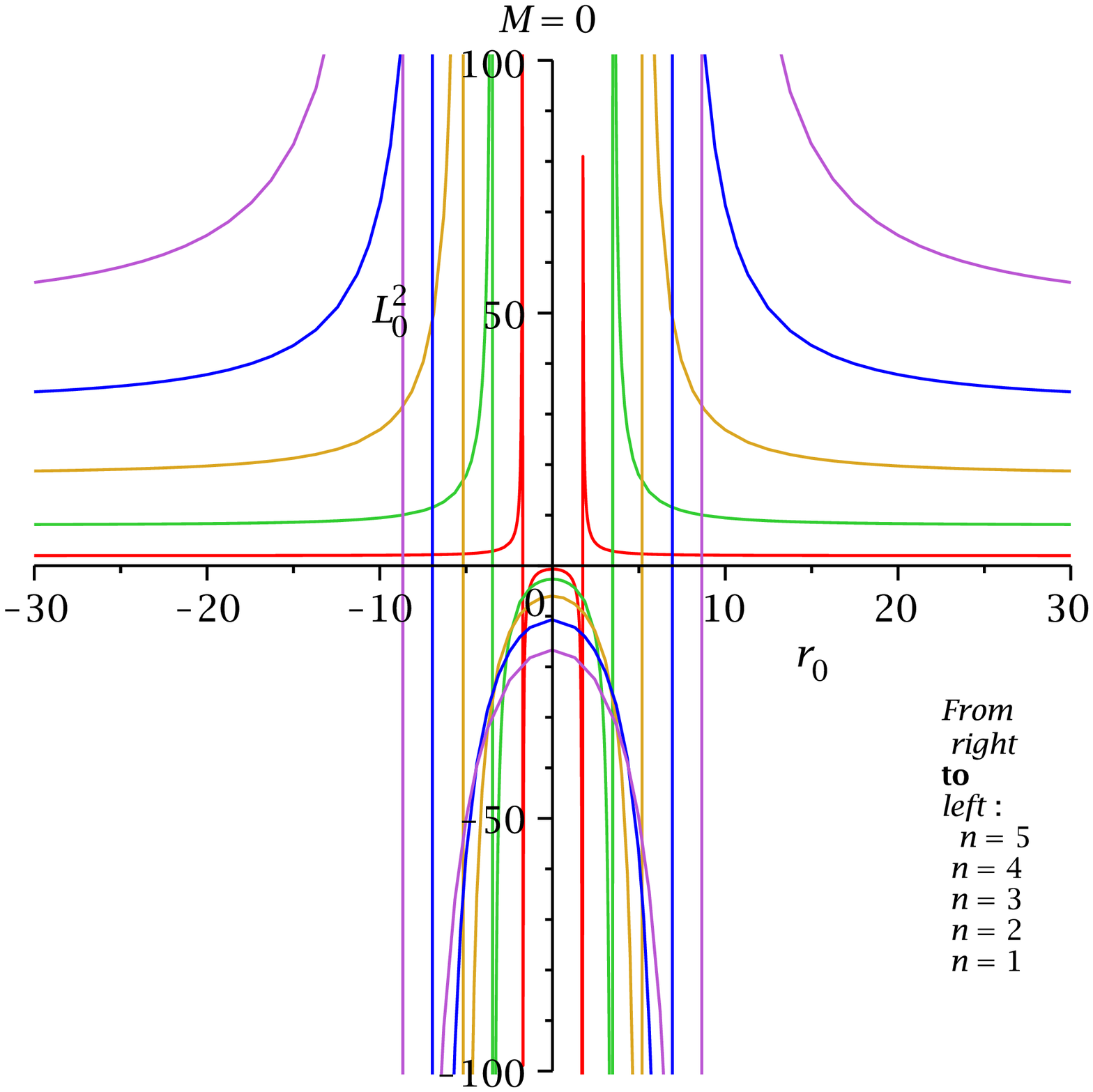}}
\end{center}
\caption{The figure depicts the variation  of $E_{0}^2$  with $r_{0}$ for TN BH and mass-less TN BH.
\label{ff1}}
\end{figure}

Now re-write the  Eq. (\ref{tnr}) for time-like geodesics as 
$$
\dot{r}^{2} \left(r^2+n^2\right)^2 = \left(E^2-1\right)\left(r^2+n^2\right)^2+2Mr\left(r^2+n^2\right)-L^2(r^2-2Mr-n^2)
$$
\begin{eqnarray}
+2n^2(r^2+n^2) ~.\label{n10}
\end{eqnarray}
For circular geodesics $r=r_{0}$, we have the following condition:
\begin{eqnarray}
\dot{r}^2|_{r=r_{0}}  &=& \frac{d{\dot{r}^2}}{dr}|_{r=r_{0}} =0 ~. \label{ef}
\end{eqnarray}
From this condition, we find the energy and angular momentum for circular orbit:
\begin{eqnarray}
E_{0}^2  &=& \frac{r_{0}\left(r_{0}^2 -2Mr_{0} -n^2\right)^2}{(r_{0}^2+n^2)\left(r_{0}^3-3Mr_{0}^2-3n^2r_{0}+Mn^2\right)}  
~. \label{ef1}
\end{eqnarray}
and 
\begin{eqnarray}
L_{0}^2  &=& \frac{\left(r_{0}^2+n^2\right)\left(Mr_{0}^2+2n^2r_{0}-Mn^2\right)}
{\left(r_{0}^3-3Mr_{0}^2-3n^2r_{0}+Mn^2\right)}  
~. \label{ef2}
\end{eqnarray}
These equations require for energy square and angular momentum square positive definite i.e. $r_{0}^3-3Mr_{0}^2-3n^2r_{0}+Mn^2>0$.
Compared with the Eq. (\ref{n6}), it implies that the minimum radius for time-like circular orbit is the radius 
of of the unstable CPO.
Interestingly for mass-less case, these values are 
\begin{eqnarray}
E_{0}^2  &=& \frac{\left(r_{0}^2 -n^2\right)^2}{(r_{0}^2+n^2)\left(r_{0}^3-3n^2r_{0}\right)}  
~. \label{ef3}
\end{eqnarray}
and 
\begin{eqnarray}
L_{0}^2  &=& \frac{2n^2\left(r_{0}^2+n^2\right)}
{\left(r_{0}^2-3n^2\right)}  ~. \label{ef4}
\end{eqnarray}
We have plotted their behaviour in the Fig. \ref{ff} and Fig. \ref{ff1}.
and 

From Eq. (\ref{n10}), and for circular orbit one obtains
\begin{eqnarray}
L_{\pm} &=& \pm \sqrt{\frac{\left(E^2-1\right)\left(r_{0}^2+n^2\right)^2+2Mr_{0}\left(r_{0}^2+n^2\right)
+2n^2(r_{0}^2+n^2)}{r_{0}^2-2Mr_{0}-n^2}}~.\label{n11}
\end{eqnarray}
It should be noted that $L_{+}=-L_{-}$. From this expression, it follows that the angular momentum parameter 
explicitly dependes upon the energy value. Therefore there must be difference in the circular orbits in the 
$L-r$ plane for $E^2>1$, $E^2<1$ and $E^2=1$. We are interested here to look the behaviour of the circular 
geodesics in this plane by incorporating these energy conditions.  

\begin{figure}
\begin{center}
{\includegraphics[width=0.45\textwidth]{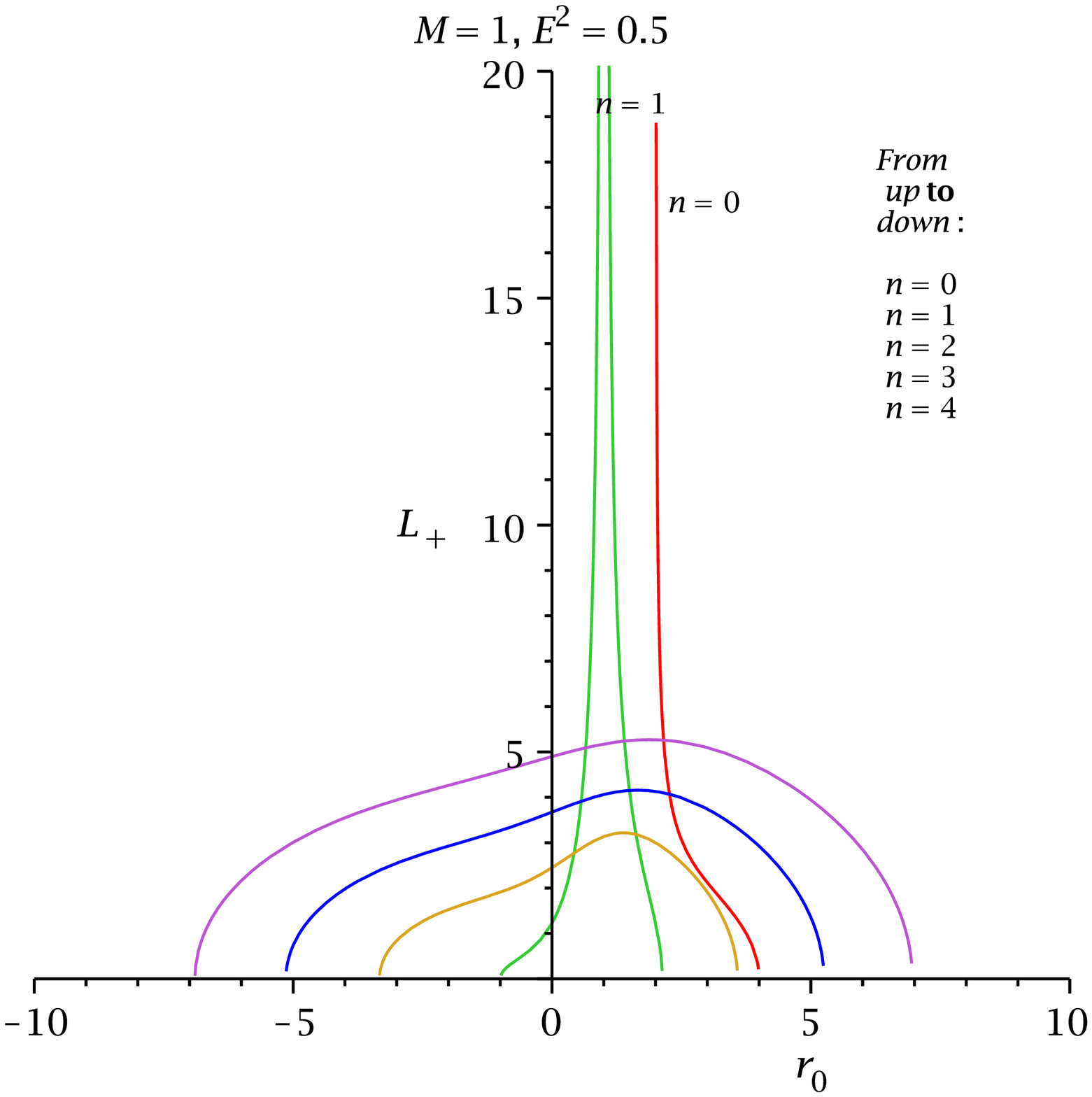}}
{\includegraphics[width=0.45\textwidth]{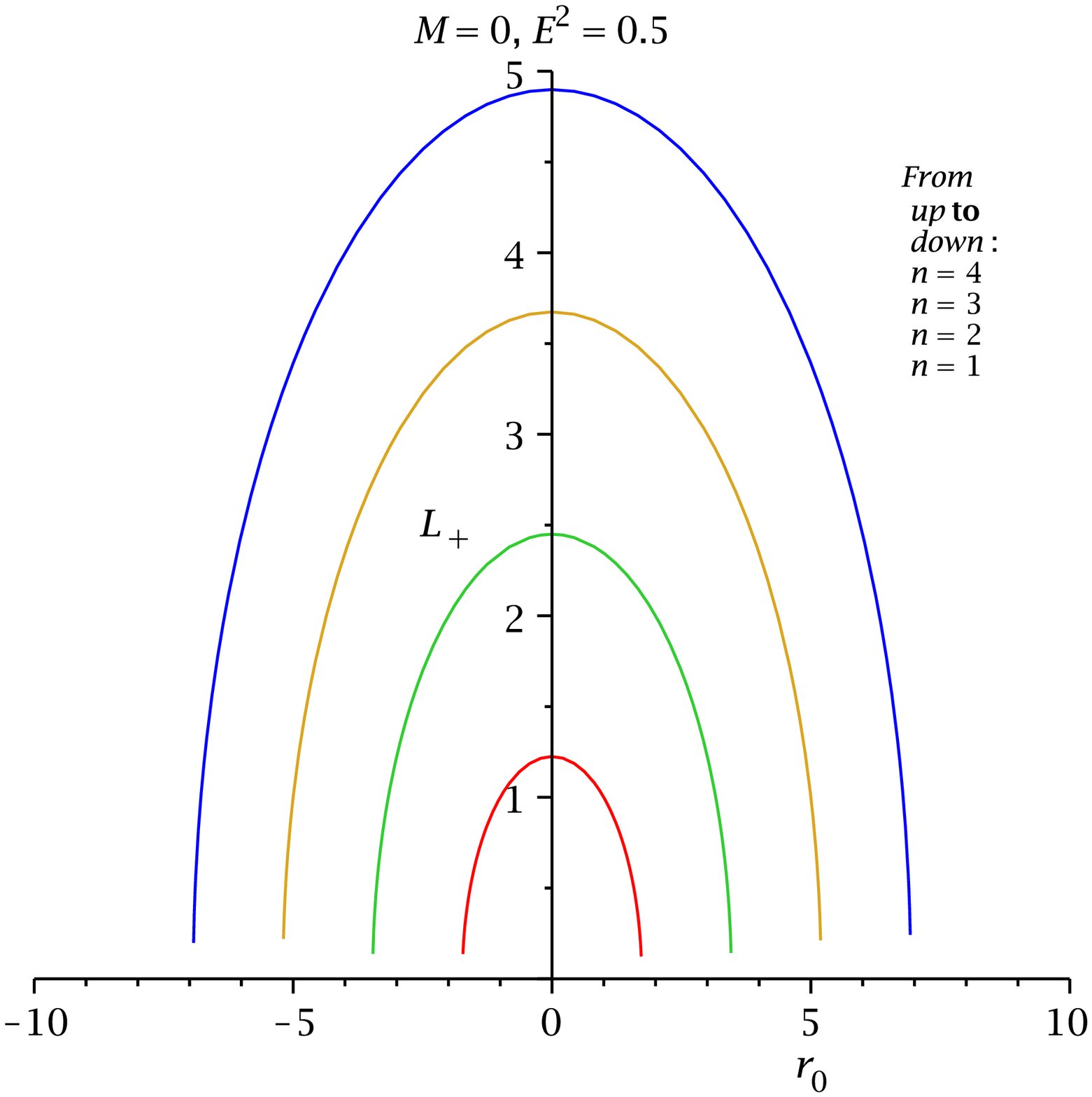}}
\end{center}
\caption{The figure shows the variation  of $L_{+}$  with $r_{0}$ for TN BH and mass-less TN BH.
\label{t10}}
\end{figure}

\begin{figure}
\begin{center}
{\includegraphics[width=0.45\textwidth]{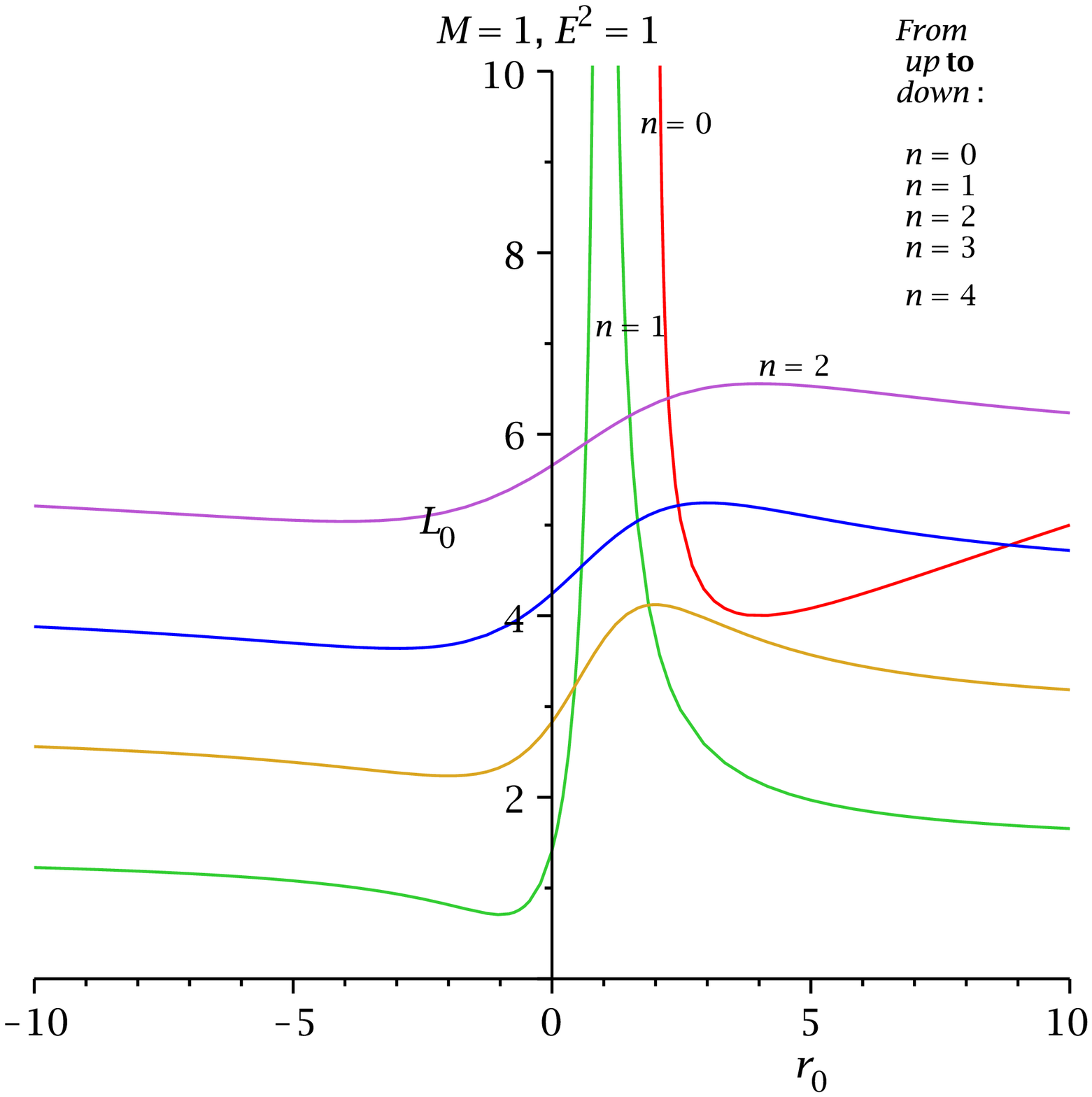}}
{\includegraphics[width=0.45\textwidth]{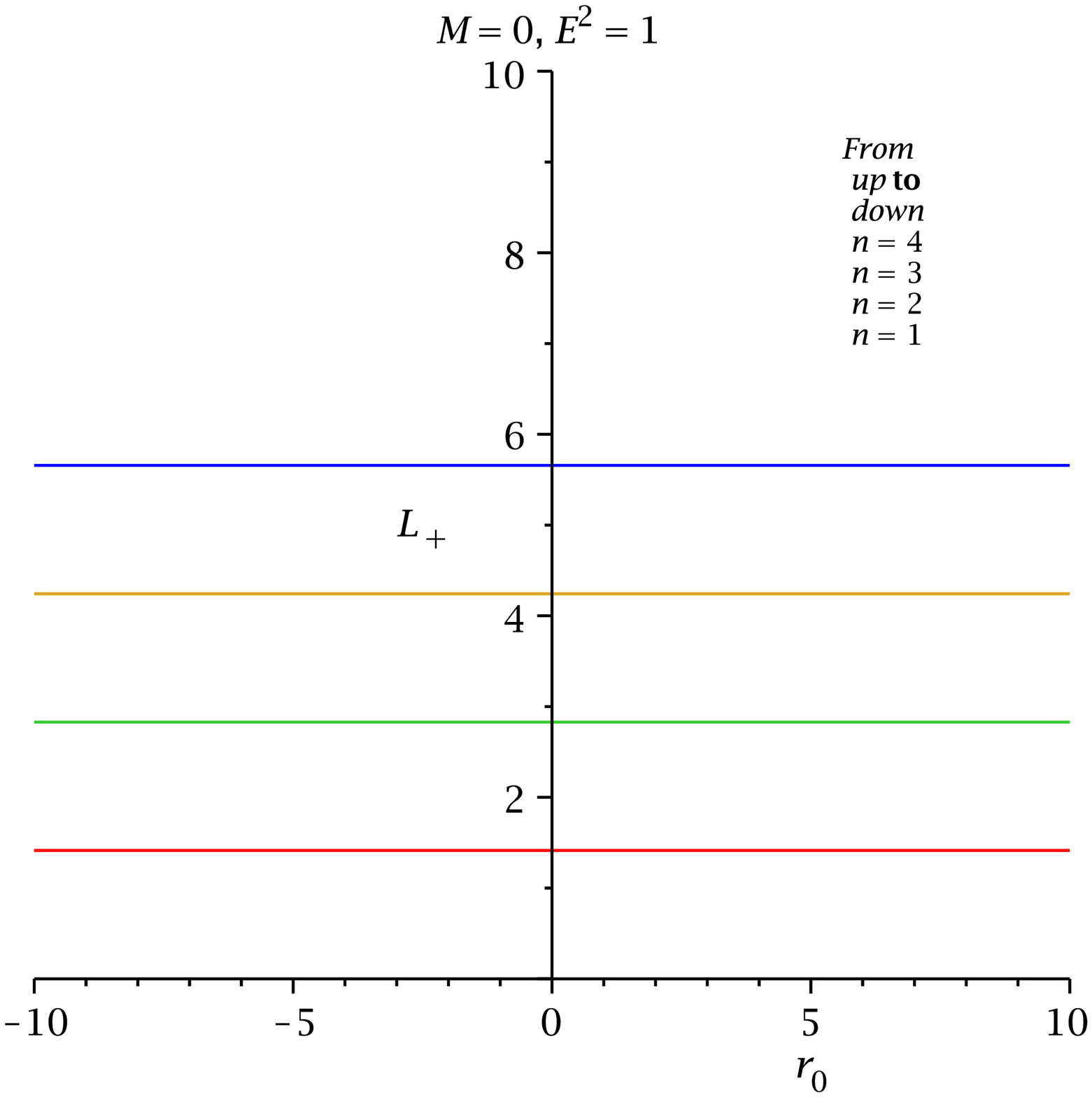}}
\end{center}
\caption{The figure shows the variation  of $L_{+}$  with $r_{0}$ for TN BH and mass-less TN BH.
\label{t11}}
\end{figure}

\begin{figure}
\begin{center}
{\includegraphics[width=0.45\textwidth]{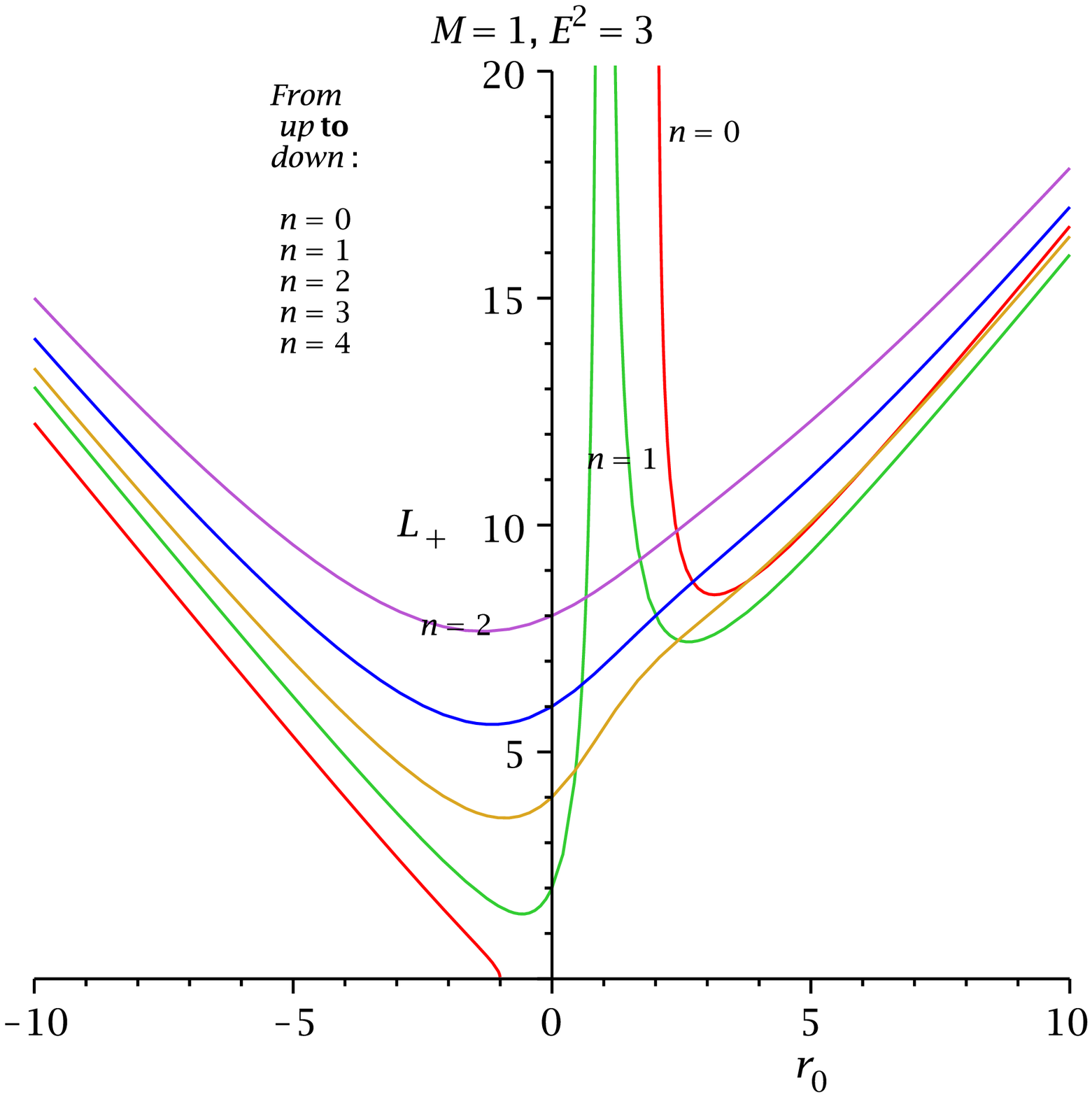}}
{\includegraphics[width=0.45\textwidth]{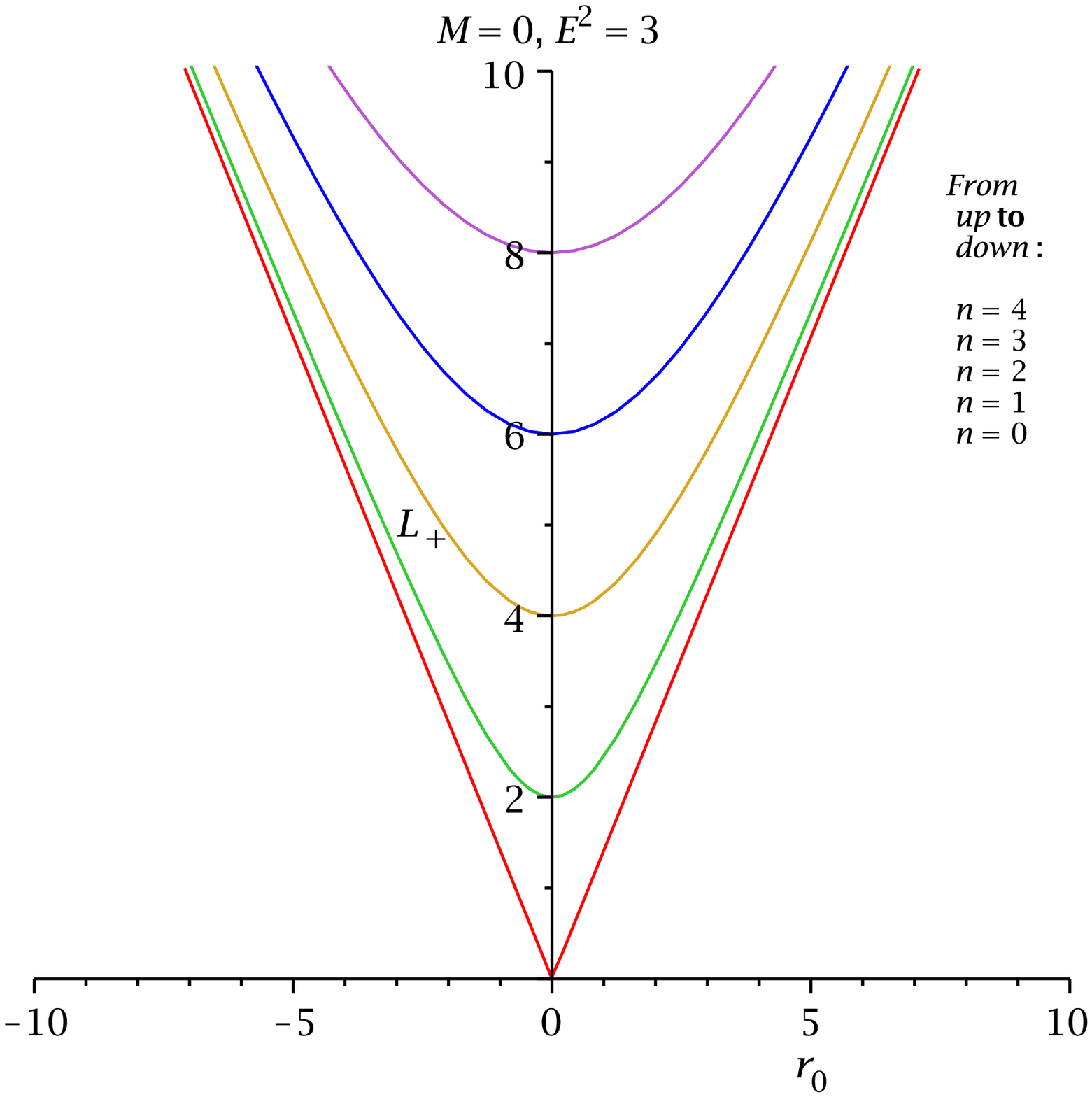}}
\end{center}
\caption{The figure shows the variation  of $L_{+}$  with $r_{0}$ for TN BH and mass-less TN BH.
\label{t12}}
\end{figure}

\begin{figure}
\begin{center}
{\includegraphics[width=0.45\textwidth]{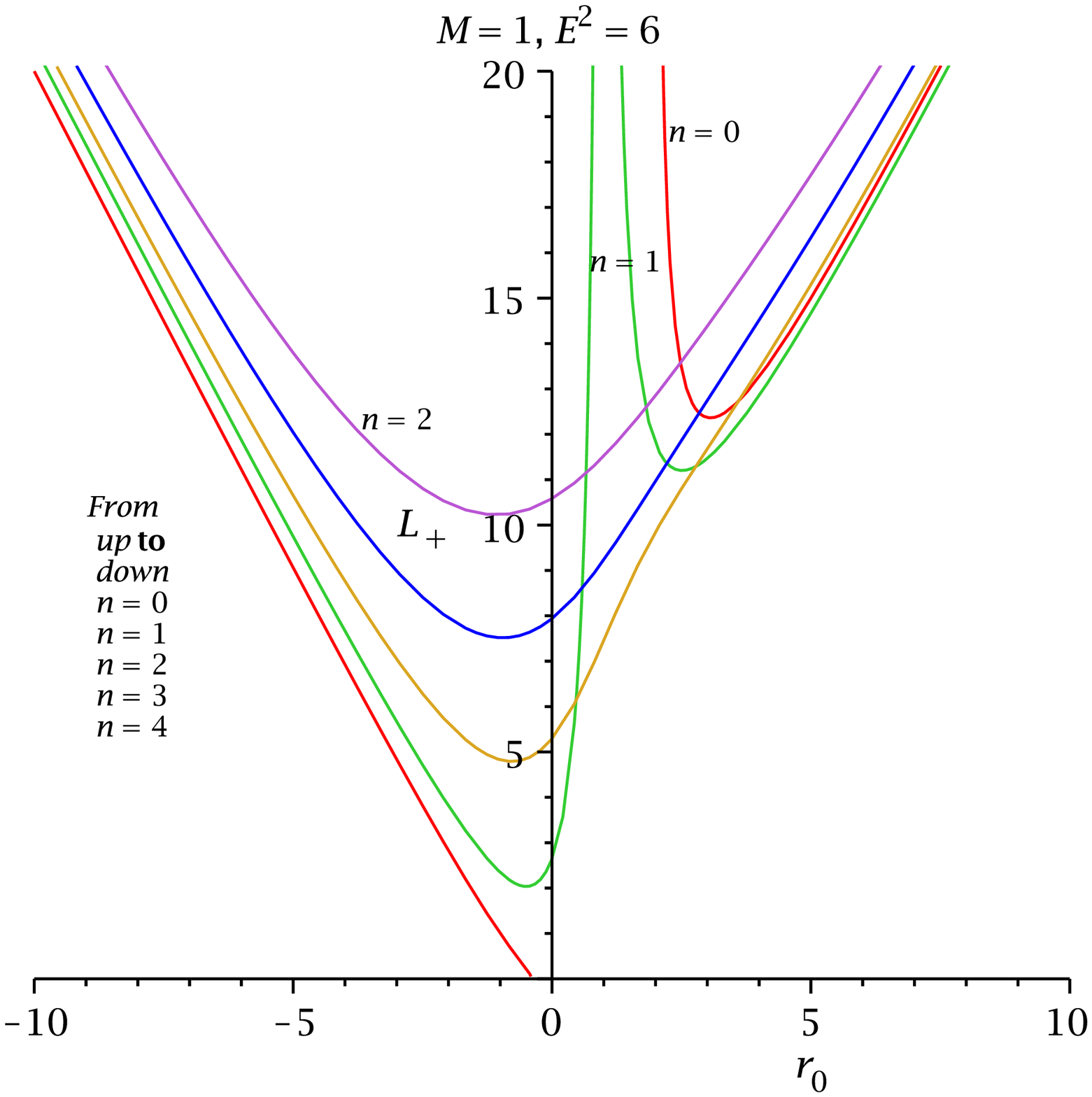}}
{\includegraphics[width=0.45\textwidth]{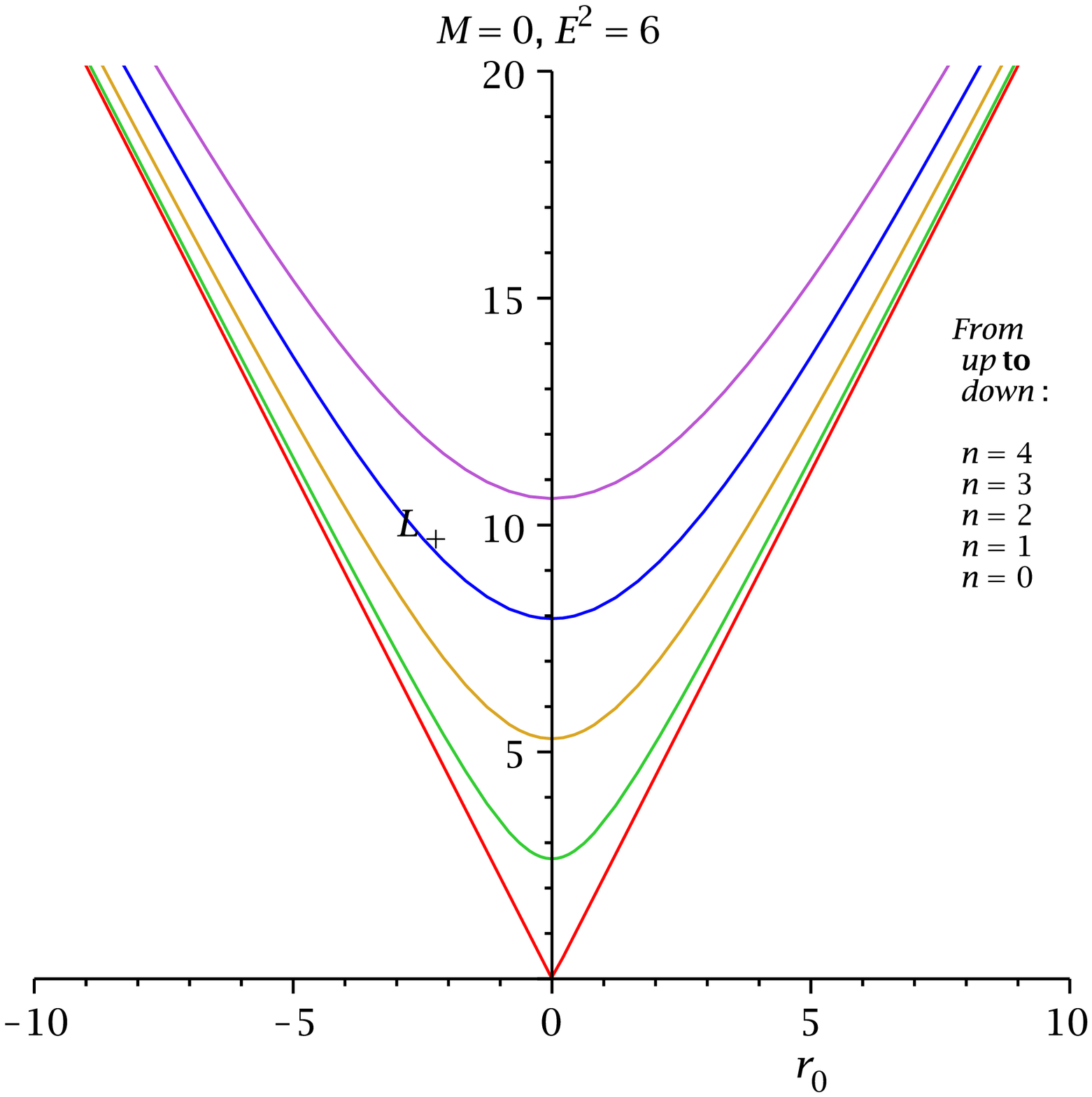}}
\end{center}
\caption{The figure shows the variation  of $L_{+}$  with $r_{0}$ for TN BH and mass-less TN BH.
\label{t13}}
\end{figure}

\begin{figure}
\begin{center}
{\includegraphics[width=0.45\textwidth]{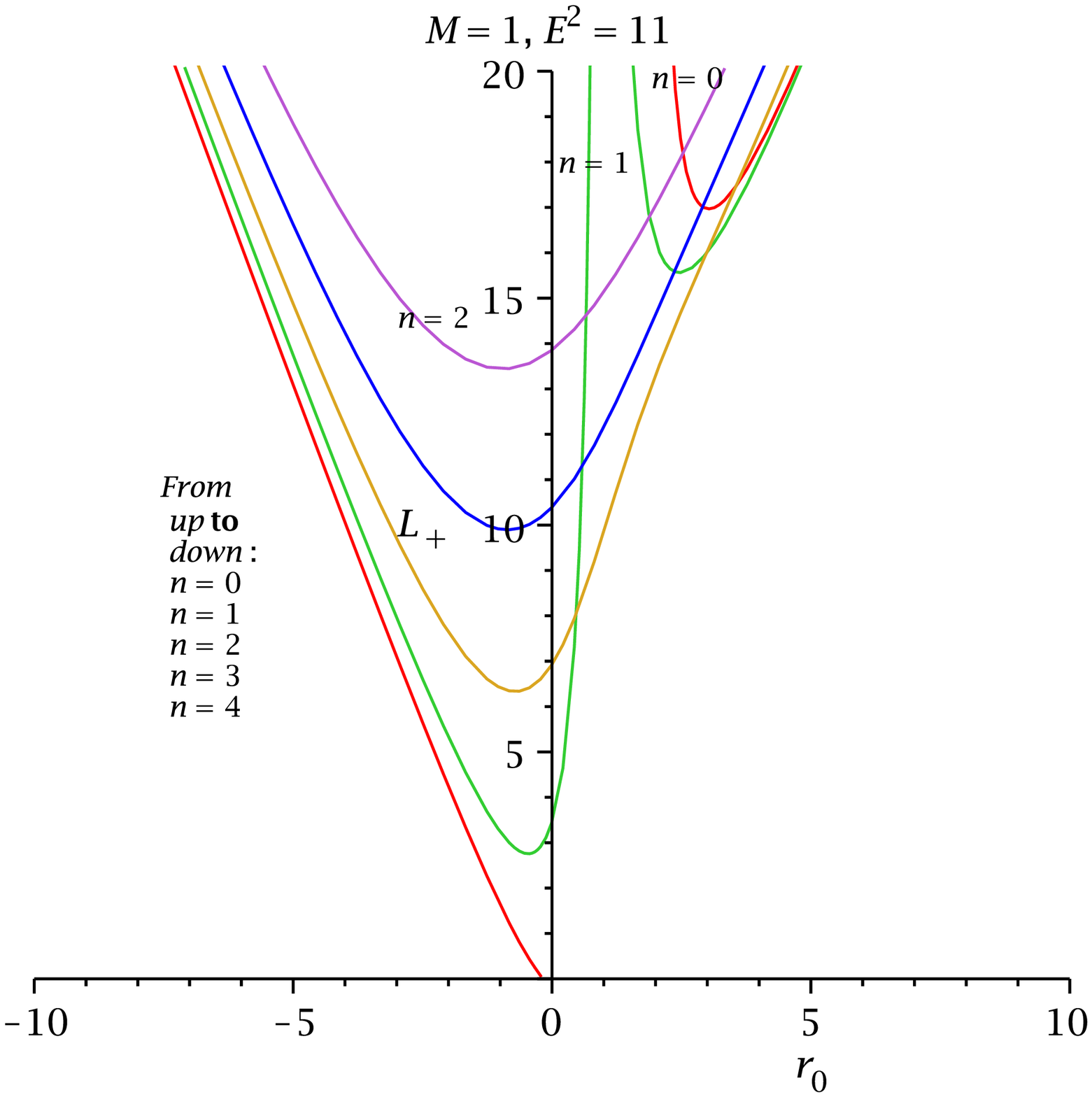}}
{\includegraphics[width=0.45\textwidth]{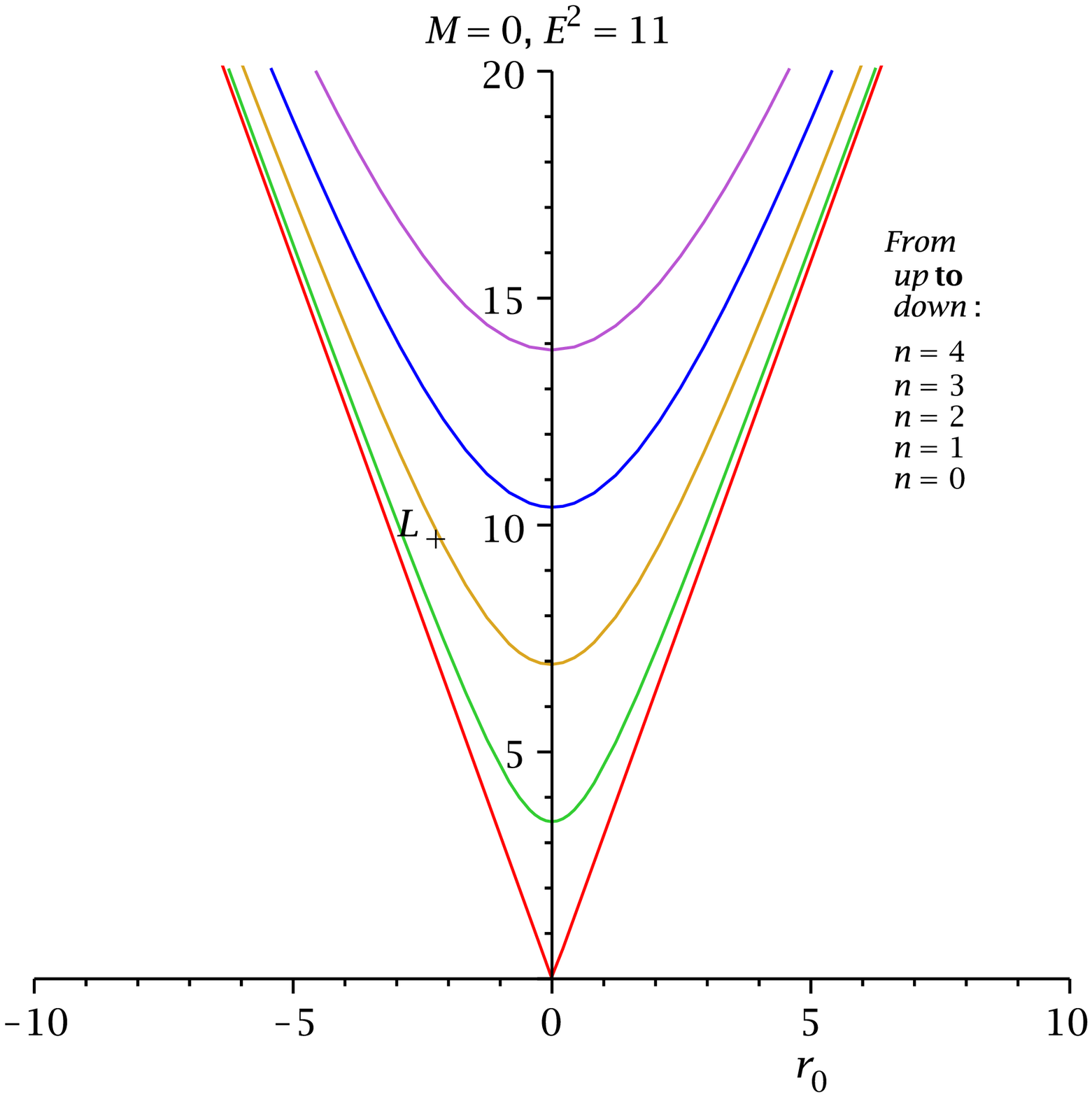}}
\end{center}
\caption{The figure shows the variation  of $L_{+}$  with $r_{0}$ for TN BH and mass-less TN BH.
\label{t14}}
\end{figure}

\begin{figure}
\begin{center}
{\includegraphics[width=0.45\textwidth]{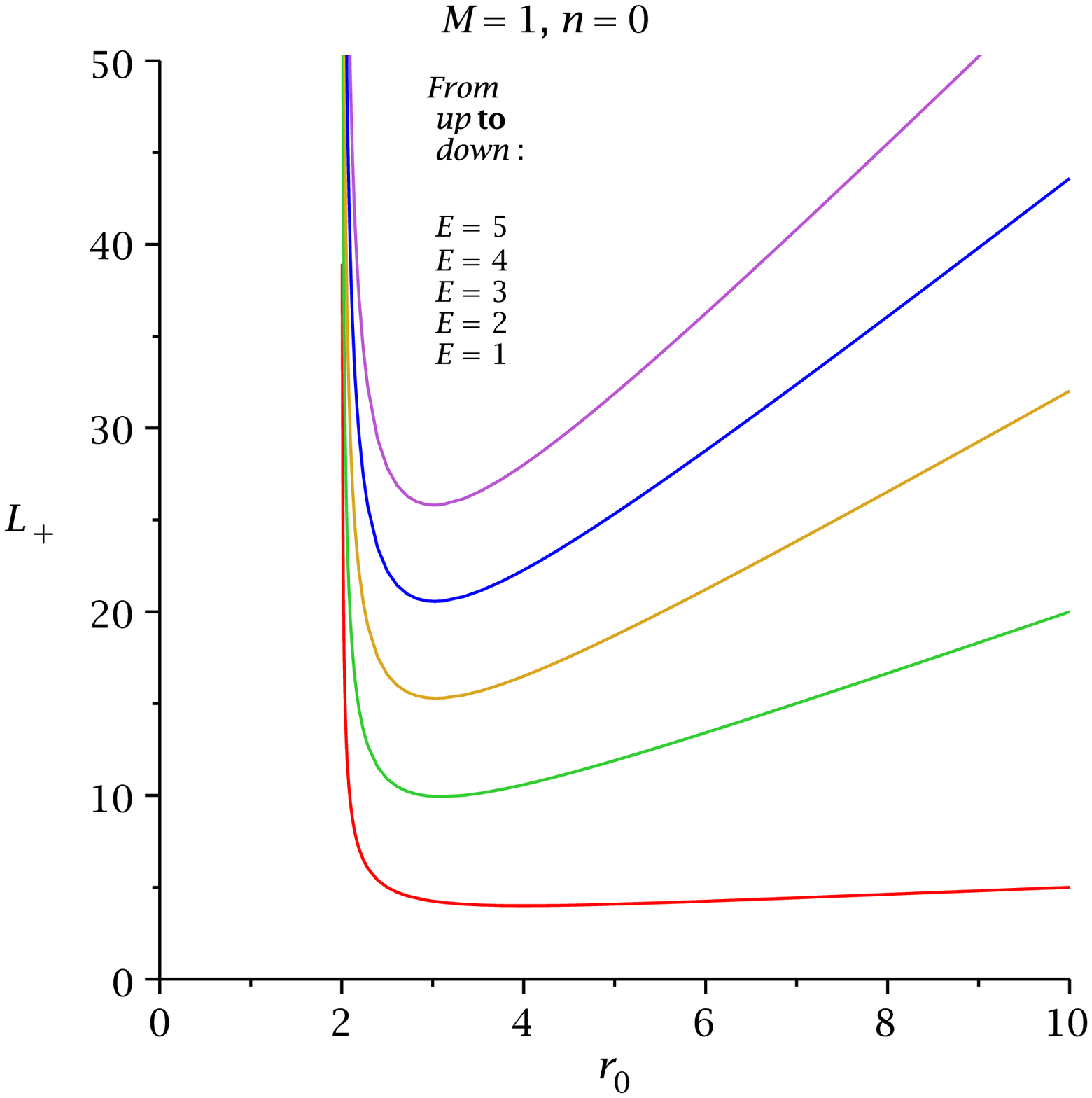}}
{\includegraphics[width=0.45\textwidth]{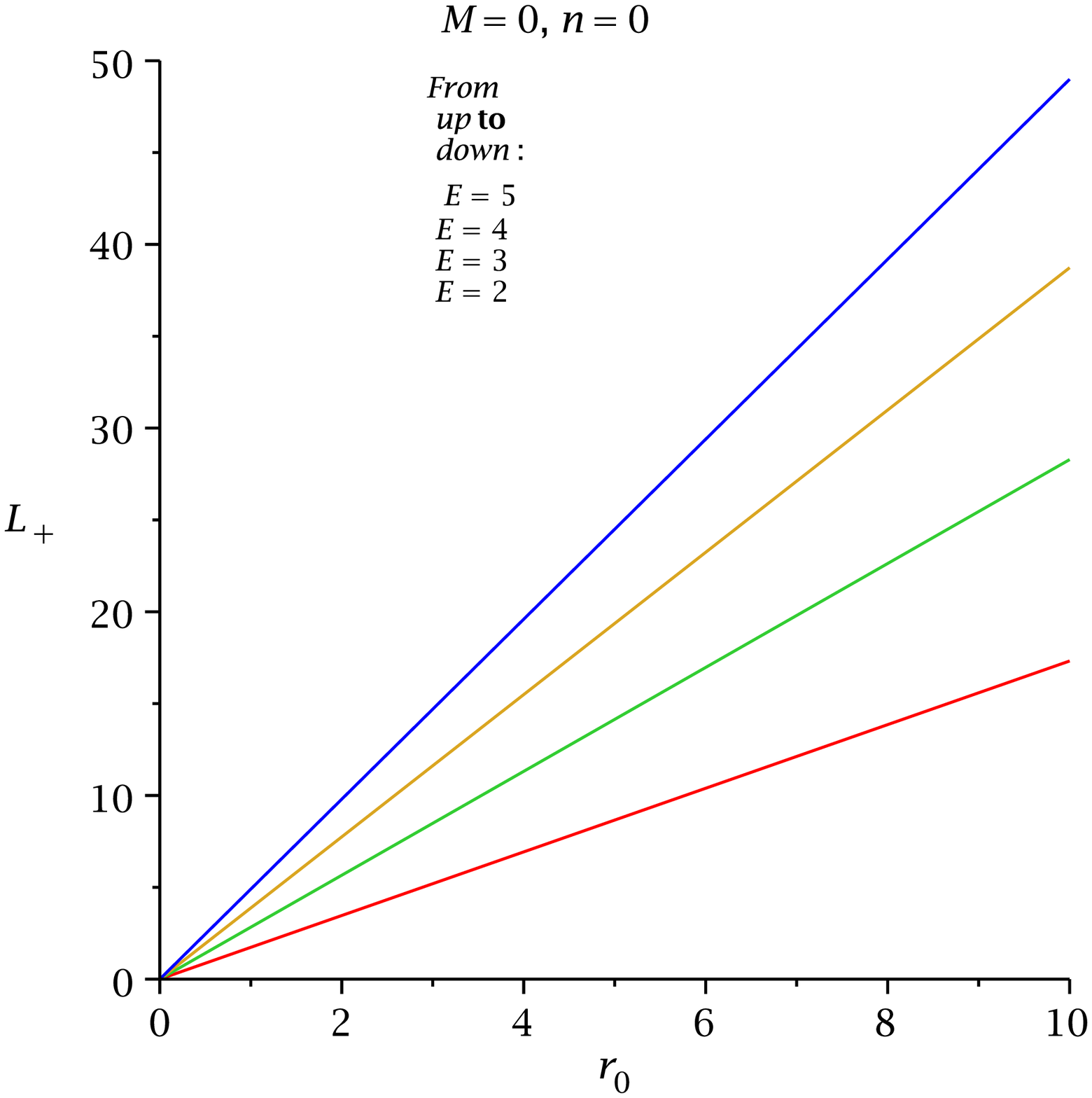}}
\end{center}
\caption{The figure depicts the variation  of $L_{+}$  with $r_{0}$ for TN BH and mass-less TN BH.
\label{t15}}
\end{figure}

\begin{figure}
\begin{center}
{\includegraphics[width=0.45\textwidth]{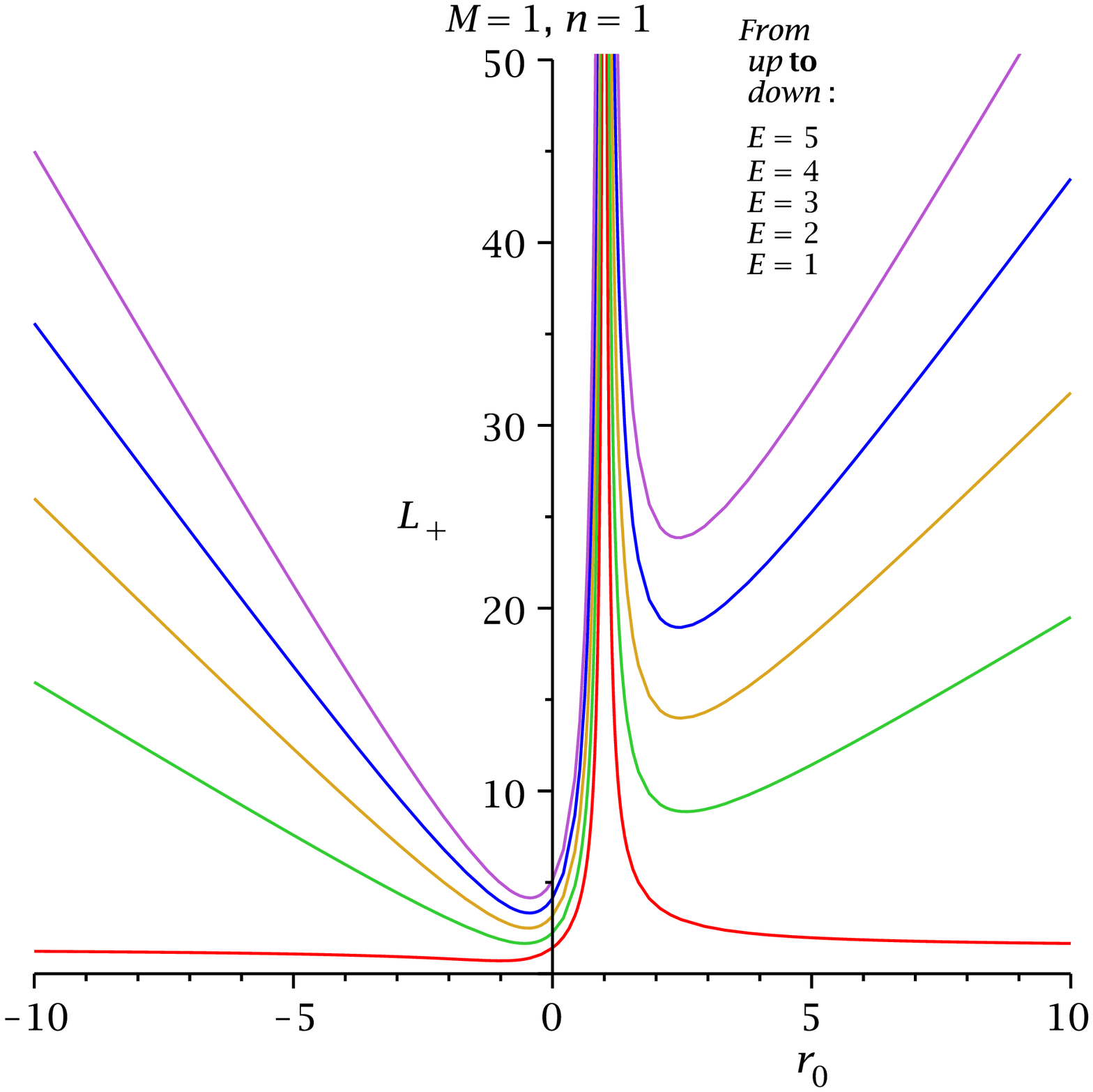}}
{\includegraphics[width=0.45\textwidth]{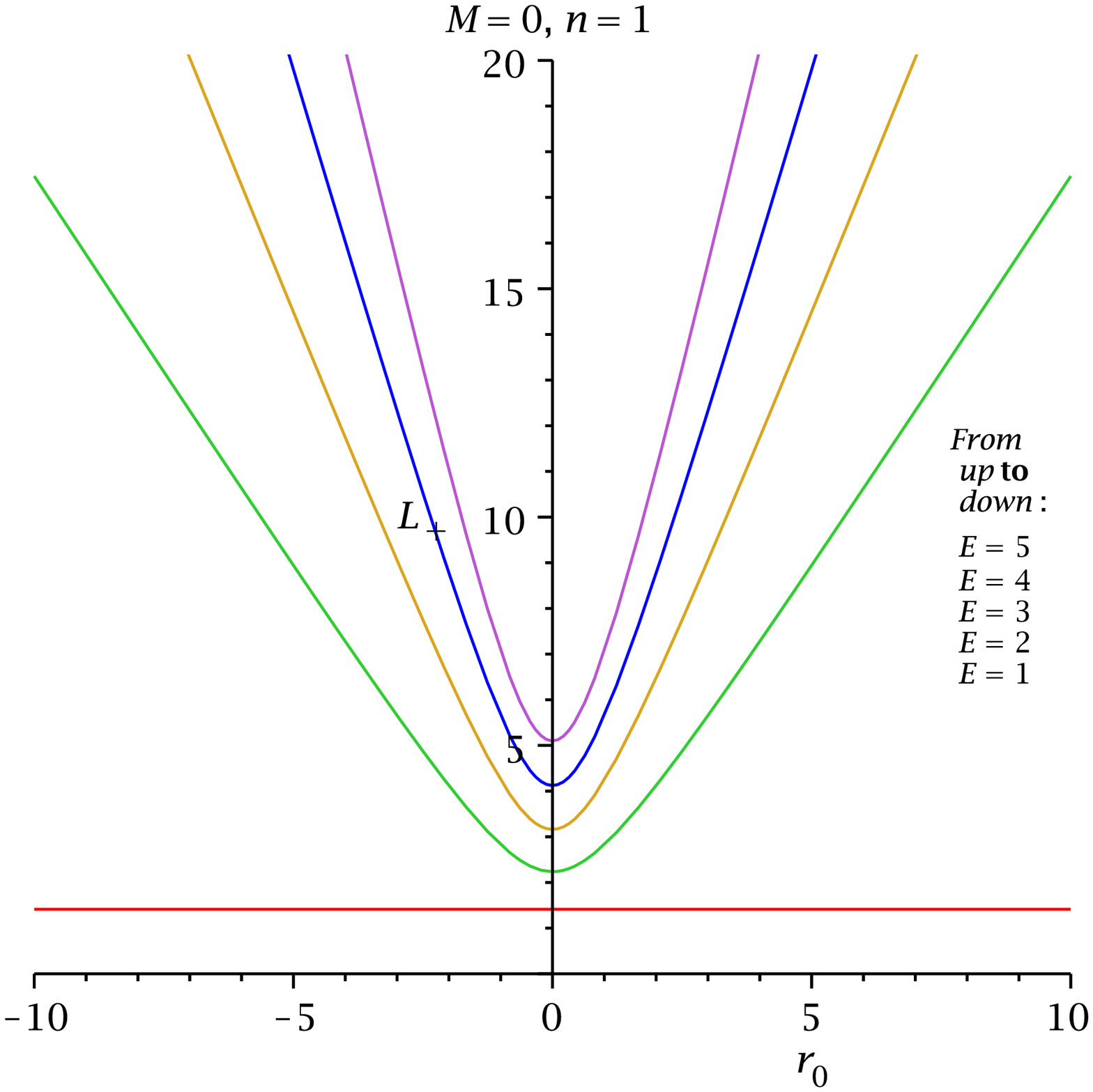}}
\end{center}
\caption{The figure depicts the variation  of $L_{+}$  with $r_{0}$ for TN BH and mass-less TN BH.
\label{t16}}
\end{figure}

\begin{figure}
\begin{center}
{\includegraphics[width=0.45\textwidth]{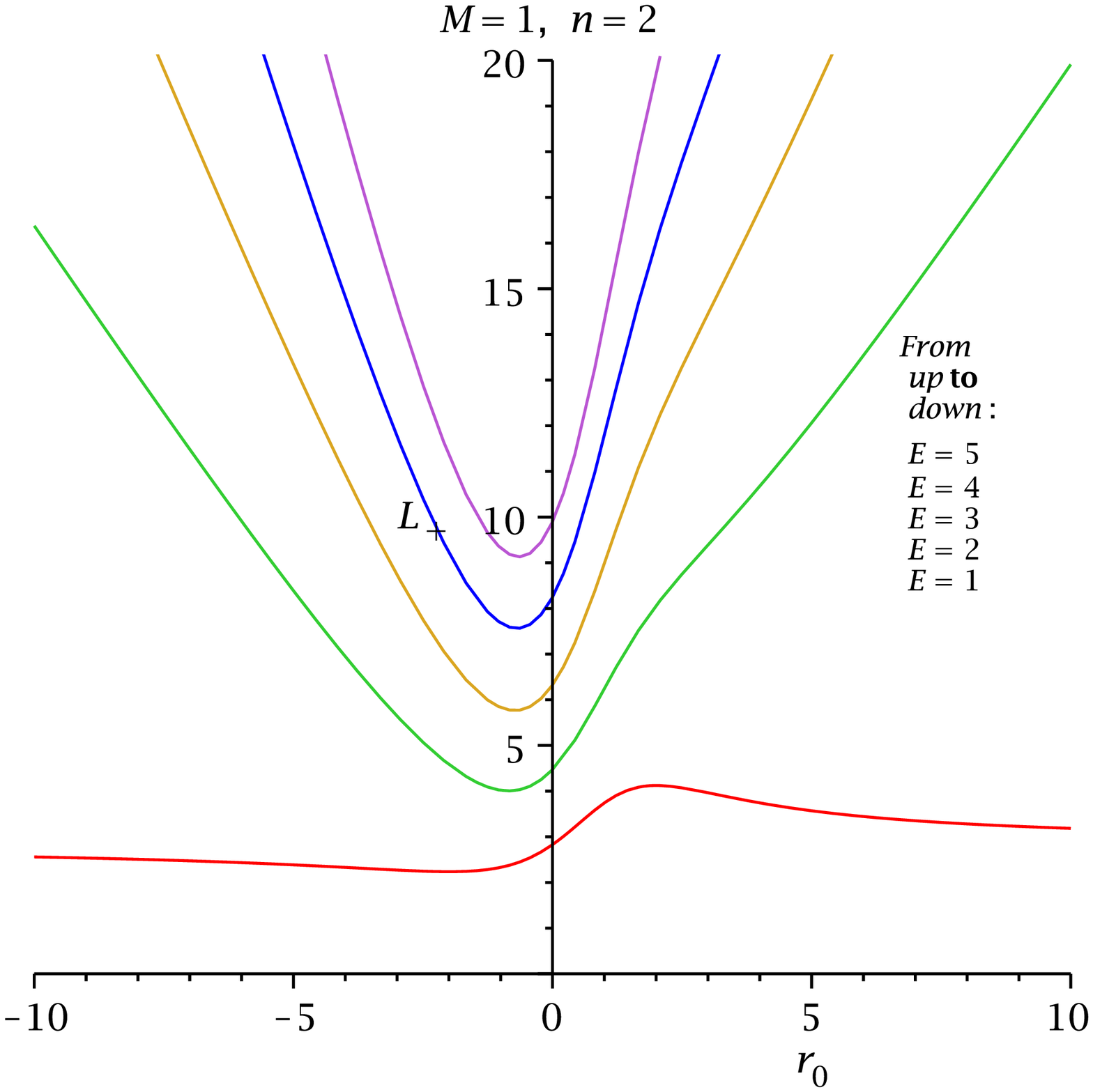}}
{\includegraphics[width=0.45\textwidth]{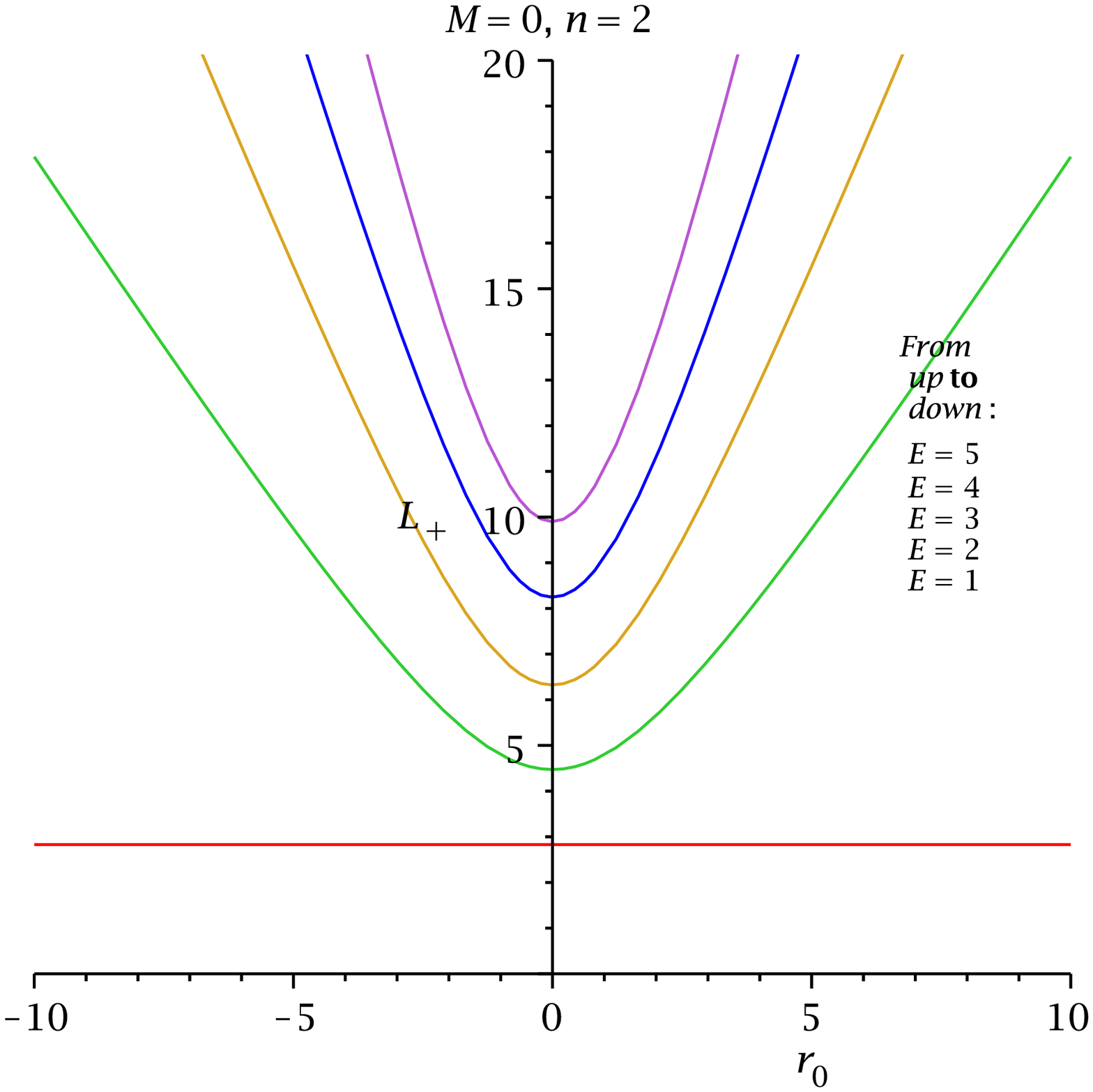}}
\end{center}
\caption{The figure depicts the variation  of $L_{+}$  with $r_{0}$ for TN BH and mass-less TN BH.
\label{t17}}
\end{figure}

\begin{figure}
\begin{center}
{\includegraphics[width=0.45\textwidth]{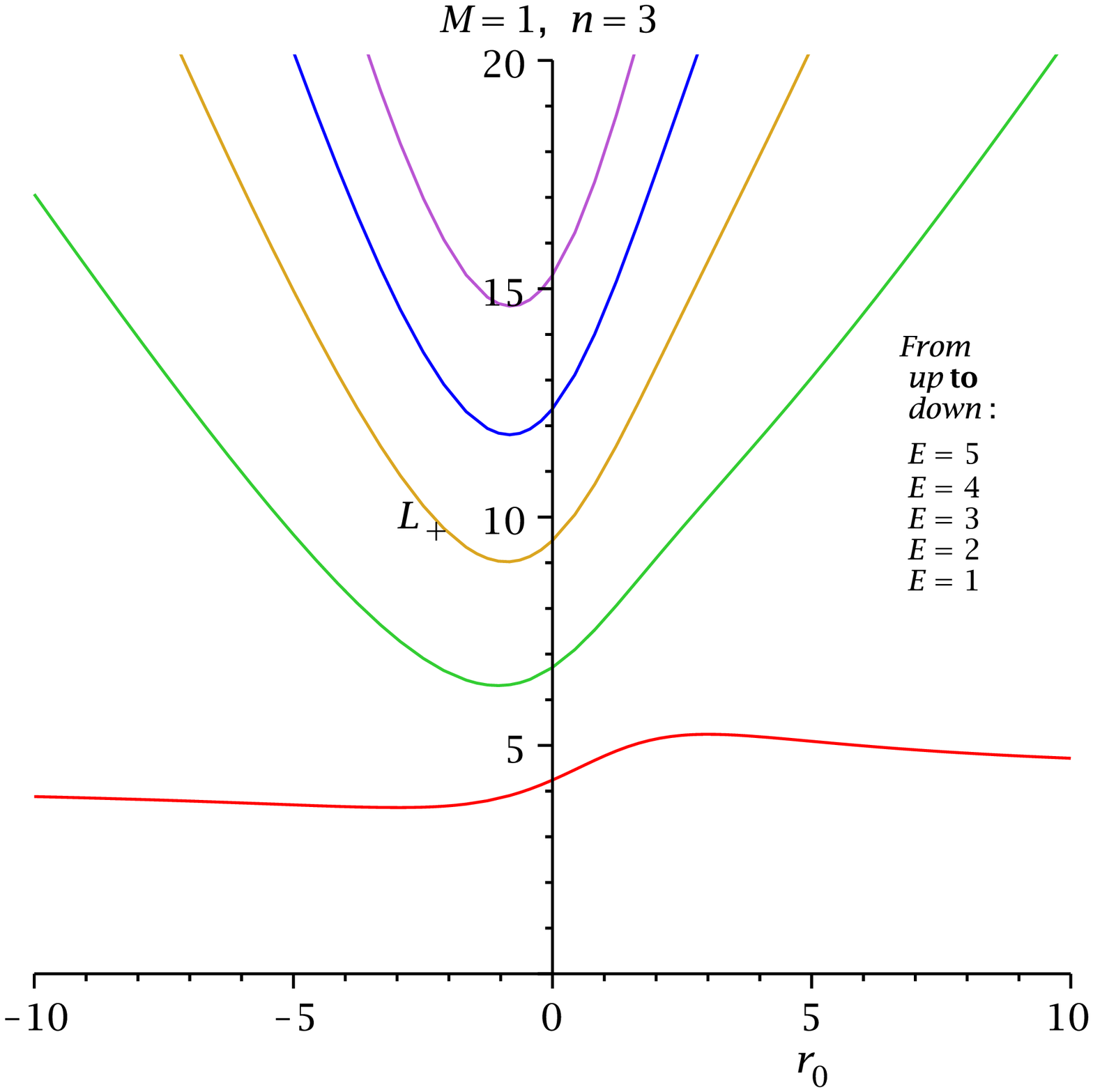}}
{\includegraphics[width=0.45\textwidth]{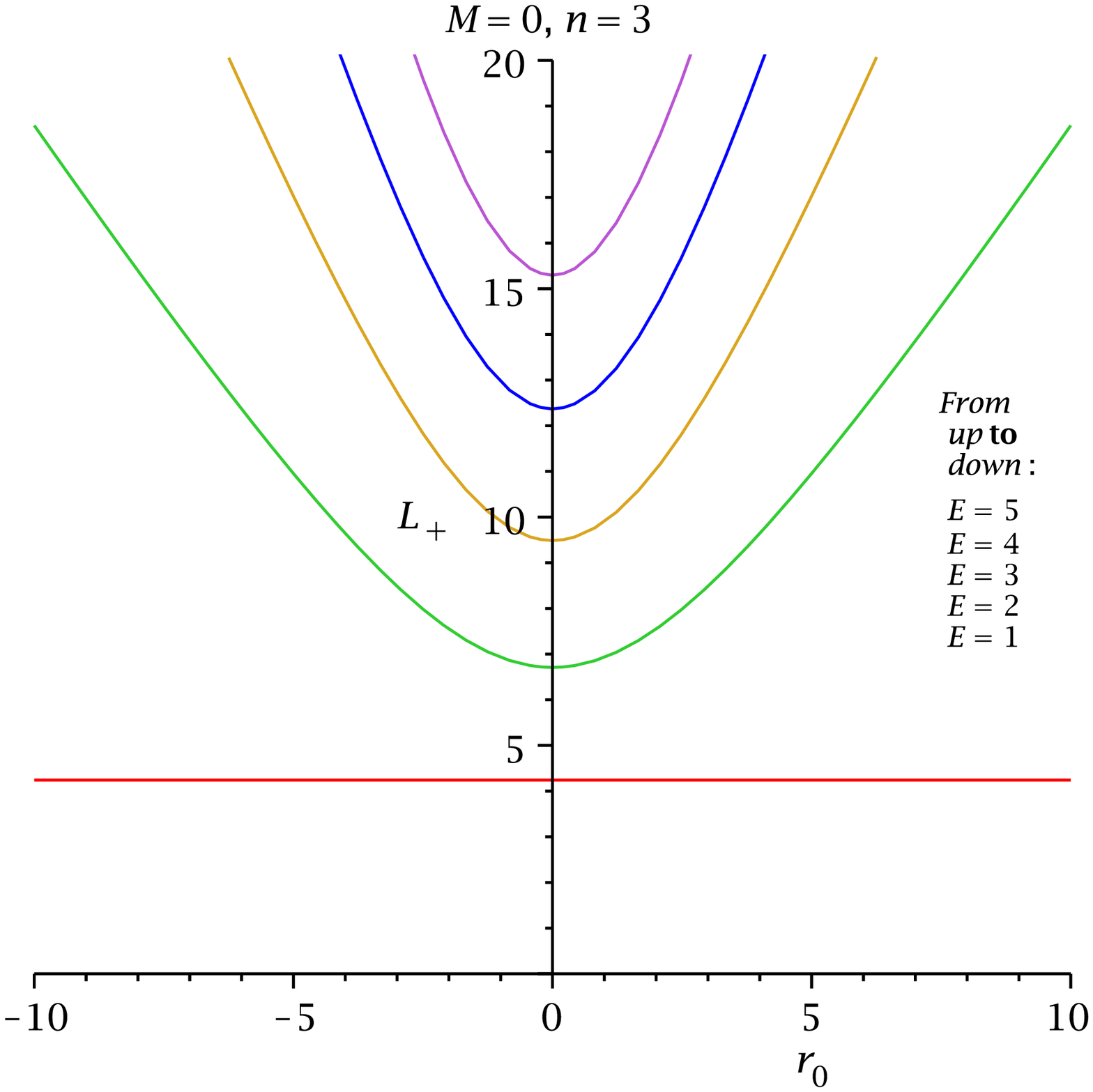}}
\end{center}
\caption{The figure depicts the variation  of $L_{+}$  with $r_{0}$ for TN BH and mass-less TN BH.
\label{t18}}
\end{figure}

\begin{figure}
\begin{center}
{\includegraphics[width=0.45\textwidth]{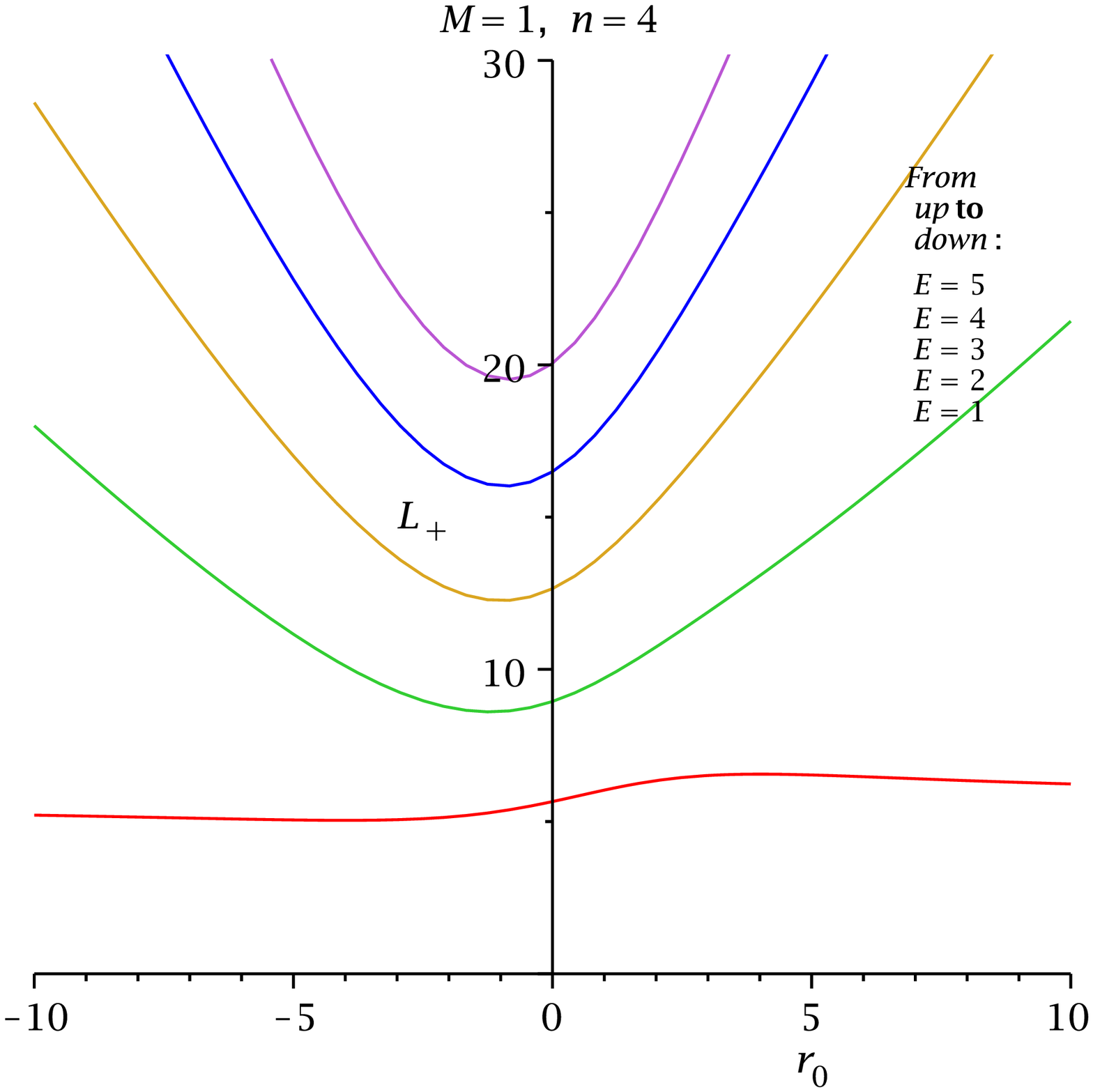}}
{\includegraphics[width=0.45\textwidth]{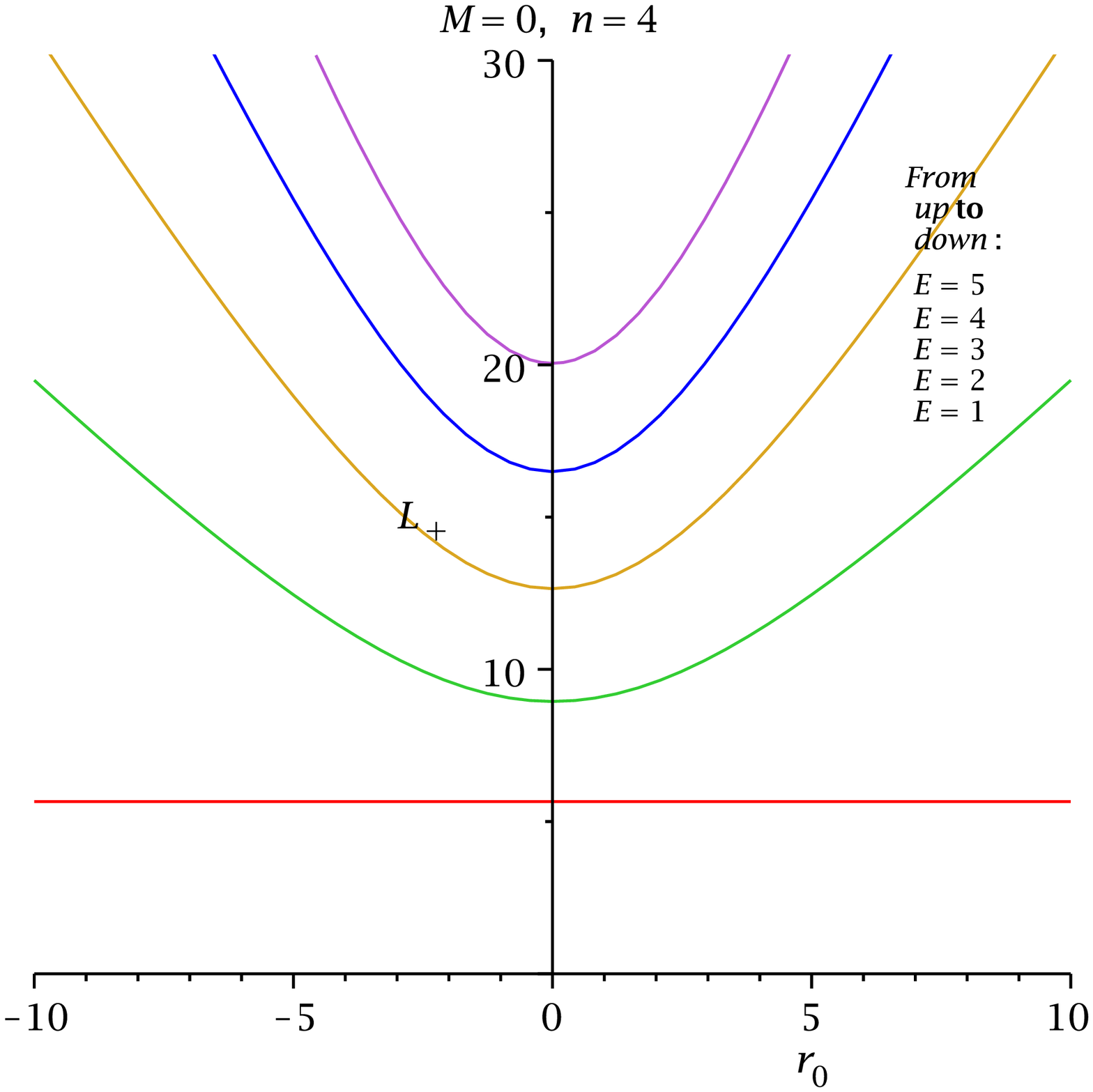}}
\end{center}
\caption{The figure depicts the variation  of $L_{+}$  with $r_{0}$ for TN BH and mass-less TN BH.
\label{t19}}
\end{figure}

When $E^2<1$, we obtain the bound orbits, this can displayed from the $L_{+}-r_{0}$ diagram. It has been shown 
in the  Fig. \ref{t10}.
For $E^2=1$, we find the marginally escape orbits. This has been seen from the Fig. \ref{t11}.
Finally for $E^2>1$, we find the unbound orbits. This could be seen from the Fig. \ref{t12}, Fig. \ref{t13} and 
Fig. \ref{t14}.
From Fig. \ref{t15}, Fig. \ref{t16}, Fig. \ref{t17}, Fig. \ref{t18} and 
Fig. \ref{t19}, one can observe that the variation 
of $L_{+}$ with respect to $r_{0}$ for various values of NUT parameter in the presence of mass parameter and without 
mass parameter.

\begin{figure}
\begin{center}
{\includegraphics[width=0.45\textwidth]{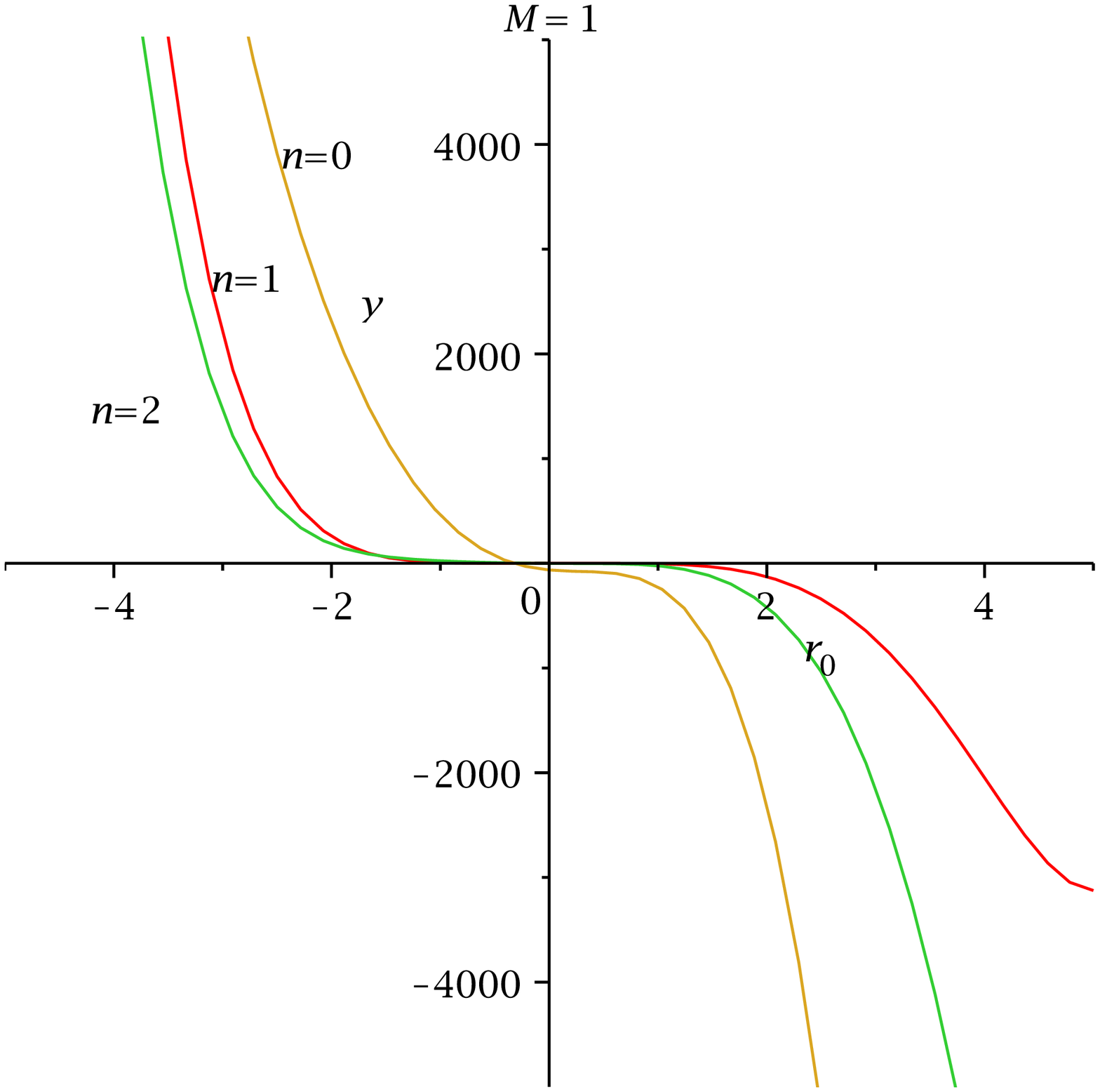}}
\end{center}
\caption{The figure depicts the variation  of ISCO  with $r_{0}$ in the presence of NUT parameter and without NUT 
parameter. Where $y=Mr_{0}^6-6M^2r_{0}^5-15 Mn^2r_{0}^4+(4M^2n^2-16n^4)r_{0}^3+15 M n^4r_{0}^2-6M^2 n^4 r_{0}-M n^6$.
\label{t20}}
\end{figure}

Now we see the behaviour of the $L$ with $r$ for different values of energy for a fixed value of the 
dual mass parameter. 
Now the ISCO radius can be obtain by using the condition  Eq. (\ref{ef}) with additional condition:
\begin{eqnarray} 
\frac{d^{2}\dot{r}^2}{dr^{2}}|_{r=r_{0}} &=& 0 ~. \label{n15}
\end{eqnarray}
Although it has been derived earlier in \cite{cqg}. We are interested here to  see the behaviour of ISCO in the presence 
of NUT parameter and with out NUT parameter in comparison with mass-less cases in more graphically which has been not shown 
there. Therefore the ISCO equation for TN BH could be obtain by applying the condition given in Eq. (\ref{ef}) 
and Eq. (\ref{n15}) (or by putting $a=Q=0$ in Eq. (81) in \cite{cqg}):
$$
Mr_{0}^6-6M^2r_{0}^5-15 Mn^2r_{0}^4+(4M^2n^2-16n^4)r_{0}^3+
$$
\begin{eqnarray}
15 M n^4r_{0}^2-6M^2 n^4 r_{0}-M n^6 &=& 0  ~.\label{n16}
\end{eqnarray}
and for the mass-less TN BH, it is 
\begin{eqnarray}
r_{0} &=& 0  ~.\label{n17}
\end{eqnarray}
and it also indicates there is \emph{no ISCO for massless TN BH}. It is a curious result although there exists mass-less CPO. 
The variation of ISCO in the presence of the NUT parameter and without NUT parameter could be found in the 
Fig. \ref{t20}.

\subsection{MBCO:}
\begin{figure}
\begin{center}
{\includegraphics[width=0.45\textwidth]{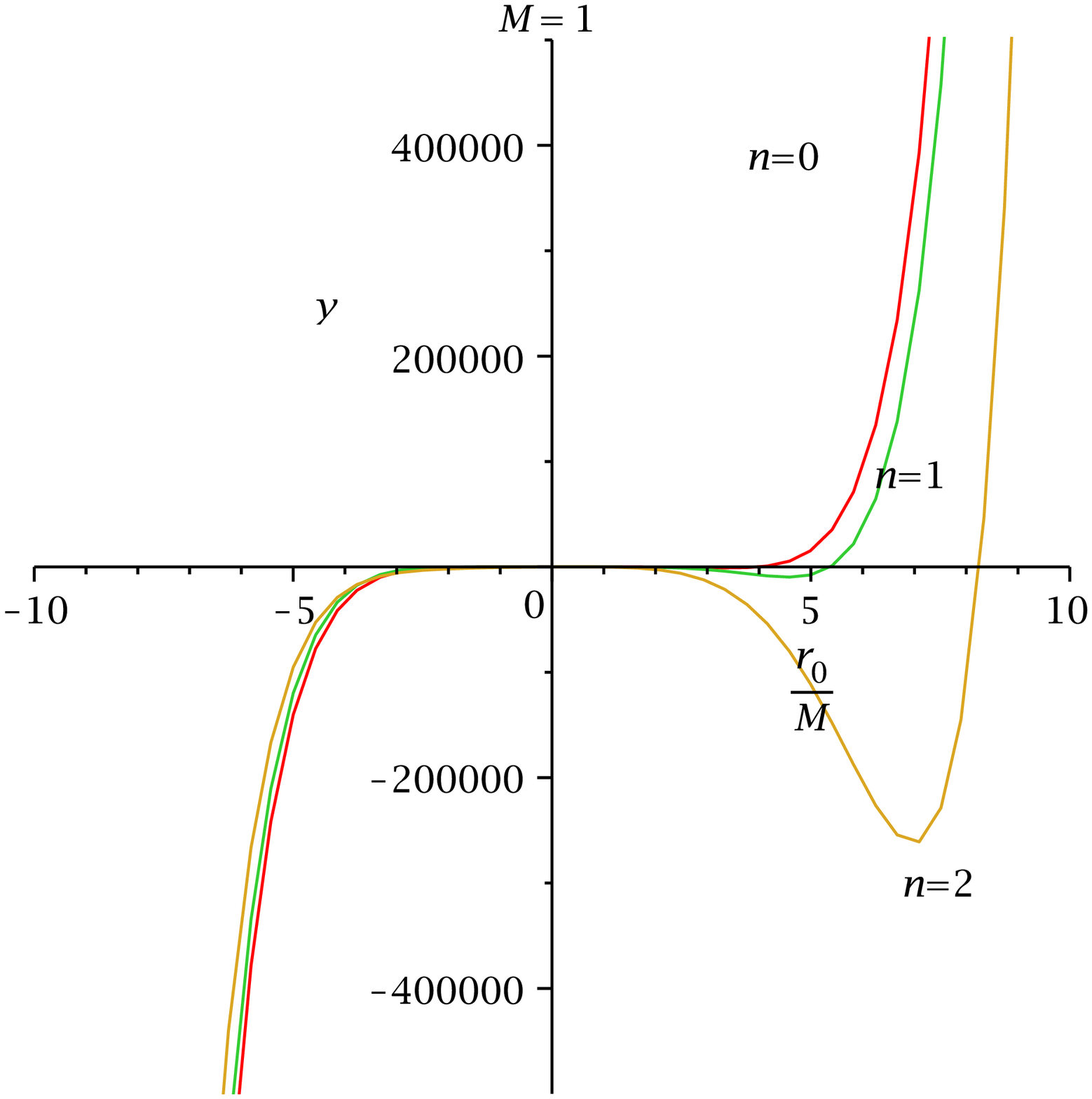}}
\end{center}
\caption{Here $y=Mr_{0}^7-4M^2r_{0}^6-7 Mn^2r_{0}^5+(2M^2n^2-4n^4)r_{0}^4+ (M^2 n^4-4n^6)r_{0}^2+4M n^6 r_{0}-M^2 n^6$.
\label{t21}}
\end{figure}

Another interesting orbit that has not been considered previously in \cite{chur} but considered in \cite{cqg} for KNTN BH. 
So we have just set $E_{0}^2=1$ in Eq. (\ref{ef1}) (or putting the parameters $a=Q=0$ in Eq. (98) in \cite{cqg} ), we find the 
MBCO for TN BH:
$$
Mr_{0}^7-4M^2r_{0}^6-7 Mn^2r_{0}^5+(2M^2n^2-4n^4)r_{0}^4+ (M^2 n^4-4n^6)r_{0}^2+
$$
\begin{eqnarray}
4M n^6 r_{0}-M^2 n^6 &=& 0  ~.\label{n18}
\end{eqnarray}
and for the mass-less TN BH, it is 
\begin{eqnarray}
r_{0}^2+n^2 &=& 0  ~.\label{n19}
\end{eqnarray}
It implies that there is no existence of MBCO for mass-less case. From Eq. (\ref{n18}), in the limit $n=0$, we get the MBCO 
for Schwarzschild BH \cite{sch}. In Fig. \ref{t21}, we have plotted the MBCO with $r_{0}$ in comparison with Schwarzschild 
BH.

\section{\label{ex} Geodesics in Extreme TN Spacetime:}
\begin{figure}
\begin{center}
{\includegraphics[width=0.45\textwidth]{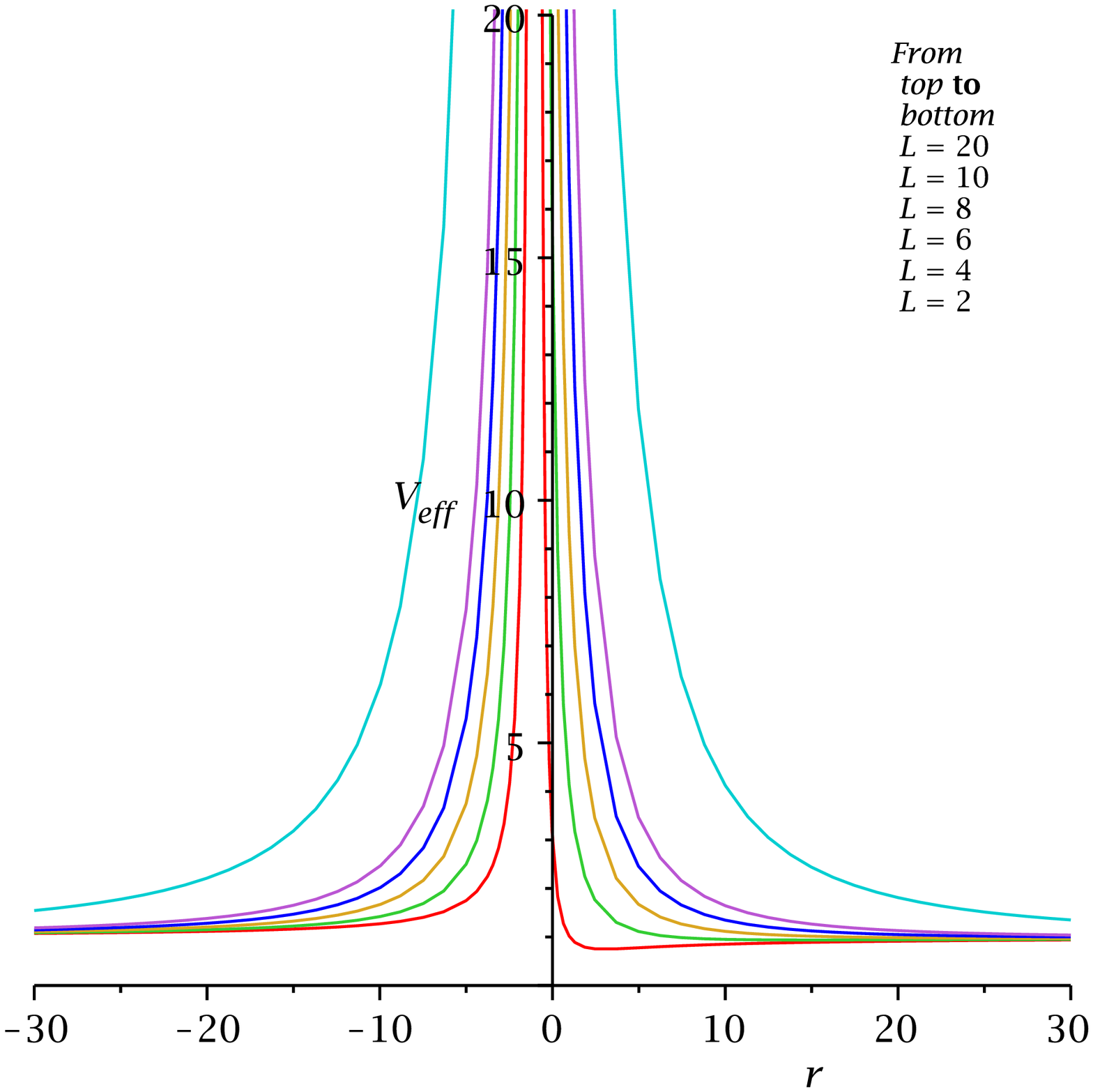}}
\end{center}
\caption{ The variation of $V_{eff}$ with $r$ for extreme TN BH, Here $M=1$.
\label{t22}}
\end{figure}

In the previous section, we have discussed the complete geodesic structure for non-extreme TN BH. Moreover, we have 
clearly explained the difference of geodesic structure between TN spacetime and mass-lees TN spacetime in graphically.
In the present section we shall discuss more interesting case i.e. \emph{extreme TN} spacetime.  What happens the 
geodesic structure in the extreme limit i.e. $r_{+}=r_{-}$. This is the main aim in this section. 
Proceeding similarly, we can write the effective potential for massive particles in the extreme limit 
[using Eq. (\ref{n8})] given by  
\begin{eqnarray}
V_{eff} &=& \left(\frac{r-M}{r+M}\right) \left(1+\frac{L^{2}}{r^2-M^2}\right) ~.\label{x1}
\end{eqnarray} 
The qualitative behaviour of the test particle may be seen from the effective potential diagram (See Fig. \ref{t22}). 

\begin{figure}
\begin{center}
{\includegraphics[width=0.45\textwidth]{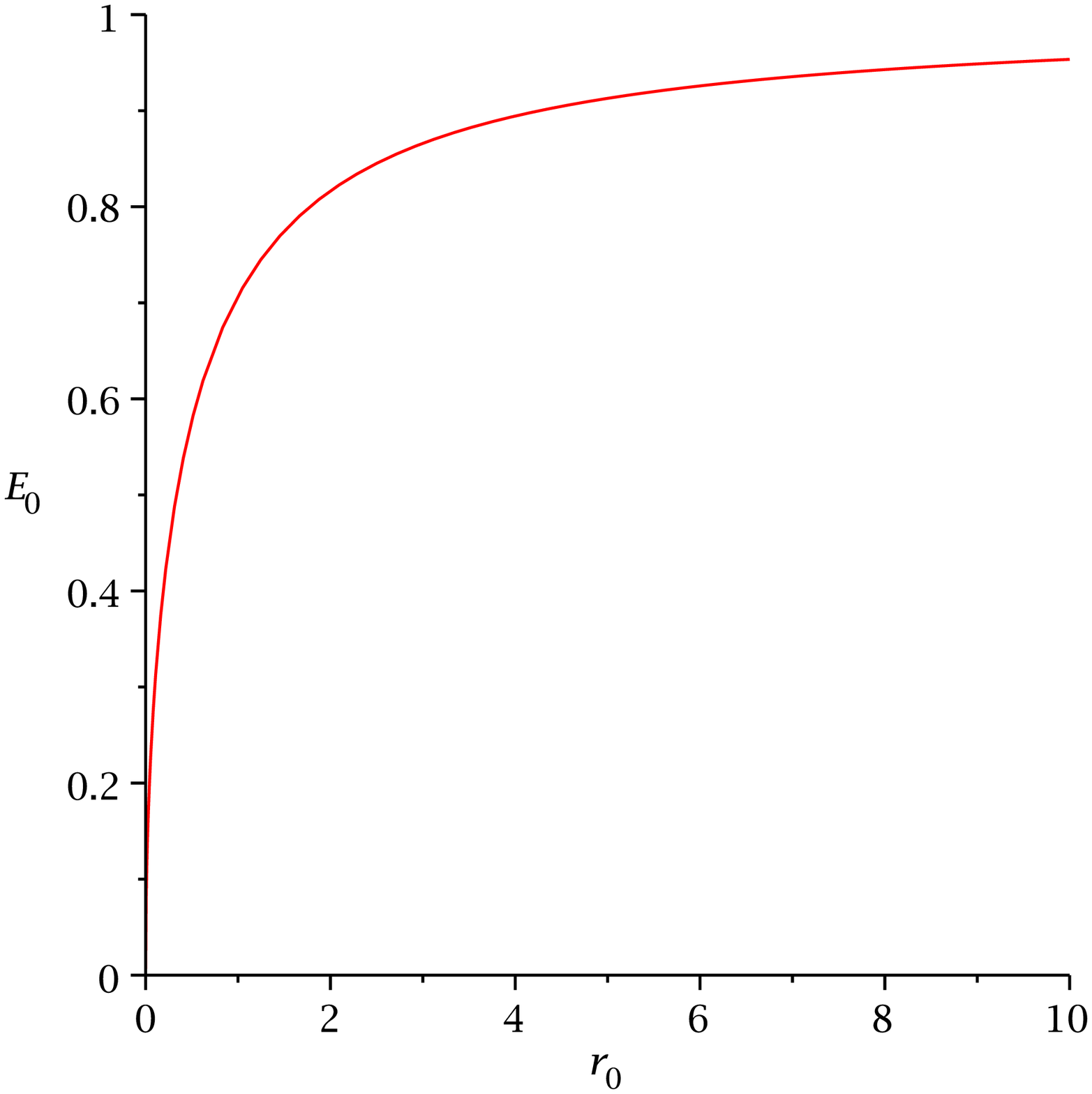}}
{\includegraphics[width=0.45\textwidth]{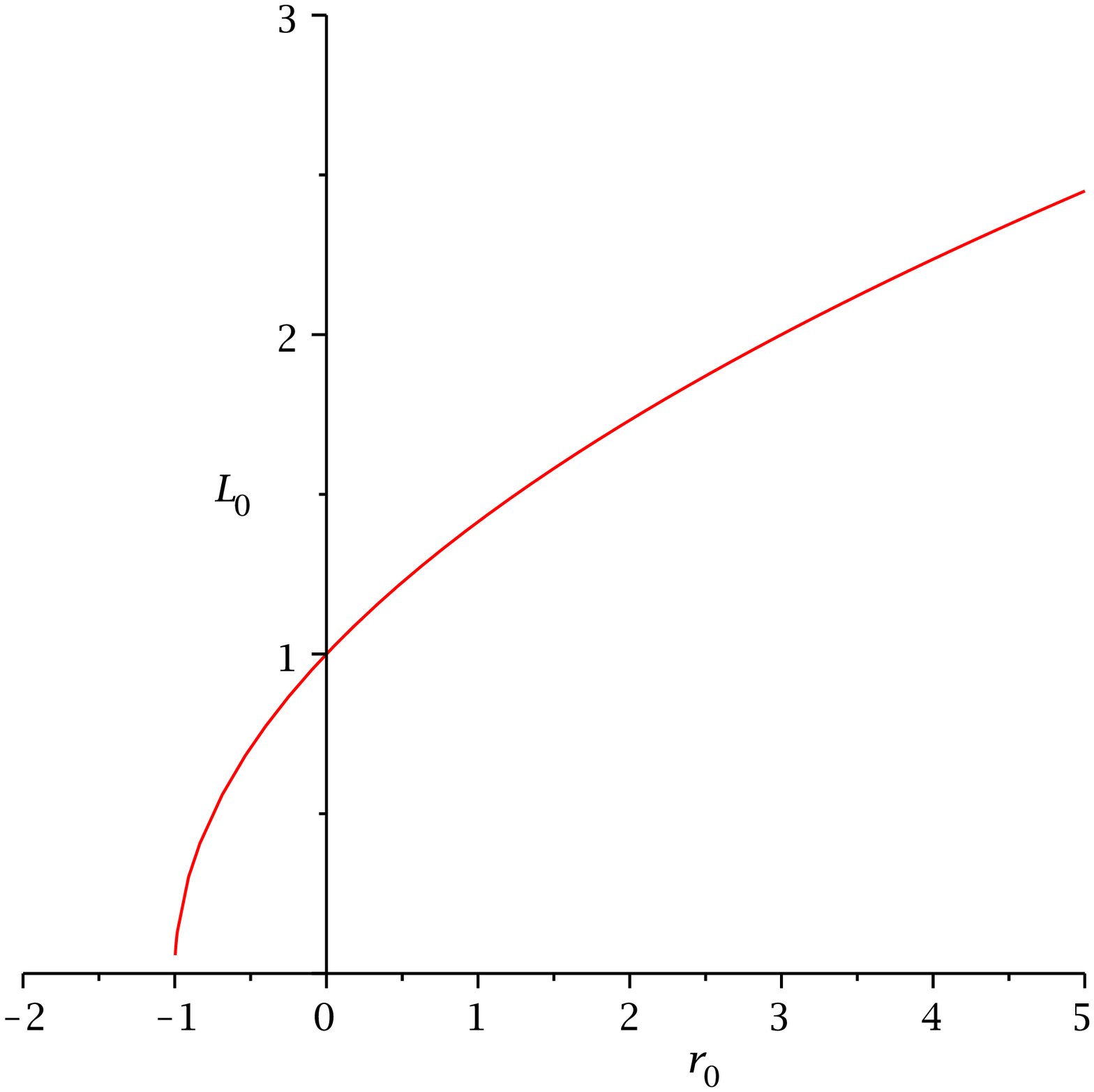}}
\end{center}
\caption{ The variation of energy and angular momentum with $r_{0}$ for extreme TN BH, Here $M=1$.
\label{t23}}
\end{figure}

Proceeding analogously, we apply for circular geodesics $r=r_{0}$, one obtains the energy and angular momentum 
for extreme TN BH [using Eq. (\ref{ef1}) and Eq. (\ref{ef2})]:
\begin{eqnarray}
E_{0}  &=& \sqrt{\frac{r_{0}}{r_{0}+M}}  ~. \label{ex1}
\end{eqnarray}
and 
\begin{eqnarray}
L_{0}  &=& \sqrt{M(r_{0}+M)} ~. \label{ex2}
\end{eqnarray}
How $E_{0}$ and $L_{0}$ are varied with $r_{0}$, it can be seen from the Fig. \ref{t23}.
Now we have derived the ISCO equation for non-extreme TN BH but here we will see what would be the ISCO radius in the 
extremal limit which is most important class of circular orbit in astrophysics. We find for extreme TN BH the ISCO 
radius [using Eq. (\ref{n16})] should be
\begin{eqnarray}
r_{0} &=& r_{isco} = M ~. \label{ex3}
\end{eqnarray}
The corresponding ISCO energy and ISCO angular momentum are 
\begin{eqnarray}
E_{isco} &=& \frac{1}{\sqrt{2}}  ~. \label{ex4} \\
L_{isco} &=& \sqrt{2} M  ~. \label{ex5}
\end{eqnarray}

\begin{figure}
\begin{center}
{\includegraphics[width=0.45\textwidth]{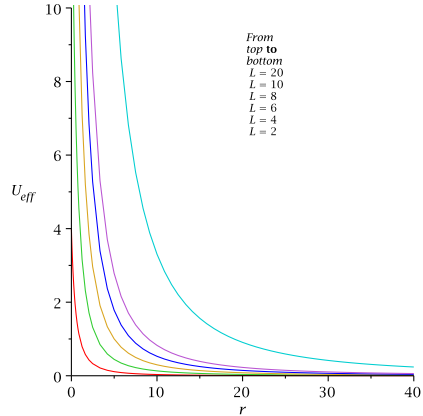}}
\end{center}
\caption{ The variation of $U_{eff}$ with $r$ for extreme TN BH, Here $M=1$.
\label{t24}}
\end{figure}

For photons, the effective potential in the extreme limit [using Eq. (\ref{n1})] reduces to
\begin{eqnarray}
U_{eff} &=& \frac{L^{2}}{(r+M)^2} ~.\label{ex6}
\end{eqnarray} 
Its behaviour for different values of angular momentum parameter can be seen from the Fig. \ref{t24}.
Therefore, one obtains the CPO [using Eq. (\ref{n6})] for extreme TN BH:
\begin{eqnarray}
r_{c} &=& r_{cpo} = M ~. \label{ex7}
\end{eqnarray}
Similarly, we can find the radius of MBCO [using Eq. (\ref{n18})] for extreme TN BH:
\begin{eqnarray}
r_{0} &=& r_{mbco} = M ~. \label{ex8}
\end{eqnarray}
Interestingly, we observed that for extreme TN BH the three radii namely ISCO, MBCO and CPO 
coincides with the horizon i.e. 
\begin{eqnarray}
r_{isco} = r_{mbco}=r_{cpo}= M ~. \label{ex9}
\end{eqnarray}
This has not been previously examined in the literature that for extreme TN BH three orbits are coincident with the 
Killing horizons i.e. null geodesic generators of the horizon. For spherically symmetric extreme string BH, this result 
has been observed in \cite{ijmpd}. Probably, this is a \emph{first} time we have seen that such type of features 
for any spherically symmetric, stationary and non-asymptotic flat extreme BH.

\section{\label{dis} Discussion:}
In this paper we examined the geodesic motion of test particles in the background geometry of TN spacetime in 
comparison with mass-less (zero mass) TN spacetime. We considered the both massive and mass-less  
cases. We differentiated the ISCO, MBCO and CPO in graphically between TN spacetime and mass-less TN spacetime. 
From effective potential diagram, we showed the presence of the NUT parameter changes the shape of the potential 
well in comparison with zero NUT paprameter and mass-less spacetime. We studied the circular orbits in the $L-r$ 
plane for different values of energy i.e. $E^2>1$, $E^2<1$ and $E^2=1$ in case of TN and mass-less TN spacetime.

We also derived the geodesic motion of test particles (both massive and mass-less) in case of extreme TN BH. 
Interestingly, we showed that the three radii i.e. the radius of ISCO, the radius of MBCO and the radius of 
CPO are coincident with the Killing horizon radius. This is  a very surprising result because probably we first 
obtained this result for any spherically symmetric, stationary and non-asymptotic flat spacetime.  
In Appendix-A, we derived the CM energy for extreme TN BH and we found that the 
diverging value of CM energy.

\appendix

\section{\label{cm} CM Energy of Particle Collision near the horizon of TN BH:}
In this section, we should study what is the role of NUT parameter in the Ba\~{n}ados, Silk and West (BSW) effect which was 
predicted by  Ba\~{n}ados, Silk and West \cite{bsw} several years ago. Does TN BH could be act 
as a particle accelerator with arbitrarily high energy when the BH is extremal. This is the main 
aim in this section  and motivated by the previous section from the analysis of 
geodesic structure. We have considered the particle acceleration and collision in the CM frame.
To determine the CM energy, we consider first two particles coming from
infinity with $\frac{{\cal E}_{1}}{m_{0}}=\frac{{\cal E}_{2}}{m_{0}}=1$ approaching the TN BH with different 
angular momenta $L_{1}$ and $L_{2}$ and colliding at some radius $r$. Later, we choose the 
collision point is at $r$ to approach the horizon $r=r_{+}$. Also we have assumed that the particles 
are initially to be at rest at infinity. 

The formula that we have used here suggested first by BSW \cite{bsw} should read
\begin{eqnarray}
\left(\frac{{\cal E}_{cm}}{\sqrt{2}m_{0}}\right)^{2} &=&
1-g_{\mu\nu}u^{\mu}_{1}u^{\nu}_{2}~.\label{cme}
\end{eqnarray}
Since in the previous section we have calculated total geodesic structure confined in the equatorial plane 
for TN BH, so we should not repeat it again. Due to the time-like isometry and space-like isometry the space-time 
possesses two conserved quantities one is the energy and other is the  angular momentum. Thus 
for massive particles, the components of the four velocity are
\begin{eqnarray}
  u^{t} &=& \dot{t} =\frac{{\cal E}}{{\cal H}(r)}  \\
  u^{r} &=& \dot{r}=\pm \sqrt{{\cal E}^{2}-{\cal H}(r) \left(1+\frac{L^{2}}{r^{2}+n^2}\right)}\label{eff}\\
  u^{\theta} &=& \dot{\theta} = 0 \\
  u^{\phi} &=& \dot{\phi} = \frac{L}{r^{2}+n^2} ~.\label{fv}
\end{eqnarray}
and

\begin{eqnarray}
u^{\mu}_{1} &=& \left( \frac{{\cal E}_{1}}{{\cal H}(r)},~ -X_{1},~ 0,~\frac{L_{1}}{r^{2}+n^2}\right) ~.\label{u1}\\
u^{\mu}_{2} &=& \left( \frac{{\cal E}_{2}}{{\cal H}(r)},~ -X_{2},~ 0,~\frac{L_{2}}{r^{2}+n^2}\right) ~.\label{u2}
\end{eqnarray}

Putting these values  in (\ref{cme}), we find the expression for CM energy:
\begin{eqnarray}
\left(\frac{{\cal E}_{cm}}{\sqrt{2}m_{0}}\right)^{2} &=&  1 +\frac{{\cal E}_{1}
{\cal E}_{2}}{{\cal H}(r)}-
\frac{X_{1}X_{2}}{{\cal H}(r)}-\frac{L_{1}L_{2}}{r^{2}+n^2} ~.\label{cm1}
\end{eqnarray}
where,
$$
X_{1} = \sqrt{{\cal E}_{1}^{2}-{\cal H}(r)\left(1+\frac{L_{1}^{2}}{r^{2}+n^2}\right)}, \,\,
\\
X_{2} = \sqrt{{\cal E}_{2}^{2}-{\cal H}(r)\left(1+\frac{L_{2}^{2}}{r^{2}+n^2}\right)}
$$
For simplicity, ${\cal E}_{1}={\cal E}_{2}=1$ and reverting back the value of ${\cal H}(r)$,
we obtain the CM energy near the event horizon ($r_{+}$) of TN space-time:
\begin{eqnarray}
{\cal E}_{cm}\mid_{r\rightarrow r_{+}} &=& \sqrt{2}m_{0}
\sqrt{\frac{4(r_{+}^{2}+n^2)+(L_{1}-L_{2})^{2}}{2(r_{+}^{2}+n^2)}} ~.\label{cmeh}
\end{eqnarray}
and for Cauchy horizon ($r_{-}$) the CM energy is 
\begin{eqnarray}
{\cal E}_{cm}\mid_{r\rightarrow r_{-}} &=& \sqrt{2}m_{0}
\sqrt{\frac{4(r_{-}^{2}+n^2)+(L_{1}-L_{2})^{2}}{2(r_{-}^{2}+n^2)}} ~.\label{cmch}
\end{eqnarray}
where $r_{\pm}$ is defined previously. It implies that the CM energy is finite and depends upon the 
values of angular momentum parameter. It also suggests that the CM energy depends upon the NUT 
parameter. It  may also playing a key role in the BSW effect as we have seen from the above expression.

In the limit $n=0$, the above expression reduces to the following form
\begin{eqnarray}
{\cal E}_{cm} &=& \sqrt{2}m_{0}\sqrt{\frac{16M^{2}+(L_{1}-L_{2})^{2}}{8M^{2}}}~.\label{cmsch}
\end{eqnarray}
which is exactly CM energy of the Schwarzschild BH found by BSW in \cite{bsw}. 

Now see what happens in case of mass-less case and \emph{extreme} case? First we consider the 
mass-less case, in this case the CM energy is given by 
\begin{eqnarray}
{\cal E}_{cm}\mid_{r\rightarrow r_{\pm}} &=& \sqrt{2}m_{0}\sqrt{\frac{8n^{2}+(L_{1}-L_{2})^{2}}{4n^{2}}} ~.\label{cmn}
\end{eqnarray}
It indicates that the CM energy is finite depends on the NUT parameter.
More interesting case i.e. the extreme case where the CM energy is given by 
\begin{eqnarray}
{\cal E}_{cm}\mid_{r\rightarrow M} &=& \sqrt{2}m_{0}
\sqrt{\frac{4(M^{2}+n^2)+(L_{1}-L_{2})^{2}}{2(M^{2}+n^2)}} ~.\label{cmx}
\end{eqnarray}
Since we know for extreme case, $r_{+}=r_{-}=M$ and $M^2+n^2=0$. Therefore we find for 
extreme TN BH: 
\begin{eqnarray}
{\cal E}_{cm}\mid_{r\rightarrow M} \rightarrow \infty  ~.\label{cmx1}
\end{eqnarray}
i.e the CM energy is \emph{divergent}. This is an amusing result. Because we first reported a 
non-asymptotic flat, spherically symmetric and stationary extreme BH spacetime possesses such 
properties.


\end{document}